\documentstyle[11pt]{article}
\def\lg{{\mathchoice{~\raise.58ex\hbox{$<$}\mkern-14.8mu\lower.52ex\hbox{$>$}~}
	            {~\raise.58ex\hbox{$<$}\mkern-14.8mu\lower.52ex\hbox{$>$}~}
		    {\raise.59ex\hbox{{$\scriptscriptstyle <$}}\mkern-12.8mu%
		     \lower.01ex\hbox{{$\scriptscriptstyle >$}}}   {} 	}}
\def\gl{{\mathchoice{~\raise.58ex\hbox{$>$}\mkern-12.8mu\lower.52ex\hbox{$<$}~}
                    {~\raise.58ex\hbox{$>$}\mkern-12.8mu\lower.52ex\hbox{$<$}~}
		    {\raise.62ex\hbox{{$\scriptscriptstyle >$}}\mkern-12.0mu%
		     \lower.05ex\hbox{{$\scriptscriptstyle <$}}}  {} 	}}

\date{}

\begin{document}

\setcounter{page}{0}
\thispagestyle{empty}

\begin{flushright}
CERN-TH. 95-181\\
\end{flushright}
\vspace{0.5cm}

\begin{center}
{\LARGE
{\bf QUANTUM FIELD KINETICS OF QCD:}
}
\bigskip

{\Large
{\bf Quark-Gluon Transport Theory for Light-cone Dominated Processes}
}

\end{center}
\bigskip

\begin{center}

{\Large
{\bf  Klaus Geiger}
}

{\it CERN TH-Division, CH-1211 Geneva 23, Switzerland}
\end{center}
\vspace{0.5cm}

\begin{center}
{\large {\bf Abstract}}
\end{center}
\smallskip

\noindent
{\small
A quantum kinetic formalism is developed to study the dynamical
interplay of quantum and statistical-kinetic properties of
non-equilibrium multi-parton systems produced in high-energy QCD
processes. The approach provides the means to follow the quantum
dynamics in both space-time and energy-momentum, starting from an
arbitrary initial configuration of high-momentum quarks and gluons.
Using a generalized functional integral representation and adopting
the `closed-time-path' Green function techniques, a self-consistent
set of equations of motions is obtained: a Ginzburg-Landau equation
for a possible color background field, and Dyson-Schwinger equations
for the 2-point functions of the gluon and quark fields. By exploiting
the `two-scale nature' of light-cone dominated QCD processes, i.e. the
separation between the quantum scale that specifies the range of
short-distance quantum fluctuations, and the kinetic scale that
characterizes the range of statistical binary interactions, the
quantum-field equations of motion are converted into a corresponding
set of  `renormalization equations' and `transport equations'. The
former describe renormalization and dissipation effects through the
evolution of the spectral density of individual, dressed partons,
whereas the latter determine the statistical occurrence of scattering
processes among these dressed partons. The renormalization equations
and the transport equations are coupled, and hence must be solved
self-consistently. This amounts to evolving the multi-parton system,
from a specified initial configuration, in time and full 7-dimensional
phase-space, constrained by the Heisenberg uncertainty principle.
This quantum-kinetic description provides a probabilistic interpretation
and is therefore of important practical value for the solution of the
dynamical equations of motion, suggesting for instance the possibility
of simulating the multi-particle dynamics with Monte Carlo methods.
}

\vspace{0.5cm}

\rightline{klaus@surya11.cern.ch}
\leftline{CERN-TH. 95-181, July 1995}

\newpage

\noindent {\bf 1. INTRODUCTION}
\bigskip

In this paper I attempt to formulate an approach towards
a fundamental and consistent description of the  statistical properties
of non-equilibrium quantum systems produced in high-energy QCD processes,
which allows to follow the quantum dynamics in time and complete
phase space starting from an initial configuration.
It provides a flexible framework for a systematic analysis of typical
problems associated with the quantum dynamics of such systems, including,
e.g., multi-particle transport phenomena of gluons, quarks
and hadrons, or, critical dynamics of phase-transition phenomena and
spontanous symmetry breaking, or, quantum dissipation,  entropy
generation and multi-particle production.

More specifically, the intentions are aimed towards a
practically applicable description of the space-time evolution
of a general  initial system of gluons and quarks, characterized by
some large energy or momentum scale, that expands, diffuses and dissipates
according to the self- and mutual-interactions, and eventually
converts dynamically into excited hadronic matter and a final state
hadron system by a ``phase transition''.
This scenario frames a wide class of
QCD processes of both fundamental and phenomenological interest.
For instance, the evolution of parton showers in the mechanism
of parton-hadron conversion in elementary high-energy processes
($e^+e^-$-annihilation into hadrons, deep-inelastic lepton-nucleon scattering,
or non-diffractive hadronic collisions), or, the description of
formation, evolution and freeze-out of a quark-gluon plasma in
ultra-relativistic heavy-ion collisions, or,
the study of the dynamics of the QCD phase-transition from the
deconfined, high-temperature partonic phase to a
low temperature hadronic phase with the simultanous breakdown of
chiral symmetry and the condensation of gluons and quarks
in the vacuum, as it occurred during the early evolution of the
Universe.

In the present paper I will confine myself to the first stage,
the high-energy quark-gluon phase, and develop a
quantum kinetic formalism that allows to describe both the
dissipative and dispersive dynamics of a multi-parton system
in real time. This description is exclusively based on the
fundamental QCD Lagrangian and its firmly established principles.
The second stage, the parton-hadron conversion and
phase transition, on the other hand, requires
supplementary phenomenological input to model the details of the confinement
mechanism that are not known at present \cite{webber}.
Such a phenomenological approach to the real-time dynamics of
parton-hadron conversion that models the transition within an
effective field theory description has been proposed recently in Ref.
\cite{ms37}.
It is preferable, however,  to keep the fundamental description of the first
stage distinct
from the less understood phenomenological aspects of the second stage,
and therefore I will address the latter in a separate paper.

In general, the study of a high-energy multi-particle system and its
quantum dynamics involves three essential aspects:
first, the aspect of space-time, geometry and
the structure of the vacuum; second, the quantum
field aspect of the particle excitations; and third, the statistical
aspect of their interactions. These three elements are generally
interconnected in a non-trivial way by their overall dynamical
dependence.
Therefore, in order to formulate a  quantum description
of the complex non-equilibrium dynamics, one needs to find a
quantum-statistical and kinetic formulation of field theory that unifies
the three aspects self-consistently.
With this paper I take steps towards this goal by combining
three corresponding theoretical methods, namely, first, the
{\it closed-time-path} (CTP) formalism
\cite{schwinger,keldysh,baym,chou,calzetta,rammer} (for treating initial
value problems of irreversible systems), second, the {\it non-local
source theory}  \cite{dominicis,cornwall,jordan}
(for incorporating quantum fluctuations),
and third, {\it transport theory} based on Wigner function techniques
\cite{wigner,hillery,degroot}
(for a kinetic description of inhomogenous non-equilibrium systems).
In principle, a dynamical theory of non-equilibrium multi-particle
systems as the above mentioned, should be described by an
exact quantum kinetic theory of QCD. Over the past 10 years, elaborate
works \cite {elze86,elze89,elze90}
have put great effort
into deriving a general QCD transport theory rigorously
from first principles. Unfortunately, due to a
number  of unresolved problems arising from the complexities
of the non-abelian gauge structure of QCD, the derived
gauge-covariant formalism remains an academic theory up to date.
It is of little practical value,
unless it is boiled down to the quasi-classical limit by
a series of approximations yielding a mean-field description,
which however cannot describe the production of physical particles
and their spectra.

I am  less ambitous here in what concerns the generality, and
instead put emphasis on applicability to realistic physical situations,
in particular to the type of lightcone-dominated processes
that I classified above.
This class of high-energy processes allows a clear distinction between
a short-distance quantum field theoretical scale and a
larger distance statistical-kinetic scale.
When described in a reference frame, in which the particles move close
to the speed of light,
the effects of time dilation and Lorentz contraction
separate the intrinsic quantum motion of the individual
particles from the  statistical correlations among them.
On the one hand,
the quantum dynamics is
determined by the self-interactions of the bare quanta, and by the possible
presence of a coherent background field (or mean field in the
Hartree-Fock sense),
in case one desires to go beyond a description in the pure vacuum.
This requires a fully quantum theoretical analysis including renormalization.
On the other hand, the kinetic dynamics
can well be described statistical-mechanically
by the motion of the quasi-particles
which arise from the `dressing' of the bare quanta by
their self-interactions and by the background field, plus the binary
interactions between these quasi-particles.
Such a distinct description of quantum and kinetic dynamics
is possible, because the quantum
fluctuations are highly concentrated around the light cone,
occurring at very short distances, and decouple to very
good approximation from the kinetic evolution
which is dictated by comparably large space-time scales.
As mentioned,
the natural two-scale separation is just the consequence of time dilation
and Lorentz contraction, and is true for any lightcone
dominated process. In fact, at asymptotic energies the
quantum fluctuations are exactly localized on the lightcone, and
so the decoupling becomes perfect.
This observation is the key to formulate a quantum kinetic description
in terms of particle phase-space densities, involving
a simultanous specification of momentum space and space-time,
because at sufficiently high energy,
the momentum scale $\Delta p$ of the individual particles' quantum fluctutions
and the scale $\Delta r$ of space-time variations of the system of particles
satisfiy $\Delta p \Delta r \gg 1$, consistent with the uncertainty principle.

With this physical input
and utilizing the aforementioned theoretical tools, the analysis
proceeds as follows.
In the first step, covered in Sec. 2,
I obtain, starting from the QCD Lagrangian,
the CTP generating functional for the gluon and quark Green functions,
being defined on a closed-time contour and incorporating initial state
correlations.  From the associated effective action, one gets the quantum
dynamical equations of motion,  which are the CTP version of
the Ginzburg-Landau equation and the Dyson-Schwinger equations.
In the second step, described in Sec. 3, I make the transition from
quantum field description to kinetic theory, by exploiting
the two-scale nature of lightcone dominance, and
moreover, choosing a  ghost-free axial gauge for the gluon fields.
As a result one obtains from the Dyson-Schwinger equations a set of
kinetic equations, consisting of a {\it renormalization equation}, that
describes
the quantum dynamics in terms of short-distance
self-interactions of gluons and quarks, plus a {\it transport equation}
that describes the  kinetic dynamics of
relaxation and collision processes in terms of the statistical
interactions of the renormalized, dynamically dressed partons among each other.
The renormalization equation and the transport equations
are coupled, and hence must be solved
self-consistently. This amounts
to evolving the system under consideration from its
initial configuration simultanously in position- and momentum-space,
constrained by the Heisenberg uncertainty principle.
Finally, Sec. 4 closes with some concluding remarks,
and the Appendices summarize,
for each of the above aspects, the technical details which are
only indicated in the text.

The main findings can be  summarized as follows.
The dynamics of high-energy multi-parton systems
can, under reasonable conditions, be described in a
semi-classical manner: the partons can be considered as dressed
quanta with a dynamical substructure and a corresponding form factor
arising from the self-interactions.
The space-time evolution of a system of many such dressed partons
is then governed by their propagation along classical
trajectories and mutual binary collisions, as determined by
their density, cross-sections and by quantum statistics.
This emerging picture is of great practical value for
formulating a systematic  calculation scheme $-$
in a sense the space-time generalization of the `jet calculus'
\cite{konishi,basetto}.
In Sec. 3.6, I outline such a scheme.
One of the greatest advantages of this kinetic description
is that it provides a probabilistic interpretation
of the time evolution in full 7-dimensional phase-space, which suggests
the opportunity to simulate the multi-particle
dynamics as sequential Markov processes with Monte Carlo methods.
\bigskip

Finally let me comment on placing this work in relation
to existing literature.
\begin{description}
\item[(i)]
The general ideas and techniques
of the CTP functional integral formalism have originally been
introduced mainly by Schwinger \cite{schwinger},
Keldysh \cite{keldysh}, and by Kadanoff and Baym \cite{baym},
more than thirty years ago.
The most extensive review that sums up the current state of
the art is probably the work of Chou {\it et al.} \cite{chou},
with diverse exemplification of the wide class of
physics applications. Further
pedagogically excellent presentations have been published by
Calzetta and Hu \cite{calzetta}, and by Rammer and Smith \cite{rammer}.
In the particular field of relativistic nuclear physics,
the concepts have been pragmatically applied , e.g.,
by  Li and McLerran \cite{mclerran83},
and by Zhang and Wilets \cite{wilets92}. Important
contributions of fundamental studies have been made
in the last years by Danielewicz \cite{danielewicz},
and Mrowczinsky and Heinz \cite{mrow}.
The goal to establish a quantum kinetic theory for QCD
was pioneered by
the ambitious efforts of Elze, Gyulassy, Heinz, and Vasak
\cite{elze86,elze89,elze90},
which resulted in a rigidly general, gauge covariant formalism.
However, the prize to pay is an intractable complexity that, without
specific physics input, is essentially of aestethic value without much
practical use.
The new achievement of the present work from this perspective may be stated
as the adaption of the general CTP formalism, applied to QCD, but with
focus on situations where the multi-parton dynamics is characterized by
a large energy scale and can be described reliably
within perturbation theory in a physical gauge.
\item[(ii)]
The most related recent works from the
viewpoint of attempting to tackle  evolution of multi-parton systems
at high energy are probably the innovative works of
McLerran, Venugopalan, {\it et al.} \cite{mclerran94}, and of Makhlin
\cite{makhlin95},
in which the issue of calculating parton distributions in the context of
ultra-relativistic nuclear collisions is addressed.
The former authors use a classical non-abelian field description of QCD
to compute coherent initial state properties of colliding large nuclei,
whereas the latter focusses on a quantum field description
of final state correlations of produced particles. Both
approaches however do not attempt to address explicitly
the space-time evolution of the multi-parton ensemble emerging from
the nuclear collision,
which is the main goal of the present paper.
\item[(iii)]
The key elements to address the space-time evolution are
provided by the Wigner function techniques, which date back to
Wigner's work on transport phenomena \cite{wigner}, and
are reviewed in in e.g. \cite{hillery,degroot}.
Although widely exploited in condensed matter and plasma physics,
these tools for quantum kinetics of many-body systems have
hardly been applied to describe high energy non-equilibrium
dynamics in QCD.
New in the present work is the synthesis of
quantum dynamics on the basis of the renormalization group of QCD, and
quasi-particle kinetics within relativistic transport theory.
The combination of these two aspects forms the foundation of the
self-consistent treatment that entails a thorough
consideration of the renormalization problem, which is
commonly avoided in other applications (see however Ref. \cite{calzetta}).
\item[(iv)]
The machinery of perturbative QCD for light-cone
dominated high energy processes is nowadays well founded.
Most of the techniques used in the perturbative analysis to describe the
parton evolution, adopt the tools developed by Dokshitzer {\it et al.}
\cite{dok80}, Amati {\it et al.} \cite{amati80}, Mueller \cite{amueller81}, and
numerous others
(for an overwiew see \cite{muellbook,dokbook}).
The new component here is the extension to incorporate a space-time
description on top of this formalism, which is commonly considered only in
momentum space.
\end{description}
It is evident that this paper attempts to join  theoretical
tools and concepts from rather different fields. Such a synthesis
is necessarily a difficult task, and the the present initiative
should be viewed as a first step in this direction. However, I believe that it
is a
promising approach towards a well founded and consistent description
of the statistical properties of non-equilibrium
parton systems. From the phenomenological
perspective, it is an inevitable necessity to
address this problem, since
the experiments carried out at the HERA, RHIC and LHC accelerators
will penetrate increasingly the physics of high-density QCD, where
quark-gluon transport phenomena are of fundamental importance.
\bigskip
\bigskip

\noindent {\bf 2. FUNCTIONAL FORMALISM}
\bigskip

The aim is  to describe the time evolution of
a general non-equilibrium quantum system consisting of an ensemble
of quarks and gluons in phase space, starting from some
given inital state at time $t_0$.
Since I am interested in the state of the system at finite times $t>t_0$,
without a priori knowledge of the asymptotic final state at $t=\infty$,
the usual $S$-matrix formalism of quantum field theory, based on {\it in-out}
matrix elements, cannot be applied.
For initial value problems as I want to describe here,
the appropriate approach is provided by the functional integral
formalism of the {\it in-in} generating functional for the
Green functions of quarks and gluons, also referred to as
{\it closed-time-path} (CTP) Green functions.
The CPT formalism is a powerful Green function formulation, originally
introduced by Schwinger \cite{schwinger} and Keldysh \cite{keldysh}
for describing
general non-equilibrium phenomena in field theory
\cite{baym,chou,calzetta,rammer}.
In combination with the so-called  non-local source theory
and the loop expansion techniques developed
by de Dominicis and Martin \cite{dominicis},
and  Cornwall, Jackiw, and Tomboulis  \cite{cornwall},
one obtains generalized Dyson-Schwinger equations which incorporate
the initial state correlations and provide
a systematic treatment
of the quantum correlations to any order in $\hbar$. Furthermore
it allows to describe
phase-transition phenomena and dynamical symmetry breaking, issues that
I will not address here, but which are of central interest when
studying the confinement dynamics, as intended to be presented elsewhere.
In this Section, I will first review the concept of the
{\it in-in} generating functional and
the effective action for the CTP Green functions, and then derive
the dynamical equations of motion. For additional reading on
these techniques I refer to the extensive review of Chou {\it et al.}
\cite{chou}
and to the instructive work of Calzetta and Hu \cite{calzetta}.
\bigskip

\noindent {\bf 2.1  Preliminaries}
\bigskip

Starting point is the QCD Lagrangian given in terms of
the gluon fields $A_a^\mu$ and the  quark fields $\psi$, $\overline{\psi}$
(which are vectors in flavor space, $\psi\equiv (\psi_u,\psi_d,\ldots)$),
\begin{equation}
{\cal L}[A^\mu,\psi,\overline{\psi}] \;=\; -\frac{1}{4}\,\,F_{\mu\nu, a}
F^{\mu\nu}_a
\;+\;  \overline{\psi}_i \left[\frac{}{}\,(i \gamma_\mu \partial ^\mu
- \hat{m}) \delta_{ij}
- g_s  \gamma_\mu A^\mu_a T_a^{ij} \right]\, \psi_j
\;+\;
\xi_a(A^\mu)
\label{LQCD}
\;,
\end{equation}
where
$F_a^{\mu\nu}= \partial^\mu A_a^\nu -\partial^\nu A_a^\mu + g_s f_{abc} A^\mu_b
A^\nu_c$
is the gluon field-strength tensor. The subscripts $a, b, c$ label the
color components of the gluon fields,
and $g_s$ denotes the color charge related to  $\alpha_s =g_s^2/(4\pi)$.
The $T_a$ are the generators of the $SU(3)$ color group, satisfying
$[T_a,T_b] = i f_{abc} T_c$ with the structure constants $f_{abc}$.
The indices $i,j$ label the color components of the quark fields and
$\hat{m}\equiv \mbox{diag}(m_u,m_d,\ldots)$.
I will in the following exploit the fact
that high energies the quark current masses $m_f$ can be neglected,
which corresponds to the chiral limit where they are exactly zero.

In general the Lagrangian (\ref{LQCD}) must also include
the Fadeev-Popov ghosts as independent
field degrees of freedom. However,
I will work exclusively in a class
of ghost-free gauges, namely the so-called space-like axial gauges, which are
defined by the gauge condition \cite{gaugebook}
\begin{equation}
n_\mu\;A_a^\mu(x) \;\equiv\; n\cdot A_a \;=\; 0
\label{Enn}
\;,
\end{equation}
where $n^\mu$ a constant four-vector in the $x^0-x^3$ plane near the
forward lightcone such that
$n^2 > 0$.
It may be parametrized, e.g., as $n^\mu=(a+b,0,0,a-b)$, with
the condition $n^2 = 4ab \ll 1$.
The associated gauge-fixing term is denoted by a  function
$\xi_a(A^\mu)$ which I take as
\begin{equation}
\xi_a(A^\mu)\;=\;
-\;\frac{1}{2\alpha n^2}\;\partial_\lambda (n\cdot A_a)\partial^\lambda (n\cdot
A_a)
\;.
\label{gauge}
\end{equation}
Here $\alpha$ is the gauge parameter that specifies the  type of
axial gauge.
In particular, I will henceforth set $\alpha=1$ which is known as the planar
gauge.
In contrast to covariant gauges where $\xi_a(A)=-1/(2\alpha) (\partial\cdot
A_a)^2$,
the class of gauges (\ref{gauge}) is well known to have a number of
advantages \cite{dok80,gaugebook}.
First, the ghost fields decouple from the gluon field and drop out.
Second, the so-called Gribov ambiguity is not present in this gauge.
Third, the gluon propagator involves only the
two physical transverse polarizations, which will simplify the analysis
considerably.
Furthermore, it allows for a rigorous resummation of the perturbative series
at high energies in terms of the leading logarithmic contributions and
consequently leads to a
simple probabilistic description of the perturbative parton evolution within
the (Modified) Leading Log approximation (MLLA) \cite{dok80,amati80,dokrevmod}
in QCD.

The classical action corresponding to (\ref{LQCD}) is represented as
\begin{equation}
I[A^\mu,\psi,\overline{\psi}] \;=\;
\;\,\equiv\;\,
I^{(0)}[A^\mu] \;+\; I^{(0)}[\psi,\overline{\psi}]\;+\;
I^{(int)}[A^\mu,\psi,\overline{\psi}]
\;,
\label{I1}
\end{equation}
where
\begin{eqnarray}
I^{(0)}[A^\mu] &=&
\int d^4x d^4 y \left\{\frac{}{}
-\frac{1}{2}\, A_a^\mu(x)\,  \left[ D_{(0)\,\mu\nu}^{ab}(x,y)\right]^{-1}\,
A_b^\nu(y)
\right\}
\nonumber \\
I^{(0)}[\psi,\overline{\psi}]&=&
\int d^4x d^4 y \left\{\frac{}{}
\overline{\psi}_i (x)\, \left[S_{(0)}^{ij}(x,y)\right]^{-1}\, \psi_j(y)
\right\}
\nonumber \\
I^{(int)}[A^\mu,\psi,\overline{\psi}]&=&
-\;\int d^4x \left\{\frac{}{}
g_s\,  \gamma_\mu T_a^{ij} \overline{\psi}_i(x)A^\mu_a(x)\psi_j(x)
\;+\;
g_s\, f_{abc} [\partial_\mu A_{\nu,a}(x)] A^\mu_b(x) A^\nu_c(x)
\right.
\nonumber \\
& &
\left.\frac{}{}
\;\;\;\;\;\;\;\;\;\;\;
+\;
g_s^2\, f_{abc} f_{ab'c'}
A_{\mu,b}(x) A_{\nu, c}(x) A^\mu_{b'}(x) A^\nu_{c'}(x)
\right\}
\;,
\label{I11}
\end{eqnarray}
with the kernels of the free parts $I^{(0)}[A^\mu]$ and
$I^{(0)}[\psi,\overline{\psi}]$
given by
\begin{eqnarray}
\left[ D_{(0)\,\mu\nu}^{ab}(x,y)\right]^{-1}&=&
\delta_{ab} \,\delta^4(x-y)
\;\Box^{\mu\nu}_x
\nonumber
\\
\left[S_{(0)}^{ij}(x,y)\right]^{-1} &=&
\delta_{ij} \,\delta^4(x-y)
\;i \gamma\cdot \partial_x
\;,
\label{prop1}
\end{eqnarray}
where the quark current masses are set to zero here and in the following.
The operator $\Box^{\mu\nu}_x$ is a generalized D'Alembertian
containing the remnant of the gauge fixing term
of (\ref{LQCD}), which for the gauge (\ref{gauge})
with $\alpha=1$ reads
\begin{equation}
\Box^{\mu\nu}_x
\;\equiv\;
\left( g^{\mu\nu} \;-\;
\frac{n^\mu\partial_x^\nu+ n^\nu\partial_x^\mu}{n\cdot\partial_x}
\right)
\, \Box_x
\;,
\label{Box}
\end{equation}
with
$\Box_x = \partial_x\cdot \partial_x$, $\partial_x^\mu = \partial/\partial
x^\mu$.
The inverses of (\ref{prop1}) are the free gluon and quark Feynman propagators,
i.e. the expectation values of the time-ordered products of the free fields
$-i\langle T A_\mu(x)A_\nu(y)\rangle_{(0)}$  and
$-i\langle T \psi(x)\overline{\psi}(y)\rangle_{(0)}$,
\begin{eqnarray}
D_{(0)\,\mu\nu}^{ab}(x,y)
&=&
\int \frac{d^4 k}{(2\pi)^4} \,
e^{-i \,k\cdot (x-y)}\;
\;\delta_{ab}\;\frac{-d_{\mu\nu}(k)}{k^2+ i \epsilon}
\;\;, \;\;\;\;\;\;\;\;\;
d_{\mu\nu}(k)=
g_{\mu\nu} - \frac{n_\mu k_\nu+n_\nu k_\mu}{n\cdot k}
\nonumber
\\
S_{(0)}^{ij}(x,y)
&=&
\int \frac{d^4 p}{(2\pi)^4} \,
e^{-i \,p\cdot (x-y)}\;
\;\delta_{ij}\;\frac{1}{\gamma \cdot p + i \epsilon}
\;.
\label{prop2}
\end{eqnarray}
It is noteworhty that the form of the gluon propagator
\footnote{
The apparent singularity of $d_{\mu\nu}(k)$ at $n\cdot k \simeq k^+ = 0$
must be dealt with the usual $i\epsilon$-prescription, or by
taking the principal value.}
arises from
the sum over the two transverse gluon polarizations,
$d_{\mu\nu}(k)=\sum_{s=1,2}\varepsilon_\mu(k,s) \cdot
\varepsilon_{\nu}^\ast(k,s)$,
having the properties \cite{dok80,gaugebook}
\begin{equation}
d_\mu^\mu(k)\;=\;2
\;,\;\;\;\;\;\;\;\;
k_\mu\; d^{\mu\nu}(k)\;=\;-\,\frac{n^\nu\,k^2}{n\cdot k}
\;\stackrel{k^2\rightarrow 0}{\longrightarrow} \;0
\;,
\label{transpol}
\end{equation}
meaning that only the two physical polarization states propagate,
with $\varepsilon_\mu k^\mu =0$. For comparison, in the covariant Feynman
gauge, $d^{\mu\nu}=g^{\mu\nu}$, $d_\mu^\mu=4$, and
$k_\mu d^{\mu\nu}=k^\nu \ne 0$.
\medskip

In going over from the classical action (\ref{I1}) to a quantum field
formulation, the fields become Heisenberg operators.
Let me introduce a compact notation
for the different field degrees of freedom $f$:
\footnote{
Since the quarks and antiquarks are treated as massless here, the
different quark flavors are, with respect to the strong interaction,
merely copies of each other.
}
\begin{equation}
\phi_f\;:=\; (\;A^\mu, \psi, \overline{\psi}\,)
\;=\; (\;A^\mu, \psi_u, \overline{\psi}_u ,\psi_d, \overline{\psi}_d, \ldots\,)
\;,\;\;\;\;\;\;\;\;\;\;\;\;\;
f\;=\; g, u,\bar u, d, \bar d, \ldots
\label{notation1}
\;.
\end{equation}
The state of the system
may be characterized by the Heisenberg field operator $\Phi_H(x)$, where
$\Phi_H \equiv \Phi_H[\phi_f]$
and $x=(t,\vec x)$. Its time evolution is determined by the Hamiltonian
$H= H^{(0)}+H^{(int)}$
of the system ($\partial_t\equiv \partial /\partial t$),
\begin{equation}
\partial_t \,\Phi_H(x) \;=\; i\,\left[ H,\,\Phi_H(x)\right]_-
\;.
\end{equation}
Defining $t = t_0$ as the initial point for the time evolution of the system,
the associated Heisenberg state vectors obey
\begin{equation}
|\,\phi (t)\,\rangle \;=\; U_J(t,t_0)\; |\,\phi (t_0)\,\rangle
\label{U1}
\;,
\end{equation}
where
\begin{equation}
U_J(t,t_0)\; \;\equiv\; T \,\exp\left[ -i \int_{t_0}^t dt'd^3x'\,J(x')
\Phi_H(x')\right]
\label{U2}
\;
\end{equation}
$T$ denotes the usual time ordering operator, and the external source
$J$ is
understood as a sum over sources for the various degrees of freedom.
Note that the adjoint
$
U_J^\dagger(t,t_0)= T^\dagger \exp\left[ i \int_{t_0}^t d^4x'J(x')
\Phi_H(x')\right]
$
involves an anti-temporal ordering  $T^\dagger$.
In the absence of external sources, the state vectors are time
independent: $|\phi(t)\rangle=|\phi (t_0)\rangle$.

Upon switching from the Heisenberg picture to the interaction
picture, the time evolution of the corresponding interaction
picture field $\Phi_I(x)$ is determined by the interaction Hamiltonian
$H^{int}$ alone,
\begin{equation}
\partial_t \,\Phi_I(x) \;=\; i\,\left[ H^{(int)},\,\Phi_I(x)\right]_-
\;,
\end{equation}
where $\Phi_I$ is related to the Heisenberg $\Phi_H$ field by
\begin{equation}
\Phi_I(x) \;=\; S(t,t_0)\; \Phi_H(x) \; S^\dagger(t,t_0)
\label{S1}
\; ,
\end{equation}
and  evolves explicitly in time through
\begin{equation}
S(t,t_0)\; \;\equiv\; T \,\exp\left[ -i \int_{t_0}^t dt'H^{(int)}(t')\right]
\label{S2}
\; .
\end{equation}

According to (\ref{U1})-(\ref{S2}),
at $t=t_0$ the Heisenberg picture and the interaction picture
coincide, $\Phi_H(t_0,\vec x) = \Phi_I(t_0,\vec x)$.
Hence, the interaction picture field $\Phi_I(x)$ can be expanded
at $t_0$ in terms of a Fock basis of free particle states,
the {\it in}-basis,
\begin{equation}
\Phi_I(x)\;=\;
\sum_{f=g,q,\bar q}
\int \frac{d^4p}{(2\pi)^4}\,\theta(p^0)\,(2\pi)\delta(p^2)\;
\sum_s
\left(e^{-ip\cdot x}\,a_f(p,s)\;+\;e^{ip\cdot x}\,a^\dagger_f(p,s)\right)
\end{equation}
\begin{equation}
|\,n^{(1)}, n^{(2)}, \ldots , n^{(\infty)}\,\rangle \;=\;
\prod_{f}\;
\prod_{i}\;
\frac{1}{\sqrt{n_f^{(i)} !}}\;
\left(\frac{}{}a_f^\dagger (p_i,s_i)\right)^{n_f^{(i)}}
\;|\,0\,\rangle
\;,
\end{equation}
where
the $a_f^\dagger$ ($a_f$) are the corresponding creation
(destruction) operators for the particle types $f=g,q,\bar q$
with definite momentum $p_i$ and spin $s_i$, and
the $n_f^{(i)}$ are the occupation numbers of the particle states,
and
\begin{equation}
a_f (p_i,s_i)\,|\,0\,\rangle \;=\;0\;\;,\;\;\;\;\;\;
n_f^{(i)}\;=\;
\langle\,n_f^{(i)}\,|\; a_f (p_i,s_i)\, a^\dagger_f (p_i,s_i)\,|\,n_f^{(i)}\,
\rangle
\;.
\end{equation}
Thus, a general multi-parton state $|\phi\rangle$ at time $t_0$ is given by a
superposition
of such states,
\begin{equation}
|\,\phi (t_0) \,\rangle\;=\;
\sum_{n^{(i)}} \,
C(n^{(1)}, n^{(2)}, \ldots , n^{(\infty)})\;
|\,n^{(1)}, n^{(2)}, \ldots , n^{(\infty)}\,\rangle
\;,
\end{equation}
with scalar coefficients $C$.
Alternatively,
the initial state of the system at  $t_0$ can be characterized by
the statistical operator, or {\it density matrix},
\begin{equation}
\hat{\rho}(t_0) \;=\; |\, \phi (t_0)\,\rangle \,\langle \,\phi (t_0)\,|
\;\;\;\;\;\;\;\;\;
\left(\,\hat{\rho}_0\,\right)_{ij} \;\equiv\; \langle \,n^{(i)}\,|\,
\hat{\rho}(t_0) \,|\,n^{(j)} \,\rangle
\label{rho}
\;,
\end{equation}
which in the Heisenberg representation is time independent, but in the
interaction
picture evolves with time according to
\begin{equation}
\partial_t \,\hat{\rho} \;=\; i\,\left[ H^{(int)},\,\hat{\rho}\right]_-
\;,
\end{equation}
so that
\begin{equation}
\hat{\rho}(t) \;=\; S^\dagger(t,t_0) \,\hat{\rho}(t_0)\,S(t_0,t)
\;,
\end{equation}
where $S$ is defined by (\ref{S2}).
For instance, a general density matrix that describes any form
of a single-particle density distribution at $t_0$ is
\begin{equation}
\hat{\rho}(t_0)\;=\; N\;\exp\left[ \sum_{f,\,s} \int_\Omega d^3x \int \frac{d^3
p}{(2\pi)^3 2p^0}
\,F_f(t_0,\vec x,p)\;
a_f^\dagger (p,s) a_f (p,s)
\right]
\;,
\label{rho1}
\end{equation}
where $\Omega$ denotes the hypersurface of the initial values and $F_f$ is
a $c$-number function related to the single-particle phase-space density of
particle species $f$ at $\vec x$ with four-momentum $p$, and $N$
a normalization factor.
\bigskip

\noindent {\bf 2.2  The CTP generating functional}
\medskip

After these preliminaries let me turn now to describe the time development
of the multi-parton state from the initial state
$|\phi^{in}\rangle = |\phi(t_0)\rangle$,
continously through finite intermediate times $t_0 < t < t_\infty$,
to some final state
$|\phi^{out}\rangle = |\phi(t_\infty)\rangle$ in the remote future (see Fig.
1).
In the usual $S$-matrix formalism of quantum field theory
one calculates the {\it in}-vacuum to {\it out}-vacuum
amplitude $Z[J] = \langle 0^{in}|0^{out}\rangle_J$, and from this,
physical quantities corresponding to {\it in-out} $S$-matrix elements of
certain operators, assuming that the Fock space of the asymptotic $out$-states
is
the same as for the $in$-states (Fig. 1a), as e.g. in scattering theory.
In the present case, however, the  system evolves forward through finite points
of time,
and so the asymptotic {\it out}-basis
$|\phi^{out}\rangle$ is not known before the solution to the problem.
There is an arrow of time, leading to an irreversible evolution.
Moreover, in general $|0^{in} \rangle \ne|0^{out}\rangle$,
as for instance in the case of a phase transition or spontanous symmetry
breaking
where {\it in}- and {\it out}-vacua are of different nature.

These problems can be overcome by using the CTP-formalism based on {\it in-in}
rather
than {\it in-out} matrix elements \cite{chou,jordan}, but otherwise uses the
familiar
techniques of the path-integral method for quantizing the theory.
The {\it in-in} generating functional
is defined as the $in$-vacuum to $in$-vacuum amplitude
$Z[J,\hat{\rho}] = Tr \sum_\varphi\langle 0^{in}|\varphi\rangle_J
\langle \varphi|\hat\rho|0^{in}\rangle_J$,
including possible initial state correlations represented by the density matrix
$\hat{\rho}$ at $t_0$, and a sum over a complete set of states
$\varphi$ at $t_\infty$ (Fig. 1b).
With reference to Appendix A, where
the relevant concepts are reviewed and applied
to the case of QCD, I merely state here the resulting path-integral
representation
for the {\it in-in}, or {\it CTP  generating functional}. It is given by the
following path integral representation in 2-point source approximation:
\begin{eqnarray}
Z_P[J^\mu,j,\overline{j}, K^{\mu\nu}, k] &=&
e^{i \,W_P[J^\mu,j,\overline{j}, K^{\mu\nu}, k]}
\nonumber \\
&=&
\int \,{\cal D} A_\alpha^\mu{\cal D} \psi_\alpha{\cal D}
\overline{\psi}_\alpha\,
\;\exp \left[i \left( \frac{}{}
I[A^\mu_\alpha,\psi_\alpha,\overline{\psi}_\alpha]
\right.
\right.
\label{Z000}
\\
& &
\left.
\left.
\frac{}{}
\;+\;\;
J_\mu^\alpha A_\alpha^\mu \;+\; j^\alpha
\overline{\psi}_\alpha\;+\;\overline{j}^\alpha \psi_\alpha
\;+\;
\frac{1}{2} \,A_\alpha^\mu K_{\mu\nu}^{\alpha\beta} A_\beta^\nu
\;+\;
\overline{\psi}_\alpha k^{\alpha\beta} \psi_\beta
\right)
\right]
\nonumber
\;.
\end{eqnarray}
where I introduced a shorthand notation for the integration over the space-time
variables
to be understood in the functional sense,
\begin{equation}
J\,\phi \;\equiv \; \int_P d^4x \,J(x)\;\phi(x)
\;\;,\;\;\;\;\;
\phi\,K\,\phi \;\equiv \; \int_P d^4x \,\phi(x)\;K(x,y)\;\phi(y)
\;.
\label{conv0}
\end{equation}
The CTP generating functional (\ref{Z000})
differs from the usual generating functional of QCD in two essential
aspects:

First, it contains both, {\it local} sources ($J,j,\overline{j}$) and
{\it non-local} 2-point sources ($K,k$). The former represent
not only the usual external source contribution $J_{ext}(x)$, but also
the source term local source term $\tilde{K}(x)$ for a possible dynamical
background field present already
at initial point $t=t_0$, that is, $J=J_{ext}+\tilde{K}$, and similarly for
$j,\bar j$.
The non-local sources $K(x,y), k(x,y)$, on the other hand,
represent the 2-particle initial state correlations at $t=t_0$.
Both these source contributions
\footnote{
In the 2-point sources approximation, the actually infinite series of
non-local $n$-point sources that generate $n$-particle correlations,
is truncated beyond $n=2$ (c.f. Appendix A).
}
stem from the general non-trivial density matrix $\hat{\rho}(t_0)$ that
defines the initial ground state. In the usual field theory formulation
both these source terms are absent.
As a consequence,
the {\it connected} generating functional
$W_P= -i\,\ln Z_P$ in (\ref{Z000}) gives both
the non-local {\it connected} Green functions gluons
and quarks, including initial state correlations
(denoted by $D_{\mu\nu}(x,y)$, respectively $S(x,y)$),
as well as  possible local  mean fields
which physically can arise either through non-vanishing external sources,
or, in the case of gluons, may be  generated
dynamically by the system itself depending on the initial conditions
(denoted in the following by $\tilde{A}_\mu(x)$).

Second, the CTP functional $Z_P$ is defined on a
{\it closed-time path} in the
complex $t$-plane (indicated by the subscript $P$).
This path $P$ for the time integration is illustrated in Fig. 2a:
the path goes forward from $t_0$ to $t_\infty$ on the positive branch,
and then back from $t_\infty$ to $t_0$ on the negative branch.
Accordingly the generalized time-ordering $T_P$ is defined such that
any point on the negative branch is understood at a later instant
than any point on the positive branch.
This is not merely a mathematical trick
to restore analogy with usual quantum field theory, but
provides the means to compute expectation values for physical observables at
finite time
in contrast to the $S$-matrix formalism.
The interpretation of this closed-time path is simple:
although for physical observables the time values
are on the positive branch, both positive and negative
branches will come into play at intermediate steps in
a self-consistent calculation, corresponding to a
quantum mixing of positive and negative energy solutions.
Therefore, in contrast to the usual path-integral formulation of quantum field
theory,
the non-local 2-point Green functions $D_{\mu\nu}(x,y)$ and $S(x,y)$
for gluons and quarks, respectively,
come each in four different forms
corresponding to the possible time orderings
$\alpha\beta=++,+-,-+,--$ along the closed-time path  $P$, as illustrated in
Fig. 2b.
In as much as the propagators $D_{\mu\nu}(x,y)$ and $S(x,y)$
can have values $x$ and $y$ on
either the positive  branch or the negative  branch on the
contour $P$, it is convenient to represent them
2$\times$2 matrices $G(x,y)\equiv D_{\mu\nu}, S$ with components
$G^{\alpha\beta}$
(a convention which holds for any 2-point function defined
along the closed-time path $P$),
\begin{equation}
G(x,y)\;=\;
-i\,\langle \,T_P \phi(x) \phi^\dagger (y) \,\rangle
\;\equiv\;
\left(
\begin{array}{cc}
G^{++}\;  & \; G^{+-} \\
G^{-+}\;  & \; G^{++}
\end{array}
\right)
\;\,\equiv \;\,
\left(
\begin{array}{cc}
G^{F}\;  & \; G^{>} \\
G^{<}\;  & \; G^{\bar F}
\end{array}
\right)
\;,
\label{G22}
\end{equation}
where
$\langle \ldots \rangle \equiv\langle\phi^+,t_0| \ldots |\phi^- ,t_0\rangle$
denotes the vacuum expectation value,
if $| \phi, t_0 \rangle = | 0\rangle$, or else
the appropriate ensemble average. The
generalized time-ordering operator $T_P$ is defined as
$ T_P\,A(x)B(y):=
\theta (x_0,y_0) A(x) B(y) \pm \theta(y_0,x_0) B(y) A(x) $,
where the  $+$($-$) sign refers to  boson (fermion) operators,
and
$\theta(x_0,y_0)\equiv 1\,(0)$ if $x_0 > y_0$ ($x_0 < 0$) on $P$.
Hence, $T_P$ coincides with the usual temporal ordering $T$ on the positive
branch
$(t_0\rightarrow t_\infty)$ of the closed time path in Fig. 2, but represents
anti-temporal ordering $T^\dagger$ on the negative branch $(t_\infty\rightarrow
t_0)$.
The notation on the right hand side expresses that
$G^F$ is the usual Feynman causal propagator, $G^{\overline{F}}$
is the corresponding anti-causal propagator, and $G^>$ ($G^<$) is
the correlation function for $x_0 > y_0$ ($x_0 < y_0$). Explicitly,
\begin{eqnarray}
D_{\mu\nu}^F(x,y)&=&
-i\,
\langle \;T\, A_\mu(x)\, A_\nu(y) \;\rangle
\;\;\;\;\;\;\;\;\;\;\;\;\;\;\;
D_{\mu\nu}^>(x,y)\;=\;
+i\,
\langle \; A_\nu(y)\, A_\mu(x) \;\rangle
\nonumber \\
D_{\mu\nu}^<(x,y)
&=&
-i\,
\langle \; A_\mu(x)\, A_\nu(y) \;\rangle
\;\;\;\;\;\;\;\;\;\;\;\;\;\;\;\;\;
D_{\mu\nu}^{\overline{F}}(x,y)\;=\;
-i\,
\langle \;T^\dagger\, A_\mu(x)\, A_\nu(y) \;\rangle
\label{D22}
\;,
\end{eqnarray}
and
\begin{eqnarray}
S^F(x,y)&=&
-i\,
\langle \;T\, \psi(x)\, \overline{\psi}(y) \;\rangle
\;\;\;\;\;\;\;\;\;\;\;\;\;\;\;\;\;\;\;\;
S^>(x,y)\;=\;
-i\,
\langle \; \overline{\psi}(y)\, \psi(x) \;\rangle
\nonumber \\
S^<(x,y)
&=&
-i\, \langle \; \psi(x)\,\overline{\psi}(y) \;\rangle
\;\;\;\;\;\;\;\;\;\;\;\;\;\;\;\;\;\;\;\;\;\;
S^{\overline{F}}(x,y)\;=\;
-i\,
\langle \;T^\dagger\, \psi(x)\, \overline{\psi}(y) \;\rangle
\;.
\label{S22}
\end{eqnarray}

The CTP generating functional $Z_P = \exp( i\,W_P)$,
eq. (\ref{Z000}), is the fundamental starting point
for deriving the dynamical equations of motion for
both  gluon mean field $\tilde{A}^\mu$, and the
dressed gluon- and quark propagators, $D^{\mu\nu}$
and $S$, using the matrix representations (\ref{G22})-(\ref{S22}).
Formally, this Green function formalism on the closed-time path is completly
analogous to usual quantum field theory, except that all propagators,
self-energies,
etc., are now 2$\times$ 2 matrices, as diagramatically represented in Fig. 3.
Correspondingly, the Feynman rules
remain the same, but each propagator line of a Feynman diagram can be
either of the four components of the Green functions.
\bigskip

\noindent {\bf 2.2  The CTP effective action}
\medskip

To proceed, it is convenient to work
with the CTP effective action $\Gamma_P$, the
{\it two}-particle irreducible vertex functional, which determines
the equations of motion for
the physically relevant Green functions and the mean field, rather than with
$Z_P$ or $W_P$ of (\ref{Z000}) which involve
the sources $J,K$ that do not have any immediate physical interpretation.
The {\it CTP effective action} $\Gamma_P$ is defined as the multiple Legendre
transform
of $W_P$ \cite{chou,calzetta}, which with respect to the 2-point source
representation (\ref{Z000}) is given by
\begin{eqnarray}
\Gamma_P[\tilde{A}^\mu, D^{\mu\nu},S]
&=&
W_P[J^\mu,j,\overline{j}, K^{\mu\nu}, k] \;-\;
\left(\frac{}{}
J_\mu \,\tilde{A}^\mu \;+\;
j \,\overline{\psi}\;+\;\overline{j}\, \psi
\right)
\nonumber \\
& &
\;\;\;\;\;\;\;\;\;\;\;\;
\;\;\;\;\;\;\;\;\;\;\;\;
\;\;\;\;\;\;\;
\;-\;
\left(\frac{}{}
\frac{1}{2} \,A^\mu \,K_{\mu\nu}\, A^\nu
\;-\;
\overline{\psi}\, k\, \psi
\right)
\;.
\label{Gamma0}
\end{eqnarray}
Note that $\Gamma_P$ reduces to the usual effective action
for the {\it one}-particle irreducible vertex functions
in the limit of vanishing mean field $\tilde{A}^\mu=0$ and
absence of initial state correlations $K=k=0$.
In the general case, one obtains
\begin{eqnarray}
\Gamma_P[\tilde{A}^\mu, D^{\mu\nu},S]&=&
\;\tilde{I}[\tilde{A}^\mu]
\;-\; \frac{i}{2}\,Tr\left[\ln(D_{(0)}^{-1} D)\,-\, \tilde{D}_{(0)}^{-1} D
\,+\, 1 \right]
\nonumber \\
& &
+\;
i\,Tr\left[\ln(S_{(0)}^{-1} S)\,-\, \tilde{S}_{(0)}^{-1}  S \,+\, 1 \right]
\;\;+\;\; \Gamma_P^{(2)}[\tilde{A}^\mu, D^{\mu\nu}, S]
\;.
\label{Gamma1}
\end{eqnarray}
The first term is of order $\hbar^0$ and is given by the classical action
(\ref{I1})
(and eq. (\ref{I3}) of Appendix A)
with
\begin{equation}
\tilde{I}[\tilde{A}^\mu]
\;\equiv\;
I[A^\mu_\alpha,\psi_\alpha,\overline{\psi}_\alpha]\left.\frac{}{}\right|_{
A^\mu_\alpha=\tilde{A}^\mu, \;\psi_\alpha=\overline{\psi}_\alpha=0}
\;.
\label{I3}
\end{equation}
The second and third terms are of order $\hbar^1$ and correspond to the
gluon and quark contributions in which the bare propagators $D^{\mu\nu}_{(0)}$
and $S_{(0)}$
(\ref{prop2}) are modified by the  presence of a local gluon mean field
$\tilde{A}^\mu$ leading to `mean field dressed' propagators
$\tilde{D}^{\mu\nu}_{(0)}$
and $\tilde{S}_{(0)}$
with an effective screening mass $\tilde{\mu}\equiv \tilde{\mu}[\tilde{A}^\mu]$
(see Fig. 4).
In analogy to (\ref{prop1}):
\begin{eqnarray}
(\tilde{D}_{(0)}^{-1})^{\mu\nu}(x,y) &=& (D_{(0)}^{-1})^{\mu\nu}(x,y)
\;-\;\tilde{\mu}_g^{\mu\nu}(x,y)\,\delta^4(x-y)
\nonumber \\
\tilde{S}_{(0)}^{-1}(x,y) &=& S_{(0)}^{-1}(x,y)
\;-\;\tilde{\mu}_q(x,y)\,\delta^4(x-y)
\label{prop4}
\;.
\end{eqnarray}
Thus, the effect of the mean field
is to shift the pole in the bare Green functions (\ref{prop1})
by a dynamical mass function $\tilde{\mu}$.

\noindent
The  last term $\Gamma_P^{(2)}$ in (\ref{Gamma1})
represents the sum of all two-particle irreducible graphs of order
$\hbar^2,\hbar^3,\ldots$
\cite{cornwall},
with full propagators $D^{\mu\nu}$ and $S$, dressed by both local mean field
and
non-local self-interactions (see Fig. 4). As will become clear,
the real (dispersive) part of $\Gamma_P^{(2)}$ contains
the virtual loop corrections associated with the self-interactions of gluons
and quarks,
whereas the imaginary (dissipative) part contains the real emission,
absorption, and
scattering processes.
In other words, $\Gamma_P^{(2)}$ embodies all the interesting quantum dynamics
that is connected with renormalization group, entropy generation, dissipation,
etc..
Explicitly writing out the color indices, it is given by (see Fig. 5a)
\begin{eqnarray}
& &
\Gamma_P^{(2)}[\tilde{A}^\mu,  D^{\mu\nu}, S]
\;=\;
\nonumber \\
& &
\;\;\;\;\;\;\;
\;=\;
-\frac{g_s^2}{2}
\;\mbox{Tr} \left[
\frac{}{}
\int d^4z_1 d^4z_2 \;
\lambda^{aa'a''}_{\mu\mu'\mu''}\,\Lambda^{b'' b'b}_{\nu''\nu'\nu}(z_2,z_1;y)\;
D_{a'b'}^{\mu'\nu'}(x,z_1)\,D_{a''b''}^{\mu''\nu''}(x,z_2)\,D_{ba}^{\nu\mu}(y,x)
\right.
\nonumber \\
& &
\;\;\;\;\;\;\;\;\;\;\;\;\;\;\;\;
\;\;\;\;\;
\left.
\frac{}{}
\;+\;
\int d^4z_1 d^4z_2 \;
\gamma_\mu T_{ii'}^a \,\Xi_{jj'\;\nu}^{b}(z_2,z_1;y) \;
D_{ab}^{\mu\nu}(x,z_1)\,S_{i'j'}(x,z_2) \,S_{ij}(y,x)
\right]
\;\;\;
\label{Gamma2}
\;.
\end{eqnarray}
Here $\Lambda_{\nu\nu'\nu''}$ and $\Xi_\nu$ are
the $qqg$ and $qqg$ vertex functions, respectively,
\begin{eqnarray}
\Lambda_{\nu\nu'\nu''}^{a a' a''}(z_1,z_2;y)
&= &
\lambda_{\nu\nu'\nu''}^{aa'a''} \;\delta^4(y-z_1) \delta^4(y-z_2) \;
g[\tilde{A}^\mu(y)] \;\;+\;\; O(g_s^2)
\nonumber
\\
\Xi_{ij\;\nu}^{a}(z_1,z_2;y)
&=&
\gamma_\nu\,T_a^{ij}\;\delta^4(y-z_1) \delta^4(y-z_2)\; g[\tilde{A}^\mu(y)]
\;\;+\;\; O(g_s^2)
\label{VG}
\;,
\end{eqnarray}
with
$\lambda_{\nu\nu'\nu''}^{aa'a''}$ and
$\gamma_\mu T_a^{ij}$ the corresponding bare vertices,
and the function $g[\tilde{A}^\mu]$ describes the effect
due to the presence of the gluon mean field $\tilde{A}^\mu$
as compared to free space, where $g[0]=1$.
\bigskip

\noindent {\bf 2.3  The self-consistent equations of motion}
\medskip

The dynamical equations of motion
for the gluon mean field and the gluon and quark Green functions in
the absence of external sources are now as usual obtained from
variation of the effective action $\Gamma_P$ with respect
to its variables, and setting the external sources to zero.
Hence by functional differentiation of $\Gamma_P$ (\ref{Gamma1})
with respect to  the gluon mean field $\tilde{A}^\mu$ one gets the
{\it Ginzburg-Landau equation} \cite{GLE}
\begin{equation}
\frac{\delta \Gamma_P}{\delta \tilde{A}^\mu(x)}
\;=\;
\frac{\delta \tilde{I}[\tilde{A}^\mu]}{\delta \tilde{A}^\mu(x)}
\;+\;
2i\, \mbox{Tr}\left\{
\frac{\delta [\tilde{D}^{\mu\nu}_{(0)}]^{-1}}{\delta \tilde{A}^\mu(x)}
\;-\;
\frac{1}{2}
\frac{\delta [\tilde{S}_{(0)}]^{-1}}{\delta \tilde{A}^\mu(x)}
\right\}
\;+\;
\frac{\delta \Gamma_P^{(2)}}{\delta \tilde{A}^\mu(x)}
\;\,=\;\,0
\label{LG}
\end{equation}
Similarly, the variation of $\Gamma_P$
with respect to  the dressed propagators $D^{\mu\nu}$ and $S$ gives
the {\it CTP version of the  Dyson-Schwinger equations} \cite{DSE},
\begin{eqnarray}
i\,\frac{\delta \Gamma_P}{\delta D^{\mu\nu}(y,x)} &=&
D_{\mu\nu}^{-1}(x,y) \;-\; \tilde{D}_{(0)\,\mu\nu}^{-1}(x,y) \;+\;
\Pi_{\mu\nu}(x,y) \;=\; 0
\label{eom2a}
\\
-i\,\frac{\delta \Gamma_P}{\delta  S(y,x)} &=&
S^{-1}(x,y) \;-\; \tilde{S}_{(0)}^{-1}(x,y) \;+\; \Sigma(x,y) \;=\; 0
\label{eom2b}
\;.
\end{eqnarray}
Here
$\Pi$ and $\Sigma$ are $2\times 2$-matrices analogous to (\ref{G22}),
representing the proper self-energy parts of gluons and quarks.
They are obtained
by functional differentiation of the quantum contribution $\Gamma^{(2)}_P$
to the effective action (\ref{Gamma1}),
\begin{equation}
2\,i\,\frac{\delta \Gamma_P^{(2)}}{\delta  D_{\nu\mu}^{ba}(y,x)}
\;=\;
\Pi_{ab}^{\mu\nu}(x,y)
\;\;\;\;\;\;\;\;\;\;\;\;\;\;\;
-i\,\frac{\delta \Gamma_L^{(2)}}{\delta S_{ij}(y,x)}
\;=\;
\Sigma_{ij}(x,y)
\;.
\end{equation}
{}From (\ref{Gamma2}), one gets (c.f. Fig. 5b),
\begin{eqnarray}
\Pi_{ab}^{\mu\nu}(x,y)
&= &
-i\,g_s^2
\left[ \frac{}{}
\int d^4z_1 d^4z_2 \;
\lambda^{\mu\mu'\mu''}_{a a'a''}\Lambda^{b''b'b}_{\nu''\nu'\nu}(z_2,z_1;y)\;
D_{a'b'}^{\mu'\nu'}(x,z_1)\,D_{a''b''}^{\mu''\nu''}(z_2,x)
\right.
\label{Pi}
\\
& &
\;\;\;\;\;\;\;\;\;\;\;\;\;\;\;\;\;\;\;\;
\;+\;
\left.
\frac{}{}
\int d^4z_1 d^4z_2 \;
\gamma_\mu T_{ii'}^a \,\Gamma_{jj'\;\nu}^{b}(z_2,z_1;y) \;
S_{ij}(x,z_1)\,S_{i'j'}(x,z_2)
\right]
\nonumber
\\
\Sigma_{ij}(x,y)
&=&
+i\,g_s^2
\int d^4z_1 d^4z_2 \;
\gamma_\mu T_{ii'}^a \,\Gamma_{jj'\;\nu}^{b}(z_2,z_1;y) \;
D_{ab}^{\mu\nu}(x,z_1)\,S_{i'j'}(x,z_2)
\label{Sigma}
\;.
\end{eqnarray}
\medskip

The Dyson-Schwinger equations (\ref{eom2a}), (\ref{eom2b})
can be brought into a more familiar form by employing
the expressions for the free propagators (\ref{prop1})
\begin{eqnarray}
\!\!\!\!
\left[
\stackrel{\rightarrow}{\Box}_{x, \,\mu\rho}
\;+\;\tilde{\mu}_g^2(x,y)
\right]
\;  D^{\rho\nu}_{ab}(x,y)
&=&
\delta_{ab} \,g^{\mu\nu} \,\delta^4_P(x,y)
\;-\;\int_P d^4x' \, \Pi^\mu_{\sigma ,\;a,b'}(x,x')\, D^{\sigma
\nu}_{b'b}(x',y)
\nonumber
\\
\!\!\!\!
D^{\rho\nu}_{ab}(x,y)\;
\left[\stackrel{\leftarrow}{\Box}_{y, \,\mu\rho}
\;+\;\tilde{\mu}_g^2(x,y)
\right]
&=&
\delta_{ab} \,g_{\mu\nu} \,\delta^4_P(x,y)
\;-\;\int_P d^4x' \, D^\mu_{\sigma ,\;a,b'}(x,x')\, \Pi^{\sigma
\nu}_{b'b}(x',y)
\label{eog1}
\end{eqnarray}
and
\begin{eqnarray}
\left[
i \gamma\cdot \stackrel{\rightarrow}{\partial}_x
\;-\;\tilde{\mu}_q(x,y)
\right]
\; S_{ij}(x,y)
&=&
\delta_{ij} \delta^4_P(x,y)\;+\;
\int_P d^4x' \,\Sigma_{ik}(x,x')\,S_{kj}(x',y)
\nonumber
\\
S_{ij}(x,y)
\;
\left[
-i \gamma\cdot \stackrel{\leftarrow}{\partial}_y
\;-\;\tilde{\mu}_q(x,y)
\right]
&=&
\delta_{ij} \delta^4_P(x,y)\;+\;
\int_P d^4x' \,S_{ik}(x,x')\,\Sigma_{kj}(x',y)
\label{eog2}
\;,
\end{eqnarray}
where $\partial_x^\mu = \partial/\partial x^\mu$,
$\Box_x^{\mu\nu}$ is defined by (\ref{Box}), the time integrations
on the right hand sides are understood along the contour $P$, and
the generalized $\delta_P$-function is defined on the closed-time path $P$
(Fig. 2) as
\begin{equation}
\delta^4_P(x,y) \;\,:=\;\,
\left\{
\begin{array}{ll}
+\delta^4(x-y) & \;\; \mbox {if} \; x_0 \;\mbox{and} \;y_0 \; \mbox{from
positive
branch} \\
-\delta^4(x-y) & \;\; \mbox {if} \; x_0 \;\mbox{and} \;y_0 \; \mbox{from
negative
branch} \\
0 & \;\; \mbox {otherwise}
\end{array}
\right.
\;.
\end{equation}
Let me emphasize once more the essential difference to usual quantum field
theory:
the eqs.  (\ref{eog1}) and ({\ref{eog2}) are matrix equations and represent
four equations, one for each of the four correlators (\ref{D22}), respectively
(\ref{S22}).
In the limiting case where correlations among different partons
vanish, one has $G^>=G^<=0$, and because $G^F=G^{\overline{F}\;\dagger}$,
one recovers the standard Dyson-Schwinger equations in terms of the Feynman
propagators alone.
The first equation in (\ref{eog1}), respectively ({\ref{eog2}),
describes the change of the propagators in the argument $x$,
whereas the second equation describes the change in $y$ of the
adjoint propagators
(adjoint `$\dagger$' means hermitian conjugate with simultanous exchange of the
arguments).
A diagrammatic representation of these
Dyson-Schwinger equations for the fully dressed Green functions
$D_{\mu\nu}(x,y)$ and
$S(x,y)$ is shown in the previous Fig. 4.

\bigskip
Let me summarize the considerations of this Section.
The CTP generating funcional $Z_P$ involving a initial state correlations
of the form (\ref{rho}),
described by the density matrix $\hat \rho$ at $t=t_0$, yields
infinite hierarchy
of $n$-point Green functions, defined along the  closed time path.
As explained in Appendix A, the
truncation of this hierarchy beyond $n > 2$ assumes that the
dynamics may be described to sufficient accuracy by a possible local gluon
mean-field
and the non-local 2-point Green functions of gluons and quarks,
and that higher order correlators are negligible.
The resulting CTP effective action may then be represented by a systematic
loop expansion corresponding to an expansion in powers of $\hbar$.
Considering the pure quantum regime with zero mean field, yields
a coupled set of  equations of motion for the gluon and quark propagators,
which are $2\times 2$ matrices containing the four possible time
orderings of their arguments $x$ and $y$.
The solution of these dynamical
equations then boils down to the evaluation
of expectation values involving the propagators and vertex functions,
e.g. by using perturbation theory \cite{calzetta,cornwall}.
\bigskip
\bigskip

\newpage

\noindent {\bf 3. QUANTUM KINETIC THEORY}
\bigskip

Within the 2-point source approximation to the full theory in terms
of  2-point Green functions, the
resulting CTP Dyson-Schwinger equations (\ref{eog1}), (\ref{eog2}) )
contain the quantum dynamics in terms of the dressed
gluon and quark propagators $D_{\mu\nu}$ and $S$.
Even with the neglect of higher-order correlators, the
equations of motion
are non-linear, non-local integrodifferential equations, generally not solvable
in closed form.
To make progress, one needs to supply reasonable physical input that
allows  to make realistic approximations for
multi-parton systems of interest.

First of all, I will
confine myself for the remainder of the paper
to the pure quantum dynamics of gluons and quarks, when a
gluon mean field is absent.  That is, I choose the homogenous
initial condition $\tilde{A}^\mu(x)=0$ at $t_0$, which
in the absence of external sources implies
that $\tilde{A}^\mu$ will remain zero at all times $t>t_0$,
\begin{equation}
\frac{\delta \Gamma_P}{\delta\tilde{A}^\mu(x)}\;=\; 0
\;,\;\;\;\;\;\;\;\tilde{A}^\mu(x)\;=\; 0
\end{equation}
Consequently, in (\ref{Gamma1}),  the classical
part $\tilde{I}[\tilde{A}^\mu]=0$  (see also eq. (\ref{I3}) of Appendix A),
and $\tilde{\mu}_g=\tilde{\mu}_q=0$,
so that the mean-field propagators reduce to the bare propagators,
\begin{equation}
\tilde{D}_{(0)}^{\mu\nu}(x,y)\;=\; D_{(0)}^{\mu\nu}(x,y)
\;\;,\;\;\;\;
\tilde{S}_{(0)}^{\mu\nu}(x,y)\;=\; S_{(0)}(x,y)
\;.
\end{equation}
This step however is not an approximation, but merely serves
as a simplification in order not to overburden the following analysis.
The more general case including a dynamical gluon background field
causes in principle no severe additional complexities, and will be
addressed elsewhere.

The essential approximation now is based on the `two-scale' nature
of high-energy QCD, as mentioned in the introduction.
The dynamical evolution of a  multi-parton system can
- in a reference frame where the partons move with highly relativistic
velocities -
be characterized
by two different time- (or length-) scales,
separated by time dilation and Lorentz contraction effects:
a {\it quantum field theoretical scale} $\Delta r_{qua}$ and a
{\it statistical-kinetic} scale $\Delta r_{kin}$.
This is illustrated in Fig. 6a.
The {\it quantum length scale} $\Delta r_{qua}$, measures the spatial range of
quantum fluctuations, associated with the partons' self-interactions,
and thus specifies the Compton wavelength $\lambda_c\equiv \mu_{gq}^{-1}$
of dressed partons.
These gluon emission and absorption processes, embodied in the self-energy,
dress up the bare propagators and allow to describe
partons as quasi-particles with finite spatial extent, but with a dynamical
substructure.
This is nothing but the underlying philosophy of the usual parton description
in QCD.
The {\it kinetic length scale} $\Delta r_{kin}$, on the other hand,
measures the range of binary interactions between these quasi-particles.
These scattering processes may be
described on a semi-classical level, provided
the local density of the quasi-particles
is smaller than a critical density where the particles begin to overlap
and the separation between quantum and classical regimes breaks down.
Quantitatively one has to require that the mean free path $\lambda_{mf}$ of
particles
is large compared to the radiative corrections to the Compton wavelength
$\lambda_c$.
The crucial point is, that with increasing energy scale the latter
range becomes increasingly short-range, concentrated around the lightcone (see
Fig 6b).
Hence, in most physical situations at high energies,
the quantum and the kinetic scales separate to
very good approximation, and in the
asymptotic limit exactly.
It is important to stress that both quantum and kinetic scales define the
microscopic regime of a semi-classical particle description. It is to be
distinguished from the macroscopic domain of the dynamics of the
bulk parton matter, characterized by comparably large space-time distances of
the order
$n^{-1/3}$, or $n/(\partial_r n)$, where $n(r)$ is the density of
quasi-particles.
In this regime the system may be described by, e.g.,  hydrodynamical evolution,
which is however beyond the scope of this paper.

To exemplify this concept,
consider the simple case of a parton in a Lorentz frame in which it
moves with large momentum $k^+\equiv E+k_z$ ($k^{+\,2}\gg k^2\gg 1$ GeV$^2$)
nearly with the speed of light along the forward lightcone, $x^+\equiv t+z$.
The quantum fluctuations around this parton's classical trajectory
stem from its self-interaction with the gluon radiation field,
corresponding to gluon emissions and reabsorptions, that
smear out its energy over an interval $\Delta E \sim k^2/k^+$.
It may thus be pictured as an unstable particle with a typical
life-time $\Delta \tau_p\sim 1/\Delta E$.
On the one hand, in the direction parallel to the lightcone,
the parton's intrinsic fluctuations decouple from the soft vacuum fluctuations
with $\Delta\tau_v\sim 1/k^+ \ll \Delta \tau_p \sim k^+/k^2$ \cite{glr}.
On the other hand, in transverse $x_\perp$-direction, the partonic fluctuations
have a small spatial extent of $\Delta r_\perp \sim 1/k_\perp \ll 1$
GeV$^{-1}$.
Therefore, on kinetic scales $\Delta_{kin}>\mu^{-1}$, the parton appears as a
dressed particle which
can be considered
quasi-classically as an extended object with a small transverse
size $\Delta r_\perp$ and a comparably long life-time $\Delta \tau_p$ $-$ a
quasi-particle.
On quantum scales, however,
the dressed parton has a substructure, determined by its surrounding
cloud of gluons that it emits and reabsorbs due to its quantum nature.

In this spirit I will classify the parton dynamics with respect
to elementary and quasi-particle excitations, referring to them
by the terms {\it bare} and {\it dressed} partons, respectively:
\noindent

(i) {\it bare} partons are to be understood as pointlike, massless quanta
in the absence of radiative self-interactions, i.e. before
renormalization.

(ii) {\it dressed} partons, on the other hand, are dressed by
the quantum self-interactions with their radiation field,
which renormalize their masses and couplings.

In the field-theoretical parton language,
a dressed parton with its dynamically generated
renormalized mass can be described (in a frame where it moves close to the
speed of light)
as a bare quantum which is surrounded by a virtual cloud of
other bare gluons and quark-antiquark pairs with which it emits and absorbs.
Hence,  kinetic space-time scales, a dressed parton can be visualized as
a quasi-particle, i.e. an extended object with a dynamical substructure that is
determined by the short-distance quantum fluctuations.
\bigskip

\noindent {\bf 3.1 Definition of quantum and kinetic space-time scales}
\bigskip

The realization of the two space-time scales, short-distance quantum and
quasi-classical kinetic, allows to reformulate the quantum field theoretical
problem as a relativistic many-body problem within kinetic theory.
The key element is to establish the connection between the quantum-theoretical
Green functions
and the kinetic particle description in terms parton phase-space densities.
In particular, the aim is to describe the evolution of a multi-parton ensemble,
given at time $t_0$, with a certain spatial and a momentum distribution,
by exploiting the `two-scale' nature of high-energy QCD.
As explained before,
this requires a choice of Lorentz frame, in which the quanta move very fast
and the typical momentum scale of their binary interactions and associated
radiative
processes is sufficiently large, such that the corresponding interaction times
are small compared to the mean free time in between mutual collisions.
For example:
Imagine a high-energy reaction has produced an initial configuration
of materialized partons (e.g. a hadronic or nuclear collision with
$\sqrt{s}\,\lower3pt\hbox{$\buildrel > \over\sim$}\,100$ GeV per hadron).
If $t_0$ denotes the earliest point of time in the lab frame,
when the parton densities have evolved to satisfy $\Delta p \Delta r \gg 1$,
where $1/\Delta p$ measures the scale of the partons' intrinsic quantum motion
and $\Delta r$ the space-time variation of the system of partons,
then, for times $t > t_0$, an approximate incoherent treatment
of quantum dynamics and kinetic evolution is justified, as has been
shown by McLerran and Venugopalan \cite{mclerran94}.

With this physical scenario in mind,
now suppose, space-time is discretized into cells, with their
size chosen intermediate between quantum and kinetic scales such that
the separation between the two scales is optimal \cite{calzetta}.
Then the correlation between {\it different} cells will be negligible,
and only when two space-time points corresponding to the arguments of
the propagators or self-energies lie in the {\it same} cell, the
2-point correlation will contribute (as explained in  Appendix B).
Consequently,
in a given cell, one can by construction neglect spatial inhomogenities
of the local gluon and quark densities of the multi-parton system.
Within each cell, one may therefore describe the short-distance quantum
dynamics analogously as in vacuum or homogenous media, whereas
inhomogenities of the spatial parton distribution
and relaxation phenomena associated with binary collisions
become apparent, as one moves from cell to cell.
In continous space-time, corrections to this discretized picture can be
taken into account by a systematic expansion in terms of gradients
of the spatial inhomogenities of the parton distributions (eq. (\ref{gradexp})
below).

In order to quantify this concept, let me clearly specify
quantum and kinetic domains with respect to the
cellular space-time.
It is important to realize that both quantum and kinetic scales are
of dynamical, `internal'  nature, i.e. determined by the multi-parton evolution
itself.
The classification of the two scales only makes sense
in the presence of self- and mutual interactions.
However, the class of high-energy parton systems addressed here is
characterized by two corresponding `external' scales:
first, the large energy scale of the reaction that produces
the initial system of large momentum partons, and second,
the initial local density of partons in phase-space that depends on the
type of reaction.
The first property implies a large characteristic momentum transfer
$q_\perp^2\equiv (k_1 -k_1')^2$ of scattering
processes ($k_1 k_2 \rightarrow k_1' k_2'$) and
radiative processes ($k_1\rightarrow k_1'k_2'$).
The second property, on the other hand,  is related to the mean free path
$\lambda_{mf}$
and mean free time $\tau_{mf}$ of partons in between subsequent scatterings.
If the latter are large compared to the typical space-time extent $1/q_\perp$
of
the the scattering- and radiation processes, then an incoherent treatment
of the binary collisions among partons, and of the partons' propagation
with associated quantum fluctuations, is applicable.
This condition may be characterized by the invariant mass scale
$\mu_{gq}$, defined such that
\begin{equation}
q_\perp^2 \;\;> \;\; \mu_{gq}^2 \;\;>\;\; \lambda_{mf}^{-2}
\label{cond0}
\;.
\end{equation}
The parameter $\mu_{gq}^2$ can be interpreted as
defining the minimum virtuality of a dressed parton,
or correspondingly, its maximum size, or Compton wavelength, $\lambda_c =
\mu_{gq}^{-1}$,
such that the applicability condition of the parton description is ensured.
Consequently, the size of each space-time cell must be chosen large enough
that the spread of the dressed partons' intrinsic quantum motion
is localized inside its four-dimensional volume, but smaller
than the mean free path of dressed partons in between scatterings.
Accordingly, I define the cell size $\Delta r^\mu\equiv\Delta r^0 \Delta^3 r$
by
\begin{equation}
\mu_{gq}^{-4} \;\;<\;\;
\Delta r^\mu \;\equiv\; \mu^{-4}(r) \;\;\ll\;\;\Lambda_{QCD}^{-4}
\;,
\label{scale1}
\end{equation}
where $\Lambda_{QCD} \simeq 0.25$ GeV is the QCD renormalization scale. For
example,
a cell size $\Delta r\,\lower3pt\hbox{$\buildrel < \over\sim$}\,0.1$ $fm$
allows to resolve particles with energy-momentum
$\,\lower3pt\hbox{$\buildrel > \over\sim$}\,2$ GeV.
One can then characterize the kinetic space-time evolution
of the system by a velocity profile of cells $i$, located around the
points $r_i^\mu$, with four-dimensional cell volume
in its restframe
\begin{equation}
\Omega(r_i) \;=\;       \int_{r'\in \Omega} d^4 r'
\;=\; \int_{r_i^0-\frac{\Delta r^0}{2}}^{r_i^0+\frac{\Delta r^0}{2}}
d r^0\,
\int_{\vec{r}_i-\frac{\Delta \vec{r}}{2}}^{\vec{r}_i+\frac{\Delta \vec{r}}{2}}
d^3 r'
\;\simeq \;\mu^{-4}
\;.
\label{Vcell}
\end{equation}
Each cell carries a total momentum
\begin{equation}
P^\mu (r_i)\;:=\; \left. \sum_{j=1}^{N_{gq}} \,k_j^\mu\right|_{(r^0_j, \vec
r_j) \in \Omega(r_i)}
\end{equation}
and a total invariant virtuality (the incoherent sum of parton virtualities),
\begin{equation}
Q^2(r_i)\;:=\;
\left. \sum_{j=1}^{N_{gq}} \,k_j^2\right|_{(r^0_j, \vec r_j) \in \Omega(r_i)}
\label{Q2}
\; ,
\end{equation}
where the sums are over all dressed partons $j$ inside the cell $i$,
i.e. those that are during a time slice $\Delta r^0=\mu^{-1}$
contained within $\Delta^3 r= \mu^{-3}$  around space-time point $r_i$.
The corresponding local four-flow velocity is $u^\mu(r_i) = P^\mu/P^0$.
This cellular space-time picture is illustrated in Fig. 7a.

The validity of the above cell picture is  controlled by
the condition that the  different scales are well separated:
\begin{equation}
P^{+\;2}(r_i) \;\, \gg \,\;Q^2(r_i) \,\;\ge\;\,\mu^2(r_i)
\,\;\gg\;\,\Lambda_{QCD}^2
\;,
\label{scale2}
\end{equation}
where
$P^\mu=(P^+,P^-,\vec P_\perp)$,
$P^\pm= P_0\pm P_3$,
$P_\perp = \sqrt{P_1^2+P_2^2}$
with $P^+\,(P^-)$ the {\it lightcone momentum (energy)}, $P^2 = P^+P^- -
P_\perp^2$,
and the normalization of a cell state $|P\rangle$ is
$\langle \,P\,|\,P'\,\rangle= 2 P^+\,(2\pi)^3 \delta^3(\vec P-\vec P')$.
On the basis of (\ref{scale2}) and
in terms of these lightcone variables, the four-momentum of a
parton $j$ can be characterized by only two variables, namely, its
lightcone momentum $k_j^+=x_j P^+$ with fraction $x_j$ of the total
cell momentum,
and its off-shellness (invariant virtuality) $k_j^2=k^+_j
k^-_j-k_{j\,\perp}^2\equiv M^2(k_j)$.
Its lightcone energy is  $k_j^-=(k_j^2+k_{j\,\perp}^2)/k^+_j\simeq 0$
($k^{+\;2}\gg k^2 \,\lower3pt\hbox{$\buildrel > \over\sim$}\,k_\perp^2$),
and one has therefore
\begin{equation}
k_j \;=\;(k_j^+,\, k_j^2)
\;=\;(x_j P^+,\, k_j^2)
\;\;\;\;\;\;\;\;\;\;\;\;
\frac{d^4k_j}{(2\pi)^4}\;(2\pi)\,\delta^+\left(k_j^2 - M^2(k_j)\right)
\;=\;\frac{1}{16 \pi^2}\;\frac{dx_j}{x_j}\; dk_j^2
\;.
\label{pj}
\end{equation}
The requirement (\ref{scale2}) together with (\ref{scale1})
hence translates to the parton level as
$k^{+\;2}_j\gg k_j^2 \ge \mu^2(r_i)$, for all partons $j$ within a given cell
around
$r_i$.
\begin{equation}
Q^2(r_i) \,\; \ge \;\, k_j^2 \;\,\ge \;\,\mu_{gq}^2\;\,\ge \;\,\mu^2(r_i)
\;.
\label{scale3}
\end{equation}
Since $\lambda_c = \mu_{gq}^{-1}$ characterizes
the maximum size of dressed partons,
the ratio $\mu^4(r_i)/\mu^4_{gq}$ determines the minimum fraction of
volume occupied by dressed partons in the cell.
The quantum and kinetic space-time regions
can now be defined as
\begin{equation}
\Delta r_{qua} \;=\; [P^{+\;-1},\; \mu^{-1}]
\;,\;\;\;\;\;\;\;\;\;\;\;\;\;
\Delta r_{kin} = [\mu^{-1}, \;\Lambda_{QCD}^{-1}]
\;.
\label{micmac}
\end{equation}

For large $P^+$ and $Q^2$,
quantum and  kinetic
length scales are well separated and the
{\it parton phase-space densities} $F_{f}$ may be
locally
represented as a convolution of the kinetic, {\it statistical density of
dressed partons} ${\cal N}_f$ of type $f=g,q$,
with the {\it quantum theoretical spectral density} ${\cal P}_f$ of each
dressed parton
describing the intrinsic density of bare parton states as its quantum
substructure,
as depicted in Fig. 7b:
\begin{equation}
F_{f} (r_i, k) \;=\; F_{f} (r_i, k^+, k^2) \;\,\equiv\;\,
{\cal N}_f\;\left( r_i, \,P^+,\mu_{gq}^2\right)
\,\otimes\,
{\cal P}_f\left(r_i,\,k^+, k^2\right)
\;,
\label{Fgq}
\end{equation}
where the convolution of the statistical density ${\cal N}_f$
of dressed partons at the scale $\mu_{gq}^2$
with the spectral density ${\cal P}_f$, is defined as
the average over the local space-time volume $\Omega(r_i)$ around $r_i$ of
the densities,
\begin{equation}
{\cal N}_f\, \otimes\, {\cal P}_f
\;\,\equiv\;\,
\frac{1}{\Omega(r_i)} \; \int_{\Omega(r_i)} d^4 r'
\,\int \frac{dy}{y} \;
{\cal N}_f\left( r',\,yP^+,\mu_{gq}^2\right)\; {\cal P}_f\left(
r',\,\frac{x}{y}, k^2\right)
\label{convolution}
\;,
\end{equation}
with $\Omega(r_i)\simeq \mu(r_i)^{-4}$,
$P^+=P^+(r_i)$, $k^+=xP^+$, $z=x/y$ ($0\le z \le 1$), and
\begin{eqnarray}
{\cal N}_f(r,\,yP^+,\mu_{gq}^2)
&=&
\frac{dN_f}{d^4 r\; d\ln y}
\,\delta(k^2\, -\,\mu_{gq}^2)
\label{Ngq}\\
{\cal P}_f\left(r,\,z, k^2\right)
&=&
\sum_{f'=g,q}\,{\cal P}_f^{f'}
\;=\;
\sum_{f'=g,q}\,
\int_{\mu_{gq}^2}^{k^2}
dk'^2\,\frac{dn_f^{f'}}{d\ln z\; dk'^2}
\label{Pgq}
\;.
\end{eqnarray}
This ansatz describes
the multi-parton system on the basis of treating each individial dressed parton
as
a composite particle of type $f$ with a substructure of number of bare quanta
$n_f^{f'}$
of type $f'$, weighted
locally with the total number of dressed partons $N_f$ in a space-time cell.
The spectral density ${\cal P}$ characterizes the intrinsic structure of a
dressed parton state,
whereas the quasi-particle density  ${\cal N}$ describes the correlations and
scatterings among
those dressed partons.
As will become clear later, the spectral densities ${\cal P}_f$ can indeed
be identified with the QCD parton structure functions.
The crucial quantities that control the cellular resolution
in space-time of the partons' substructure are the characteristic cell size
$\mu^{-1}(r_i)$,
and the minimum resolvable virtuality $\mu_{gq}^2\ge \mu^2$ of dressed partons
in the cell,
or alternatively, the fractional space-time volume occupied, $\Delta
\Omega/\Omega=\mu^4/(N_{gq} \mu_{gq})^4$,
that determines how dense a cell may populated without the
partons overlapping.
Hence, the validity of the kinetic approximation, based on the separation of
quantum and kinetic scales, is controlled by the choice of these quantities,
which need not be constant but rather may be taken as space-time dependent,
i.e. variable from cell to cell chosen such that the resolution is
optimal.
A convenient choice would  be, for instance,
\begin{equation}
\mu(r_i)\;\,\simeq\;\, \mu_{gq}
\label{mumu}
\;
\end{equation}
which I will adopt in the following for lucidity, keeping in mind that
$\mu_{gq}$ is not a free external parameter, but rather is
to be understood as a dynamical, possibly space-time dependent quantity,
which in principle should be determined self-consistently from screening
effects.
I will not address this latter issue here.
\bigskip

\noindent {\bf 3.2 Wigner transformation and the kinetic equations of motion}
\bigskip

Let me proceed, referring to Appendix B for details, by
introducing
center-of-mass and relative coordinates of two space-time points $x$ and $y$,
\begin{equation}
r\;\equiv \; \frac{1}{2}\,(x\,+\,y)
\;\;,\;\;\;\;\;\;\;\;\;\;\;\;
s\;\equiv \;x\,-\,y
\;,
\end{equation}
in terms of which one can express any 2-point function
$W(x,y)\equiv D_{\mu\nu},S,\Pi_{\mu\nu},\Sigma$, as
\begin{equation}
W(x,y)\; =\; W\left(r+\frac{s}{2}, r-\frac{s}{2}\right)
\;=\; W(r,s)
\;,
\end{equation}
and introduce its
{\it Wigner transform} $W(r,k)$ as \cite{wigner}
\begin{equation}
W(x,y) \;=\;
\int \frac{d^4k}{(2\pi)^4} \, e^{-i\,k\,\cdot\, s}\;\, W\left( r,k\right)
\;\;,\;\;\;\;\;\;\;\;\;\;\;\;\;
W(r,k) \;=\;
\int d^4 s \, e^{i\,k\,\cdot\, s}\;\, W\left( r,s\right)
\;,
\label{W}
\end{equation}
i.e. one Fourier-transforms with respect to the relative coordinate $s$
being the canonical conjugate to the momentum $k$.
In the cell picture of space-time, the coordinate $r$ is
the cell index that labels the kinetic space-time dependence $O(\Delta
r_{kin}$),
whereas $s$ measures the quantum space-time distance $O(\Delta r_{qua}$),
as illustrated in Fig. 8.
In homogenous systems, such as the vacuum, translation invariance dictates that
the dependence
on $r$ drops out entirely, and the Wigner transforms then coincide with
the momentum-space Fourier transforms of the Green functions and self-energies.
In general,
spatial inhomogenities
can be systematically accounted for by performing an expansion in terms
of gradients $\partial_r \equiv \partial/\partial r^\mu$:
\begin{equation}
W(r+s, s) \;\simeq \; W(r, s) \;+ \;s \,\cdot\, \partial_r\,W(r,s)
\;\,+\;\, O[(s \cdot\partial_r)^2]
\label{gradexp}
\;.
\end{equation}
For quasi-homogenous, or moderately inhomogenous
systems, such that $s\cdot\partial_r W\ll W$, the
correlations between different cells will be small so that
the propagators and self-energies accordingly  vary only slowly with $r$.
One may then truncate the series (\ref{gradexp}) after the second term, and
convert the quantum field equations of motion (\ref{eog1}) and
(\ref{eog2}) into a set of kinetic equations by first performing the
Wigner transformation (\ref{W}) for all
Green functions and self-energies, and then taking for (\ref{eog1}) and
(\ref{eog2})
the sum and difference of the two adjoint equations in their transformed
representation.

This procedure (see Appendix B) yields
two distinct equations for each of the Wigner transforms $D_{\mu\nu}$ and $S$
with rather different physical interpretations, which I will refer to
as {\it renormalization equation} and
{\it transport equation}, respectively.
The {\it renormalization equations} are obtained as
\begin{eqnarray}
\!\!\!\!\!
\left(k^2 \,-\,\frac{1}{4} \partial_r^2 \right)
\; D^{\mu\nu}_{ab} (r,k)
&=&
-\,d^{\mu\nu}(k)\,\delta_{ab}\,\hat 1_P
\;+\;
\frac{1}{2} \;\left(\frac{}{}\left\{\Pi\,,\, D \right\}_+\right)^{\mu\nu}_{ab}
\;+\; \frac{i}{4}\;{\cal G}^{\mu\nu\, (-)}_{ab}
\label{R}
\\
\!\!\!\!\!
\frac{1}{2} \;\left\{ \gamma\cdot p\,,\, S_{ij}(r,p)\right\}_+
&=&
\delta_{ij}\,\hat 1_P
\;-\;
\frac{i}{2} \;\left(\frac{}{}\left[\gamma\cdot \partial_r\,,\, S
\right]_-\right)_{ij}
\;+\;
\frac{1}{2} \;\left(\frac{}{}\left\{\Sigma\,,\, S \right\}_+\right)_{ij}
\;+\;
\frac{i}{4} \,{\cal F}^{(-)}_{ij}
\;,
\;\;\;
\nonumber
\end{eqnarray}
where  $d^{\mu\nu}(k)$ is given by eq. (\ref{prop2}),
$\partial_r^2 \equiv \partial_r\cdot\partial_r$,
and $[A,B]_- \equiv AB-BA$, $\{A,B\}_+\equiv AB+BA$.
The {\it transport equations} are found in the form
\begin{eqnarray}
k\cdot\partial_r\; D_{ab}^{\mu\nu} (r,k)
&=&
-\;\frac{i}{2} \;\left(\frac{}{}\left[\Pi\,,\, D \right]_-\right)^{\mu\nu}_{ab}
\;+\;\frac{1}{4}\;{\cal G}^{\mu\nu\,(+)}_{ab}
\label{T}
\\
\frac{1}{2} \;\left\{ \gamma\cdot \partial_r\,,\, S_{ij}(r,p)\right\}_+
&=&
\frac{i}{2} \;\left(\frac{}{}\left[\gamma\cdot p\,,\, S \right]_-\right)_{ij}
\;-\;
\frac{i}{2} \;\left(\frac{}{}\left[\Sigma\,,\, S \right]_-\right)_{ij}
\;+\;
\frac{1}{4} \,{\cal F}^{(+)}_{ij}
\;.
\nonumber
\end{eqnarray}
In (\ref{R}) and (\ref{T}) the self-energies $\Pi$ and $\Sigma$
are explicitly given by (\ref{Pi}), (\ref{Sigma}),
and
the operator functions ${\cal G}$ and ${\cal F}$,
on the right hand sides,
which  include the effects of spatial inhomogenities to
first order in the gradient expansion (\ref{gradexp}),
are
\footnote{
Note,
$\partial_r^\mu\equiv \partial/\partial k^\mu$
acts on a function $f(r,k)$ as the derivative
with respect to the space-time coordinate, whereas
$\partial_k^\mu\equiv \partial/\partial k^\mu$
and
$\partial_p^\mu\equiv \partial/\partial p^\mu$
refer to the variation of four-momentum.
}
\begin{equation}
\!\!\!\!\!\!
{\cal G}^{\mu\nu\,(-)}\,=\,
\left[ \partial^\lambda_k \Pi^\mu_{\sigma} \, , \partial_\lambda^r
D^{\sigma\nu}\right]_-
-
\left[ \partial^\lambda_r \Pi^\mu_{\sigma} \, , \partial_\lambda^k
D^{\sigma\nu}\right]_-
\;,\;\;\;
{\cal G}^{\mu\nu\,(+)}\,=\,
\left\{ \partial^\lambda_k \Pi^\mu_{\sigma} \, , \partial_\lambda^r
D^{\sigma\nu}\right\}_+
-
\left\{ \partial^\lambda_r \Pi^\mu_{\sigma} \, , \partial_\lambda^k
D^{\sigma\nu}\right\}_+
\nonumber
\end{equation}
\begin{equation}
{\cal F}^{(-)}\;=\;
\left[ \partial^\lambda_p \Sigma \, , \partial_\lambda^r S\right]_-
-
\left[ \partial^\lambda_r \Sigma \, , \partial_\lambda^p S\right]_-
\;,\;\;\;\;\;\;\;
{\cal F}^{(+)}=
\left\{ \partial^\lambda_p \Sigma \, , \partial_\lambda^r  S\right\}_+
-
\left\{ \partial^\lambda_r \Sigma \, , \partial_\lambda^p  S\right\}_+
\;.
\end{equation}
For completness, I note that the equations  for quark Green functions
can formally also be brought in a more familiar quadratic form, similar
to the equations for the gluon Green functions,
which exhibits the mass- and drift-term on the left hand side of
the renormalization and transport equation, respectively:
\begin{eqnarray}
\!\!\!\!\!\!
\left(p^2 \,-\,\frac{1}{4} \partial_r^2 \right)
\; S_{ij} (r,p)
&=&
\left(\gamma\cdot p + \Sigma\right)
\,\delta_{ij}\,\hat 1_P
\;+\;
\frac{1}{2} \;\left(\frac{}{}\left\{\Sigma^2\,,\, S \right\}_+\right)_{ij}
\;+\; \frac{i}{4}\;{\cal A}^{(+)}_{ij}
\;-\; \frac{1}{8}\;{\cal B}^{(-)}_{ij}
\nonumber
\\
\!\!\!\!\!\!
p\cdot\partial_r\; S_{ij} (r,p)
&=&
\frac{1}{2}\,
\,\left(\gamma\cdot \partial _r\right)
\,\delta_{ij}\,\hat 1_P
\;-\;
\frac{i}{2} \;\left(\frac{}{}\left[\Sigma^2\,,\, S \right]_-\right)_{ij}
\;+\;\frac{1}{4} \;{\cal A}^{(-)}_{ij}
\;+\;\frac{i}{8}\;{\cal B}^{(+)}_{ij}
\; ,
\end{eqnarray}
where $\Sigma_{ij} =\delta_{ij}\Sigma$, and
\begin{eqnarray}
& &
{\cal A}^{(\pm)}\;=\;
\frac{1}{2}\left(
\frac{}{}(\gamma\cdot p + \Sigma)\;({\cal F}^{(-)} + {\cal F}^{(+)}) \;\pm\;
({\cal F}^{(-)} - {\cal F}^{(+)}) \;(\gamma\cdot p + \Sigma)\right)
\nonumber
\\
& &
{\cal B}^{(\pm)}\;=\;
\frac{1}{2}\left(
\frac{}{}
(\gamma\cdot \stackrel{\rightarrow}{\partial_r})\;({\cal F}^{(-)} + {\cal
F}^{(+)}) \;\pm\;
({\cal F}^{(-)} - {\cal F}^{(+)}) \; (\gamma\cdot
\stackrel{\leftarrow}{\partial_r}) \right)
\;.
\end{eqnarray}

As  will be seen in the following, the renormalization equations (\ref{R})
express the normalization conditions imposed by unitarity and
renormalization group due to the  quantum self-interactions,
and redefine the bare quanta in terms of renormalized quasi-particles.
The transport equations (\ref{T}) on the other hand describe
the kinetic space-time evolution of the system of quasi-particles and
their binary collisions.

The kinetic approximation of trading the Green functions $G(x,y)$ with
their Wigner transforms $G(r,p)$,
means in the   picture of cellular
space-time, that inside a given cell carrying the space-time coordinate
$r=(t,\vec r)$ as a label,
$G(r,p)$ equals the  translation invariant Fourier transform $G(p)$ of
$G(x-y)$,
but outside of the cell it is zero.
In another cell $r'$, the Wigner function $G(r',p')$ is determined by a
different translation invariant $G(p')$. Hence, when looking at the
short-distance quantum fluctuations within a given space-time cell around
$r=(r^0,\vec r)$,
one may approximate the spatial distribution of partons as being
homogenous and constant over the cell volume, and describe the short-range
quantum dynamics in a tranlation-invariant manner.
With the same accuracy of approximation, one can neglect
in the quantum regime binary parton collisions,
provided the mean-free-path $\lambda_{mf} = (\sigma_{gq} \,F_{gq})^{-1}$
in terms of the parton-parton cross-sections $\sigma_{gq}$ and
the local density $F_{gq}$,
is large compared to the spatial spread of the quantum fluctuations which
is typically of the order of $1/\sqrt{p^2}$.
Hence, the essential requirement $p^2 \gg \lambda_{mf}^{-2}$ can
always be realized, if the particle energies are sufficiently large.
\smallskip

On the basis of these considerations,
I  first study the
quantum theoretical aspects embodied in the renormalization equations
(\ref{R}) to obtain the  renormalized gluon and quark propagators,
and from this determine
the momentum dependence of the phase-space densities $F_{gq}$, eq. (\ref{Fgq}),
 associated with the
variation of the parton structure functions.
Subsequently, I will investigate the transport theoretical aspects
of the statistical kinetic dynamics,
described by  transport equation (\ref{T}),
which determines the space-time variation of the phase-space densities
$F_{gq}$ in terms of renormalized, dressed partons.
\bigskip


\noindent {\bf 3.3 The `physical representation' and strategy of solution}
\bigskip

Within the kinetic approximation,
the goal is to obtain the best possible approximation to the complete
propagators $G=D_{\mu\nu},S$, starting from the corresponding free Wigner
transformed
Green function.
The `free-field' solutions of the four types of correlators in
(\ref{G22})-(\ref{S22}),
namely $G^F,G^>,G^<,G^{\overline{F}}$, are in their most general form given by
\cite{chou}
\begin{eqnarray}
D_{(0)\,\mu\nu}^F(r,k)&=&
-d_{\mu\nu}(k)\;\left[\frac{1}{k^2 + i \varepsilon}
\;-\; 2\pi i \;  F_{(0)\;g}(r,k) \,\delta(k^2-\mu_{gq}^2)
\right]
\nonumber \\
D_{(0)\,\mu\nu}^>(r,k)
&=&
-\; 2\pi i \; (-d_{\mu\nu}(k))\; \left[\theta(+k_0)\,+\,F_{(0)\;_g}(r,k)
\right]
\,\delta(k^2-\mu_{gq}^2)
\nonumber \\
D_{(0)\,\mu\nu}^<(r,k)
&=&
-\; 2\pi i \; (-d_{\mu\nu}(k))\, \left[\theta(-k_0)\,+\, F_{(0)\;_g}(r,k)
\right]
\, \delta(k^2-\mu_{gq}^2)
\nonumber \\
D_{(0)\,\mu\nu}^{\overline{F}}(r,k)
&=&
+d_{\mu\nu}(k)
\left[
\frac{1}{k^2 - i \varepsilon}
\;-\; 2\pi i \; F_{(0)\;_g}(r,k)
\, \delta(k^2-\mu_{gq}^2)
\right]
\label{D00}
\;,
\end{eqnarray}
where $d_{\mu\nu}(k)$ is defined by (\ref{prop2}), and
\begin{eqnarray}
S_{(0)}^F(r,p)&=&
\frac{+1}{\gamma\cdot p + i \varepsilon}
\;+\; 2\pi i \;  F_{(0)\;q}(r,p)
\, \delta(p^2-\mu_{gq}^2)
\nonumber \\
S_{(0)}^>(r,p)
&=&
+\; 2\pi i \;  F_{(0)\;q}(r,p)
\, \delta(p^2-\mu_{gq}^2)
\nonumber \\
S_{(0)}^<(r,p)
&=&
-\; 2\pi i \;  \left[ 1\,-\, F_{(0)\;q}(r,p)\right]
\, \delta(p^2-\mu_{gq}^2)
\nonumber \\
S_{(0)}^{\overline{F}}(r,p)
&=&
\frac{-1}{\gamma\cdot p - i \varepsilon}
\;+\; 2\pi i \;  F_{(0)\;q}(r,p)
\, \delta(p^2-\mu_{gq}^2)
\;.
\label{S00}
\end{eqnarray}
The scalar functions $F_{(0)\;_g}$ and $F_{(0)\;q}$ are the
free-field analogues of (\ref{Fgq})
with the spectral densities ${\cal P}_f$ replaced by unity,
$ F_{(0)\;f}(r,k) = {\cal N}_f\left(r, k\right) \,\otimes\, 1 $,
i.e.  the
phase-space densities of gluons and quarks that measure the number of
non-interacting quanta in a phase-space element $d^3rd^4p$ at a given time
$t=r^0$.
Their presence is a direct consequence of the CTP formulation which
incorporates initial state correlations due to a non-trivial density
matrix $\hat{\rho}(t_0)$, eq. (\ref{rho}), corresponding
to $F(t_0,\vec r, p)\ne 0$, as opposed to the usual quantum field
theory description, where $\hat{\rho}(t_0)=|0\rangle \langle 0|$
and $F(t_0,\vec r, p)$ vanishes.
It is evident that in this latter case $G^>=G^<=0$ and
$G^F=G^{\overline{F}\,\dagger}$
at all times, so that the dynamics is describe by the Feynman propagators $G^F$
alone.

More suitable for practical purposes, one may
employ instead of the set
$G^F,G^>,G^<,G^{\overline{F}}$,
an equivalent  set of
the {\it retarded (advanced) propagators} $G^R$ ($G^A$) plus
the {\it correlation function} $G^C$. The latter are directly connected with
physical
observable quantities, and are commonly referred to as {\it physical
representation} \cite{chou}.
The functions $G^R,G^A,G^C$ are obtained via the relations
$$
G^R \;=\;
G^F \;-\; G^> \;=\; G^<\;-\; G^{\overline{F}}
\;,\;\;\;\;\;\;\;\;\;\;\;
G^A \;=\;
G^F \;-\; G^< \;=\; G^>\;-\; G^{\overline{F}}
\;,
$$
\begin{equation}
G^C \;=\;
G^F \;+\; G^{\overline{F}} \;=\; G^<\;+\; G^>
\;.
\label{rac}
\end{equation}
Because the fourth possible linear combination
$G^F- G^> - G^<+ G^{\overline{F}}$ is always identically zero, the three
physical functions
$G^R,G^A, G^C$ form a complete alternative set that eliminates
the overdetermination of the set
$G^F,G^>,G^<,G^{\overline{F}}$.
The `free-field' forms of $G^R, G^A$ and $G^C$, corresponding to the ones of
(\ref{D00}) and (\ref{S00}) are
$$
D_{(0)\,\mu\nu}^{R\,(A)}(r,k)
\;=\;
\frac{-d_{\mu\nu}(k)}{k^2 \pm  i \varepsilon k_0}
\;\;\;\;\;\;\;\;\;\;\;\;\;\;\;\;\;
S_{(0)}^{R\,(A)}(r,p)\;=\;
\frac{1}{\gamma\cdot p \pm  i \varepsilon p_0}
$$
\begin{eqnarray}
D_{(0)\,\mu\nu}^{C}(r,k)&=&
- 2\pi i \; \left(-d_{\mu\nu}(k)\right)\; \left[ 1+ 2\,F_{(0)\;g}(r,k)\right]
\;\delta(k^2-\mu_{gq}^2)
\label{S000}
\\
S_{(0)}^{C}(r,p)&=&
- 2\pi i \;  \left(\gamma\cdot p\right)\;\left[ 1-  2\,F_{(0)\;q}(r,p)\right]
\;\delta(p^2-\mu_{gq}^2)
\nonumber
\;,
\end{eqnarray}
where $+ (-)$ in the denominators corresponds to the $R$ ($A$).
Generally speaking, the retarded and advanced functions characterize
the quantum nature of parton states, whereas the correlation function describes
the phase-space occupation of these states.

The preceding relations (\ref{rac}) are generally valid for any 2-point
function
defined on the closed-time path, and hence apply to
not only to the free-field case, but
also to
the full Green functions $G=D_{\mu\nu}, S$, as well as to the self-energies
${\cal E}=\Pi_{\mu\nu}, \Sigma$.
In matrix form, the correspondance between the representation (\ref{G22}) in
terms of $G$ (${\cal E}$), and the physical representation denoted by
$\breve{G}$ ($\breve{{\cal E}}$), is given by a unitary transformation
${\cal U}$, being a 2$\times$2 matrix with
${\cal U}_{mn}=1/\sqrt{2}$ for $mn=11,21,22$ and
${\cal U}_{12}=-1/\sqrt{2}$:
\begin{equation}
\breve{G}
\;=\;
{\cal U} \;G\; {\cal U}^{-1} \;=\;
\left(
\begin{array}{cc}
0\;  & \; G^A \\
G^R\;  & \; G^C
\end{array}
\right)
\;\;, \;\;\;\;\;\;\;\;\;\;
\breve{{\cal E}}\;=\;
{\cal U}^{-1} \;{\cal E}\; {\cal U} \;=\;
\left(
\begin{array}{cc}
{\cal E}^C\;  & \; {\cal E}^R \\
{\cal E}^A\;  & \; 0
\end{array}
\right)
\;.
\label{GRA}
\end{equation}
where, in subtle contrast to (\ref{rac}),
$$
{\cal E}^A \;=\;
{\cal E}^F \;+\; {\cal E}^> \;=\; -\,{\cal E}^<\;-\; {\cal E}^{\overline{F}}
\;,\;\;\;\;\;\;\;\;\;\;\;
{\cal E}^R \;=\;
{\cal E}^F \;+\; {\cal E}^< \;=\; -\,{\cal E}^>\;-\; {\cal E}^{\overline{F}}
\;,
$$
\begin{equation}
{\cal E}^C \;=\;
{\cal E}^F \;+\; {\cal E}^{\overline{F}} \;=\; -\,{\cal E}^<\;-\; {\cal E}^>
\;.
\label{rac2}
\end{equation}
The great advantage of this  physical representation is that the dependence
on the partons' phase-space densities $F_g$ and $F_q$
is essentially carried by the correlation functions $G^C$, whereas the
dependence
of the retarded and advanced functions, $G^R$, $G^A$, is weak.
In the free-field case, this separation of correlations is exact,
as is evident from (\ref{S000}),
such that the retarded and advanced functions do not depend at all on $F_g$,
$F_q$.
In fact, even in the general case of interacting fields, this advantagous
property
becomes very suggestive  when rewriting the renormalization and transport
equations,
(\ref{R}) and (\ref{T}),
in generic form for the individual Green function components,
\begin{eqnarray}
\left\{ G_{(0)}^{-1}\, , \,  G^{R}-G^{A}\right\}_+ &=&
-2 \,\left(k^2 -\frac{1}{4} \partial_r^2\right) \,  \left(G^{R}-G^{A}\right)
\;=\;
\left\{ \delta {\cal E}\, , \,  {\cal P}\right\}_+
\;+\;
\left\{ \Gamma\, , \,  \delta G\right\}_+
\label{X1}
\\
\left[ G_{(0)}^{-1}\, , \,  G^{C}\right]_- &=&
-2i\,k\cdot \partial_r  \,  G^{C}
\;=\;
\left[ {\cal E}^C\, , \,  \delta G\right]_-
\;+\;
\left[ \delta {\cal E}\, , \,  G^{C}\right]_-
\nonumber \\
& &
\;\;\;\;\;\;\;\;\;\;\;\;
\;\;\;\;\;\;\;\;\;\;\;\;
\;+\;
\frac{i}{2} \,\left(
\frac{}{}
\left\{ {\cal E}^C\, , \,  {\cal P}\right\}_+
\;+\;
\left\{ \Gamma\, , \,  G^{C}\right\}_+
\right)
\;,
\label{X2}
\end{eqnarray}
where
\begin{eqnarray}
\delta G &\equiv&
\mbox{Re} G\;=\; \frac{1}{2}\, \left( G^R+G^A\right)
\;\;\;\;\;\;\;\;\;\;\;\;
\delta {\cal E} \;\equiv\;
\mbox{Re} {\cal E}\;=\; \frac{1}{2} \,\left( {\cal E}^R+{\cal E}^A\right)
\nonumber \\
{\cal P} &\equiv&
\mbox{Im} G\;=\; i \,\left( G^R-G^A\right)
\;\;\;\;\;\;\;\;\;\;\;\;
\;\;\;
\Gamma \;\equiv\;
\mbox{Im} {\cal E}\;=\; i \, \left( {\cal E}^R-{\cal E}^A\right)
\label{X3}
\end{eqnarray}
are the real and imaginary components of the
retarded and advanced Green functions and self-energies,
whereas
\begin{equation}
G^C\;=\;  G^<\,+\,G^>
\;\;\;\;\;\;\;\;\;\;\;
{\cal E}^C\;=\;  - ({\cal E}^<\,+\,{\cal E}^>)
\end{equation}
are the real correlation functions and corresponding self-energies.
The physical significance of the (\ref{X1}) and (\ref{X2}) is
the following:
Eq. (\ref{X1}) determines the state of a dressed parton with
respect to their virtual fluctuations and real emission (absorption) processes,
corresponding to the real and imaginary parts of the retarded and advanced
self-energies.
Eq. (\ref{X2}), on the other hand characterizes the correlations
mong different dressed parton states,
and the self-energies appear here in two distinct ways.
The first two terms on the right hand side account for scatterings between
quasi-particle
states, i.e. dressed partons, whereas the last two terms incorporate the
renormalization effects
which result from the fact that the dressed partons between collisions do not
behave as
free particles, but change their dynamical structure due to virtual
fluctuations,as well as real emission and absorption of quanta.
For this reason ${\cal E}^{R(A)}$ are called {\it radiative} self-energies, and
${\cal E}^C$
is termed {\it collisional} self-energy.
As shown by Kadanoff and Baym \cite{baym}, the imaginary parts of
the retarded and advanced Green functions  and self-energies (\ref{X3}) are
just the spectral density ${\cal P}$, giving the probability for finding an
intermediate
multi-particle state in the dressed parton, respectively the decay width
$\Gamma$,
describing the dissipation of the dressed parton. The general solution for
${\cal P}$ is
given by
\begin{equation}
{\cal P}(r,k)\;=\;
\frac{\Gamma}{k^2 \,-\,\delta {\cal E}\,+\,(\Gamma/2)^2}
\;\,\equiv\;\,
\,\Delta {\cal P}_{\delta {\cal E}}
\;+\;\Delta {\cal P}_{\Gamma}
\;,
\label{X4}
\end{equation}
where the second form in terms of the `wavefunction'-renormalization
($\Delta {\cal P}_{\delta {\cal E}}$) due to
virtual fluctuations, and
the dissipative parts ($\Delta {\cal P}_{\Gamma}$)
due to real emission (absorption) processes,
will prove convenient later
\footnote{
This formula holds for both,
{\it space-like} ($k^2 < 0$) and time-like ($k^2 > 0$) momenta.
If is $k^2$ space-like then the imaginary part
$\Delta {\cal P}_{\Gamma}$ vanishes, so that ${\cal P}$ is purely real.
On the other hand, if $k^2$ is time-like then both
$\Delta {\cal P}_{\delta {\cal E}}$ and
$\Delta {\cal P}_{\Gamma}$ contribute, and so ${\cal P}$ is complex.
}
{}.
The spectral density ${\cal P}$
satisfies the sum rule \cite{chou,baym}
\begin{equation}
1\;\,=\;\,
\frac{1}{P^+}
\int \frac{dk^+}{2\pi\,k^+}\,T^{++}\,{\cal P}(r,k^+,k^2)
\;=\;
\frac{1}{P^+}
\int \frac{dk^+}{2\pi}\,k^+\,{\cal P}(r,k^+,k^2)
\label{X6}\;,
\end{equation}
which is an implicit consequence of unitarity, and
requires that the total lightcone momentum of the spectral density of
the internal bare partons must be equal to the dressed partons' momentum.
For example, in the `free-field' case,
i.e. in the absence of interactions, one has
$\Delta {\cal P}_{\delta {\cal E}}=\delta(1-k^+/P^+)$
and
$\Delta {\cal P}_{\Gamma}=0$ with $k^2=\mu_{gq}^2$, so
that ${\cal P}\rightarrow {\cal P}_{(0)}$
describes a single `on-shell' parton state
\begin{equation}
{\cal P}_{(0)}(r,k^+,k^2)\;=\; \delta (k^2 - \mu_{gq}^2)\;
\delta\left(1-\frac{k^+}{P^+}\right)
\label{X5}
\;,
\end{equation}
on the `mass-shell' $k^2=\mu_{gq}^2$, and carrying the total lightcone
energy $k^+ = P^+$
\footnote{
Note that for the choice (\ref{mumu}), the fraction $z=x/y$ in the
defining equation for ${\cal P}$ (\ref{Pgq}) reduces to $z=x$.
}
{}.
This is nothing but the fact that the presence of a pole in the Green function
means the presence of a particle, stable if it occurs for real $k^2$, unstable
if
it occurs for complex $k^2$, as in the Breit-Wigner formula (\ref{X4}).
The generalization of (\ref{X5}) to the case of interactions,
in which, as advocated before, a dressed parton may be visualized as a
substructured particle
with a fluctuating number of bare quanta intermediately present in its
wavefunction,
is straightforward. A dressed parton has now a ``blurred'' mass shell,
because its internal excitations fluctuate due to virtual and real
emission (absoprtion) processes of its bare daughter partons.
The spectrum of these quantum excitations will have a finite
extension around $\mu_{gq}^2$, described by the real part
$\delta {\cal E}$ of the self-energy, with a width $\Gamma$,
described by the imaginary part and being inversely proportional to
the life-time of the particular parton state.
Hence, one may write formally instead of (\ref{X5}),
\begin{equation}
{\cal P}(r,k)\;=\;
\delta\left(\frac{}{} k^2 \;-\;{\cal M}^2(r,k^+,k^2)\right)
\;\;,\;\;\;\;\;\;\;\;\;\;
{\cal M}^2\;=\;
\delta {\cal E}\;-\;\frac{\Gamma ^2}{4}
\label{offshell}
\;,
\end{equation}
where
$\delta{\cal E}$ and $\Gamma$ are given in terms of the real and imaginary
parts of the
retarded and advanced self-energies (\ref{X3}).
This representation serves to maintain the analogy with the free-field case,
for
which one has an immediate intuition. However, instead of a
simple mass-shell condition, the argument of the $\delta$-function now
expresses a non-trivial functional dependence of the spectrum on $k^+,k^2$ and,
in general, on space-time $r$. The solution of this implicit equation
determines
the spectral density ${\cal P}$, which is subject of Sec. 3.4.

Once the spectral density is known, the correlation function $G^C$ is
given by the generic expression \cite{baym}
\begin{equation}
G^C(r,k) \;=\; -2\pi i \;\left[1\pm 2 {\cal N}(r,k)\right] \,\otimes\, {\cal
P}(r,k)
\;=\; -2\pi i\; \left[ 1\,\pm \,2 \,F(r,k)\right]\;
\delta\left( k^2 \;-\;{\cal M}^2(r,k)\right)
\;,
\label{GCF}
\end{equation}
where $+(-)$ is for gluons (quarks).
It reduces to the free-field form, when ${\cal P}$ is replaced
by ${\cal P}_{(0)}$, eq. (\ref{X5}), so that
$
G^C_{(0)}= -2\pi i\left[1\pm 2 {\cal N}\right]\delta (k^2-\mu_{gq}^2)
= -2\pi i \left[ 1\pm 2 F_{(0)}\right]\delta (k^2-\mu_{gq}^2)
$
becomes an `on-shell' distribution, as in eqs. (\ref{S000}).
The theoretical basis for the previous, more physically
motivated, ansatz (\ref{Fgq}) for the parton phase-space distributions $F$,
becomes
evident now:
It is the logical generalization of the free-field forms (\ref{S000})
to include renormalization effects and dissipation in terms of  non-trivial
spectral densities, or parton structure functions, which embody the underlying
quantum dynamics. In this sense the Wigner functions $F$ are the quantum
kinetic
extension  of the classical particle phase-space distributions.

Following this strategy, I will now proceed on the basis of
the factorized ansatz (\ref{Fgq}) for the gluon and quark densities
$F_f$ in terms
of the quasi-particle distributions ${\cal N}_f$ with the
spectral densities ${\cal P}_f$, i.e. the presumption that
the separation between quantum and kinetic scales allows
a distinct treatment of the intrinsic quantum fluctuations of dressed partons
and the kinetic correlations among them.
In contrast to (\ref{S000}), the poles
of the retarded and advanced Green functions are shifted by the
real and imaginary parts of the self-energies ${\cal E}^{R(A)}$,
and in the expression for the correlation functions,
the $\delta$-function is replaced by the spectral density ${\cal P}$.
Introducing the scalar functions for the $\hat \Pi$ for the gluon and $\hat
\Sigma$ for
the quark self-energies through
\begin{equation}
\Pi^{\mu\nu}_{ab}\;=\;
\delta_{ab}\,
\left(k^\mu k^\nu \,- \,g^{\mu\nu}\,k^2\right) \;\hat{\Pi}
\;\;\;\;\;\;\;\;\;\;\;\;\;
\Sigma_{ij}\;=\;
\delta_{ij}\;p^2\;\hat{\Sigma}
\label{hatPiS}
\;,
\end{equation}
instead of (\ref{S000}) one has now
\begin{equation}
D_{\mu\nu}^{R\,(A)}(r,k)\;=\;
\frac{-d_{\mu\nu}(k)}{k^2\,\left(1 -\hat{\Pi}^{R(A)}\right)}
\;\;\;\;\;\;\;\;\;\;\;\;\;\;\;
S^{R\,(A)}(r,p)\;=\;
\frac{\gamma\cdot p}{p^2\,\left( 1 - \gamma\cdot p
\,\hat{\Sigma}^{R(A)}\right)}
\nonumber
\end{equation}
\begin{eqnarray}
D_{\mu\nu}^{C}(r,k)
&=&
- 2\pi i \, \left(-d_{\mu\nu}(k)\right)\, \left[ 1+ 2\,{\cal N}_g(r,k)\right]
\,\otimes\, {\cal P}_g(r,k)
\nonumber \\
&=&
- 2\pi i \, \left(-d_{\mu\nu}(k)\right)\, \left[ 1\,+\, 2\,F_g(r,k)\right]
\;\delta\left( k^2 \;-\;{\cal M}_g^2(r,k)\right)
\nonumber
\\
S^{C}(r,p)&=&
- 2\pi i \,  \left(\gamma\cdot p \right)\;\left[ 1-  2\,{\cal N}_q(r,p)\right]
\,\otimes\, {\cal P}_q(r,p)
\nonumber \\
&=&
- 2\pi i \,  \left(\gamma\cdot p \right)\;\left[ 1\,-\, 2\,F_q(r,p)\right]
\;\delta\left( p^2 \;-\;{\cal M}_q^2(r,p)\right)
\;.
\label{SIII}
\end{eqnarray}
\smallskip

\noindent
The two-step strategy that I will follow in the next Sections is then:

\noindent
{\bf 1)}
In Sec. 3.4, the renormalization equations (\ref{R}) will be solved
for the retarded (advanced) Green functions
\begin{eqnarray}
\left( D_{R(A)}^{-1}\right)_{ab}^{\mu\nu}(r,k)
&=&
\left( D_{(0)\,R(A)}^{-1}\right)_{ab}^{\mu\nu}
\;-\;
\left(\Pi_{R(A)}\right)_{ab}^{\mu\nu}
\nonumber \\
\left( S_{R(A)}^{-1}\right)_{ij}(r,p)
&=&
\left( S_{(0)\,R(A)}^{-1}\right)_{ij}
\;-\;
\left(\Sigma_{R(A)}\right)_{ij}
\label{DSE5}
\;,
\end{eqnarray}
which determine
the spectral densities ${\cal P}_g$ and ${\cal P}_q$  in terms of the radiative
self-energies $\Pi_{\mu\nu}^{R(A)}$ and $\Sigma^{R(A)}$.

\noindent
{\bf 2)}
In Sec. 3.5 the transport equations will be solved for the correlation
functions
\begin{eqnarray}
D_{C\,ab}^{\mu\nu} (r,k)
&=&
-\,D_{R\,aa'}^{\mu\mu'}
\left[ \frac{}{}
\left( D_{(0)\,C}^{-1}\right)_{a'b'}^{\mu'\nu'}
-
\left(\Pi_{C}\right)_{a'b'}^{\mu'\nu'}
\right]
\,D_{A\,b'b}^{\nu'\nu}
\nonumber
\\
S_{C\,ij}(r,p)
&=&
-\,S_{R\,ii'}
\;\left[ \frac{}{}
\left( S_{(0)\,C}^{-1}\right)_{i'j'}
\;-\;
\left(\Sigma_{C}\right)_{i'j'}
\right]
\;S_{A\,i'j}
\label{GC}
\;,
\end{eqnarray}
which determine the  parton phase-space distributions $F_g$ and $F_q$ by the
collisional
self-energies $\Pi_{\mu\nu}^C$ and $\Sigma^C$,
in conjunction with the spectral densities ${\cal P}_g$ and ${\cal P}_q$.
\bigskip


\noindent {\bf 3.4 Quantum dynamics and renormalization equations}
\bigskip

As advocated above, when
addressing  renormalization effects and dissipative quantum dynamics,
it is appropriate to focus on the retarded and advanced propagators
and the imaginary parts of the self-energies, which embody the
short-distance propagation of quantum fluctuations.
Furthermore, on quantum scales, one can neglect
the $r$-dependence,
and thus ignore in this regime the functions ${\cal G}$ and ${\cal F}$ in eqs.
(\ref{R}) and
(\ref{T}). Then, by performing the transformation of the
Wigner transformed Green functions $D_{\mu\nu}$, $S$ and
self-energies $\Pi_{\mu\nu}$, $\Sigma$ to the  physical representation via
(\ref{GRA}),
one obtains simplified equations for the retarded (advanced) functions (c.f.
Appendix B).
The renormalization equations reduce to the following form:
\begin{eqnarray}
\left(k^2 \,-\,\frac{1}{4} \partial_r^2 \right)
\; D_{ab}^{\mu\nu\;R(A)} (r,k)
&=&
-\,d^{\mu\nu}(k)\,\delta_{ab}
\;+\;
\frac{1}{2} \;\left(\frac{}{}\Pi^{R(A)}\, D^{R(A)}\;+\; D^{R(A)}\,
\Pi^{R(A)}\right)^{\mu\nu}_{ab}
\nonumber
\\
\frac{1}{2} \;\left\{ \gamma\cdot p\,,\, S_{ij}^{R(A)}(r,p)\right\}_+
&=&
\delta_{ij}\;
\;+\;
\frac{1}{2} \;\left(\frac{}{} \Sigma^{R(A)}\, S^{R(A)} \;+\; S^{R(A)}\,
\Sigma^{R(A)} \right)_{ij}
\label{R2}
\end{eqnarray}
To solve the equations (\ref{R2}),
it is suggestive in view of (\ref{S000}),
to parametrize the renormalized, `dressed' propagators $D_{\mu\nu}^R$, $S^R$
on account of their Lorentz structure as  \cite{dok80,amati80}
\begin{eqnarray}
D^{\mu\nu\;R(A)}_{ab} (r,k) &=&
\delta_{ab}\,
\frac{- d^{\mu\nu}(k)}{k^2\,(1  -\hat{\Pi}^{R(A)}(r,k))}
\;\equiv\;
\Delta_g(r,k^2,\kappa_k)\;\delta_{ab}\,\frac{- d_{\mu\nu}(k)}{k^2\pm
i\varepsilon k_0}
\;\,+\;\ldots
\nonumber
\\
S^{R(A)}_{ij}(r,p) &=&
\delta_{ij}\,
\frac{\gamma\cdot p}{p^2 \,( 1   - \gamma\cdot p \hat{\Sigma}^{R(A)}(r,p))}
\nonumber \\
&\equiv&
\Delta_q(r,p^2,\kappa_p)\;\delta_{ij}\,\frac{\gamma\cdot p}{p^2\pm i\varepsilon
p_0}
\,\;+\;\, \tilde\Delta_q(r,p^2,\kappa_p)\;\delta_{ij}\,\frac{\gamma\cdot
n}{n\cdot p}
\;\,+\;\ldots
\label{ansatz1}
\;,
\end{eqnarray}
where
$d_{\mu\nu}(k)=g_{\mu\nu} - (n_\mu k_\nu+n_\nu k_\mu)/(n\cdot k)$ as before,
the scalar self-energy functions $\hat{\Pi}^R$, $\hat{\Sigma}^R$
are defined by (\ref{hatPiS}), and
$\kappa$ implicitly accounts for the dependence of $\Delta_g$ and $\Delta_q$ on
the
coordinate $r$, which is conjugate to $k^+$.
The function $\kappa$ is of the order of the  large
lightcone momentum $k^+$ squared (c.f. Appendix C)
\begin{equation}
\kappa_k^2 \;\equiv \; \frac{(n\cdot k)^2}{n^2}
\;\,\simeq\;\, k^{+\;2}
\;\;\;\;\;,\;\;\;\;\;\;\;\;
n^2 \ll 1
\;.
\label{kappa}
\end{equation}
The {\it renormalization functions} $\Delta_g$ ($\Delta_q$)
account for the modifications of the
`bare' propagators (\ref{S000}) due to
the self-interactions embodied in $\Pi$ ($\Sigma$).
The third function $\tilde{\Delta}_q$  turns out to be
proportional to $\Delta_q$ (c.f. Appendix C).
They are normalized in accord with the condition (\ref{scale3}), such that
\begin{equation}
\left. \Delta_g(r,k^2,\kappa_p)\right|_{k^2=\mu_{gq}^2} \;=\;
\left. \Delta_q(r,p^2,\kappa_p)\right|_{p^2=\mu_{gq}^2}
\;=\; 1
\label{norm1}
\;,
\end{equation}
meaning that a gluon or quark is considered as a
maximally dressed particle (in the sense of the applicability of
(\ref{scale3})), corresponding to the invariant scale
$\mu_{gq}$, which may be called the dressed partons mass-shell.

It is well known \cite{dok80} that other contributions to
the propagators in (\ref{ansatz1}), indicated by the dots, are strongly
suppressed
\footnote{
Note the benefit of the employed gauge
$n_\mu A^\mu=0$ for the gluon field, eq.  (\ref{gauge}):
by suitable choice of the vector $n^\mu$ such that $n^2 \ll 1$,
one can concentrate the short-distance
quantum fluctuations to arbitrary proximity
around the lightcone, $n^2\rightarrow 0$, i.e. $\kappa_k\rightarrow \infty$,
corresponding to the asymptotic linit $k^+\rightarrow \infty$.
In this regime the leading log singularities of the propagators
give the dominant contributions and the dotted terms
in (\ref{ansatz1}) and (\ref{ansatz2})
can be neglected because they do not generate leading logs.
}
in lightcone-dominated processes $Q^2\rightarrow \infty$ by inverse powers of
$Q^2$.
In fact, this feature is the very foundation of the QCD parton description
within the (modified) leading log approximation (MLLA) \cite{dok80,amati80},
where
the renormalization problem reduces to a multiplication of the bare
propagators and vertices by scalar functions.
That is, as depicted in Fig. 9, with respect to the Wigner transforms of the
self-energies
(\ref{Pi}) and (\ref{Sigma}),
\begin{eqnarray}
\!\!\!\!\!\!\!\!
\Pi^{\mu\nu\,R(A)}_{ab}(r,k)
&=&
\;\;\;
- \frac{g_s^2}{2}\,
\int \frac{d^4 k'}{(2\pi)^4i}
\,f_{aa'c} \lambda_{\mu\mu'\sigma}(-k,-k+k',k')\,
D_{cd}^{\sigma\tau\,R(A)}(r,k')\,
\nonumber
\\
& &
\;\;\;\;\;\;\;\;\;\;
\Lambda_{db'b}^{\tau\nu'\nu}(-k',k-k',k)\,
D_{a'b'}^{\mu'\nu'\,R(A)}(r,k-k')
\label{Pi1l}
\\
& &
\;
- \,g_s^2\,N_f
\int \frac{d^4 k'}{(2\pi)^4i}
\,\gamma_{\mu}T_{ln}^{a} \, S_{nn'}^{R(A)}(r,-k'+k)
\,\Xi_{n'l'\;\nu}^{b}(-k'+k,k,k') \, S_{ll'}^{R(A)}(r,k')
\nonumber
\\
\!\!\!\!\!\!\!\!
\Sigma^{R(A)}_{ij}(r,p)
&=&
\; \,g_s^2
\int \frac{d^4 k'}{(2\pi)^4i}
\,\gamma_{\sigma}T_{ii'}^{c} \, S_{i'j'}^{R(A)}(r,p-k')
\,\Xi_{j'j\;\tau}^{d}(p-k',p,k')
\, D_{cd}^{\sigma\tau\,R(A)}(r,k')
\nonumber
\;,
\end{eqnarray}
one can represent the  $qqg$- and $ggg$  vertex functions
$\Gamma$, respectively $\Lambda$, eq. (\ref{VG}),
as multiplicative renormalization functions,
\begin{eqnarray}
\Gamma^{\mu\,a}_{ij} (r;\,p_1,p_2,k_3)\;=\; \gamma^\mu\;T^a_{ij}\;
V_{qqg}(r;\, p_1^2,p_2^2,k_3^2, \kappa_1,\kappa_2,\kappa_3)
\;\,+\;\ldots
\nonumber \\
\Lambda^{abc}_{\mu\nu\lambda} (r;\,k_1,k_2,k_3)\;=\;
f^{abc}\;g_{\mu\lambda}\,k_\nu\; V_{ggg}(r;
\,k_1^2,k_2^2,k_3^2,\kappa_1,\kappa_2,\kappa_3)
\;\,+\;\ldots
\label{ansatz2}
\;,
\end{eqnarray}
where $k_\nu$ is understood as the momentum associated with the
intermediate of the three gluon virtualities, i.e.
$k_\nu=k_{2\,\nu}$ if $k_1^2 < k_2 ^2 < k_3^2$, etc..
Analogous to (\ref{norm1}), the normalization conditions are
\begin{equation}
\left.
V_{qqg}(r;\, p_i^2, \kappa_i) \right|_{p^2_i=\mu_{gq}^2}
\;=\;
\left.
V_{ggg}(r;\, k_i^2, \kappa_i) \right|_{k^2_i=\mu_{gq}^2}
\;=\;1
\label{norm2}
\end{equation}

Employing these definitions in the
renormalization equations (\ref{R2}) and the expressions for
self-energies given by (\ref{Pi})
and (\ref{Sigma}),
and differentiating the inverse of the propagators
$D_{\mu\nu}^{R(A)}$ and $S^{R(A)}$ with respect to the gluon and quark
virtuality, respectively,
\begin{eqnarray}
\frac{\partial}{\partial k^2}  \left(-d^{\mu\nu}(k)\,
[D^{R(A)}(r,k)]_{\mu\nu}^{-1}\right)
&=&
1\;-\; \frac{\partial}{\partial k^2} \hat{\Pi}^{R(A)}(r,k)
\nonumber
\\
\frac{\partial}{\partial p^2}  \left(\gamma\cdot p\,[S^{R(A)}(r,p)]^{-1}\right)
&=&
1\;-\; \frac{\partial}{\partial p^2} \hat{\Sigma}^{R(A)}(r,p)
\;
\end{eqnarray}
one obtains
the
following determining equations for the momentum dependence of
the scalar self-energy functions (\ref{hatPiS}),
$\hat{\Pi}^{R(A)}$ and $\hat{\Sigma}^{R(A)}$ to order $g_s^2$ in terms of the
renormalization
functions $\Delta_g$ and $\Delta_q$,
\begin{eqnarray}
& &
\!\!\!\!\!\!\!\!\!\!\!\!\!\!
k^2 \frac{\partial}{\partial k^2} \hat{\Pi}^{R(A)}(r,k)
\;=\;
\frac{g_s^2}{2}\,C_A \,(2\pi i) \;\int \frac{d^4 k'}{(2\pi)^4 i}
\int_{r_0-1/(2\mu(r))}^{r_0+1/(2\mu)} d\tau\,{\cal T}(k'\tau)
\nonumber
\\
& &
\;\times\;
\;
\frac{\partial \Delta_g(r,k')}{\partial k'^{\,2}}
\;
\frac{\partial}{\partial k''^{\,2}}
\left(\frac{}{}
V^2_{ggg}(r;\, k^2,k'^2,k''^2, \kappa_k,\kappa_{k'}, \kappa_{k''})
\,
\Delta_g(r,k'')
\right)
\times
{\cal U}_g^{gg}(k',k'',n)
\nonumber
\\
& &
\;-\;
g_s^2\,Tr \,N_f \,(2\pi i) \;\int \frac{d^4 p'}{(2\pi)^4 i}
\int_{r_0-1/(2\mu)}^{r_0+1/(2\mu)} d\tau\,{\cal T}(p'\tau)
\label{dg1}
\\
& &
\;\times\;
\;
\frac{\partial \Delta_q(r,p')}{\partial p'^{\,2}}
\;
\frac{\partial}{\partial p''^{\,2}}
\left(\frac{}{}
V^2_{gqq}(r;\, k^2,p'^2,p''^2, \kappa_k,\kappa_{p'}, \kappa_{p''})
\,
\Delta_q(r,p'')
\right)
\times
{\cal U}_g^{qq}(p',p'',n)
\nonumber
\\
& & \nonumber \\
& &
\!\!\!\!\!\!\!\!\!\!\!\!\!\!
p^2 \frac{\partial}{\partial p^2} \hat{\Sigma}^{R(A)}(r,p)
\;=\;
- \;g_s^2\,C_F \,(2\pi i) \;\int \frac{d^4 p'}{(2\pi)^4 i}
\int_{r_0-1/(2\mu)}^{r_0+1/(2\mu)} d\tau\,{\cal T}(p'\tau)
\label{dq1}
\\
& &
\;\times\;
\;
\frac{\partial \Delta_q(r,p')}{\partial p'^{\,2}}
\;
\frac{\partial}{\partial k''^{\,2}}
\left(\frac{}{}
V^2_{qqg}(r;\, p^2,p'^2,k''^2, \kappa_p,\kappa_{p'}, \kappa_{k''})
\,
\Delta_g(r,k'')
\right)
\times
{\cal U}_q^{qg}(p',k'',n)
\;,
\nonumber
\end{eqnarray}
plus terms $O(n^2)$ which can be neglected for $n^2 \ll 1$.
The constants  $C_A\delta_{ab} = f_{acd}f_{bcd}=N_c\delta_{ab}$,
$T_R\delta_{ab} = Tr(T^a\cdot T^b)= \frac{1}{2} \delta_{ab}$,
$C_F  \delta_{ij}= (T^a\cdot T^a)_{ij}= \frac{N_c^2-1}{2N_c}\delta_{ij}$,
arise from summing over color indices,
and $N_f$ is the number of quark flavors.
The time integral $\int d\tau$ on the right hand sides extends over the finite
time slice $\mu^{-1}\equiv\mu^{-1}(r)$ of the space-time cell around $r$,
weighted by the function ${\cal T}$ \cite{ms17},
which incorporates the relation between
momentum and space-time as  constrained by the uncertainty principle:
it limits the range of virtualities $k^{'\,2}$ such that
within the finite time interval $\Delta r_0=\mu^{-1}$ only those fluctuations
$k\rightarrow k'k''$ are resolvable that are sufficiently short-living,
with proper life-time $\tau_0\simeq 1/k'$ and $\gamma\tau_0 \simeq k^+/k{'\,2}
< \Delta r_0$.
Finally, the functions ${\cal U}_f^{f'f''}(k',k'',n)$ represent the
squared matrix elements for the virtual decay processes $k\rightarrow k'k''$.

In the cellular space-time picture, the momenta of partons in a given cell
around
$r$ are
per design limited by the condition (\ref{scale3}), such that
\begin{equation}
P^{+\,2}(r)\;\, \ge \,\;k^{+\,2} \,\;\ge \;\;k^2 \;\;\ge \;\,\mu_{gq}^2
\,\;\ge\;\,\mu^2(r)
\label{scale4}
\;.
\end{equation}
As explained in Appendix C,
by employing this condition and
introducing the fractional lightcone momenta of the
daughter partons in the process $k\rightarrow k'k''$,
\begin{equation}
z_{k'} \;\equiv\;
z\;=\;\frac{k'^{\,+}}{k^+}
\;\;,\;\;\;\;\;\;
z_{k''} \;\equiv\;
1-z\;=\;\frac{k''^{\,+}}{k^+}
\;,
\end{equation}
\begin{equation}
d^4 k' \;=\; \frac{\pi}{2} \,dk^{'\,2} dk^{''\,2} \, dz \;
\theta\left( k^2 - \frac{k^{' \,2}}{z} - \frac{k^{''\,2}}{1-z}\right)
\;.
\end{equation}
the integrals (\ref{dg1}) and (\ref{dq1}) are readily evaluated
to leading log accuracy.
Solving for the renormalization functions $\Delta_g$, $\Delta_q$,
and taking $\kappa_k=k^{+\,2}, \kappa_p=p^{+\,2}$ from (\ref{kappa}),
the result is (c.f. Appendix C):
\begin{eqnarray}
\Delta_g(r,k^2,k^{+\,2})
&=&
\exp\left\{
-\,\int_{k^2}^{k^{+\,2}}
\frac{d k^{'\,2}}{k^{'\,2}}
\int_0^1 dz \;
A(r,z,k^{'\,2})
\;\left(
\,\frac{1}{2} \gamma_g^{gg}(z,\epsilon)
\,+\,
\,\gamma_g^{qq}(z,\epsilon)
\right)
\right\}
\nonumber
\\
\Delta_q(r,p^2,p^{+\,2})
&=&
\exp\left\{
-\,\int_{p^2}^{p^{+\,2}}
\frac{d p^{'\,2}}{p^{'\,2}}
\int_0^1 dz \;
A(r,z,p^{'\,2})
\,\gamma_q^{qg}(z,\epsilon)
\right\}
\;,
\label{dq2}
\end{eqnarray}
where
\begin{equation}
A(r,k^2,z)\;\simeq\; \frac{\alpha_s\left((1-z)k^2\right)}{2\pi}\;
\,\theta\left( \frac{z k^2}{k^+}\,-\,\mu(r) \right)
\label{As}
\end{equation}
is the effective local coupling strength
\footnote{
In (\ref{As}) one could imagine,
instead of the $\theta$-function $\theta(k^2-\mu^2)$,  a smeared-out
probability distribution, e.g. $\propto \exp(-k^2/\mu^2)$, by
choosing a more refined form for the function ${\cal T}$ under the
time integral in eqs. (\ref{dg1}) and (\ref{dq1}).
The specific choice is ambigous at this level of calculation (see Ref.
\cite{ms17} for details).
}
averaged over a
cell centered around $r$ of
space-time extent $\Omega(r)\simeq\Delta^4 r$,
and  $\alpha_s(q^2)= b\; \ln \left(q^2/\Lambda_{QCD}^2\right)]^{-1}$, $b=11
N_c- 2 N_f$.
The  function $A(r,z,k^2)$ reflects the fact that
in cellular space-time the relevant quantum fluctuations are
restricted by the uncertainty principle, as embodied in the $\theta$-function
term: $k^2 \le \mu^2 = 1/(\Delta r)^{2}$, $k^0 \le 1/\Delta r^0$, with
$\Delta r^0 = k^+/k^2$.

The functions
$\gamma(z,\epsilon)$ in (\ref{dq2})
involve, at leading-log level, the standard DGLAP kernels \cite{dok80},
carrying an explicit $k^+$-dependence arising from the dependence on $\kappa
\simeq k^{+\;2}$,
\begin{eqnarray}
\gamma_{g}^{g g} (z,\epsilon) &=&
2\,C_A\;\left( \frac{z}{1-z+\epsilon(\kappa_k)} + \frac{1-z}{z} + z ( 1 - z )
\right)
\nonumber
\\
\gamma_{g}^{q  q} (z,\epsilon) &=&
\frac{1}{2} \, \left( z^2 + (1 - z)^2 \right)
\nonumber
\\
\gamma_{q}^{q g} (z,\epsilon) &=&
C_F\; \left( \frac{1 + z^2}{1 - z +\epsilon(\kappa_k)} \right)
\label{gamma}
\end{eqnarray}
where in the denominators the function $\epsilon$ appears,
\begin{equation}
\epsilon(\kappa_k) \;=\; \frac{k^{'\,2} n^2}{4 (k\cdot n)^2} \;\simeq \frac{
k^2}{k^{+\,2}}
\;,
\end{equation}
which arises here as a consequence of the $\partial_r^2$ term in the
renormalization equations (\ref{R2}), after Fourier transfoming
with respect to $r^-=r^0-r^3$ being the conjugate variable of $k^+$. It can be
interpreted as
manifestation of the indeterminancy principle,
which determines space-time uncertainty of the order of
the cell size $\Delta r$ that is associated with
the off-shellness of  the partons.
The presence of $\epsilon$ effectively cuts off small-angle gluon emission when
the emitted
gluon is soft, i.e. when $z_g=1-z\rightarrow 0$,
by modifying the free gluon propagator $\propto 1/z_g$ to the form
$1/(z_g+\epsilon)$
when $k/k^+ = O(1)$, that is, in branching processes with
large space-time uncertainty. This
ensures that the two daughter partons can be resolved as
individual quanta only if they are separated sufficiently by
$\Delta r \propto 1/k$ in position space,
in accord with the uncertainty principle.
Note that $\epsilon$ can be neglected in the terms
$\propto 1/(z_g+\epsilon)$ in (\ref{gamma})
for energetic gluon emission ($z_g\rightarrow 1$), but is essential in the
soft regime $(z_g\rightarrow 0$).
The effect of $\epsilon$ has been shown \cite{dok80,amati80} to result in a
natural
regularization of the infra-red-divergent behaviour of
the branching kernels (\ref{gamma}), due to destructive gluon
interference which becomes complete in the limit $z_g\rightarrow 0$.
\smallskip

As summarized in Appendix C, the renormalization functions $\Delta_{f}$
are intimatly related to the spectral densities
${\cal P}_{g}$ and ${\cal P}_q$ defined by (\ref{Pgq}) and (\ref{X4}),
\begin{eqnarray}
{\cal P}_{g}\;=\;\sum_{f'=g,q} \;{\cal P}_g^{f'}
\;=\; i\;\mbox{Tr}\left[d^{\mu\nu}\,( D_{\mu\nu}^R\,-\,D_{\mu\nu}^A ) \right]
\nonumber
\\
{\cal P}_{q}\;=\;\sum_{f'=g,q} \;{\cal P}_q^{f'}
\;=\; i\;\mbox{Tr}\left[ S^R\,-\,S^A \right]
\end{eqnarray}
where the ${\cal P}_{f}^{f'}$ have the following solution,
\begin{eqnarray}
& &
{\cal P}_{f}^{f'}(r; x,k^2) \;=\;
\delta_f^{f'}\;\delta(1-x)\,\delta\left(k^2 - \mu_{gq}^2\right)
\;\Delta_{f}(r;\mu_{gq}^2,k^{+\,2})
\nonumber
\\
& &
\;\;\;+
\;\;
\Delta_{f}(r,\mu_{gq}^2,k^{+\,2})
\;\,\sum_{f''}
\int_{\mu_{gq}^2}^{k^2}\frac{d k^{'\,2}}{k^{'\,2}}
\int_0^1 dz \,
\left\{
A(r,k^2,z) \,\gamma_f^{f''f'}(z,\epsilon)
\;{\cal P}_{f''}^{f'}\left(\frac{x}{z},k^{'\,2};k^{'\,+\,2}\right)
\right.
\nonumber
\\
& &
\left.
\;\;\;\;\;\;\;\;\;\;\;\;\;\;
\;\;\;\;\;\;\;\;\;\;\;\;\;\;
\;\;\;\;\;\;\;\;\;\;
\times\;
\Delta_f^{-1} \left(r,\frac{k^{'\,2}}{z},k^{'\,+\,2}\right)
\right\}
\;.
\label{PP}
\end{eqnarray}
Comparing this expression with eq. (\ref{X4}),
${\cal P}= \delta(k^2 - \mu^2)\Delta {\cal P}_{\delta {\cal E}}
+\Delta {\cal P}_{\Gamma}$, one sees now that the Sudakov formfactor
$\Delta_f$
together with the real emsission and absorption probabilities
$W_f\equiv \sum_{f''} \int d\ln k^{'\,2}dz\, A \,\gamma_f^{f''f'}{\cal
P}_{f''}^{f'}$
combine to play the role of the `wavefunction renormalization' part
$\Delta{\cal P}_{\delta {\cal E}}$
and the dissipative
part $\Delta {\cal P}_{\Gamma}$.
Equation (\ref{PP}) has a simple physical significance:
The first term on the right hand side represents the probability
to find a dressed parton of type $f$ in the cell around space-time point $r$
as a `classical' particle, i.e., without
any other gluons or quarks present in its wavefunction, or spectral density.
In accord with the normalization (\ref{norm1}), this means that it is
propagating `on-shell' with  $k^2=\mu_{gq}^2\ge\mu^2(r)$, where
$\mu(r)^{-1}$ is the resolution size of the cell as explained in Sec. 3.1.
Its fraction of the cell's lightcone energy is $x=k^+/P^+(r) = 1$.
The probability for finding such a rare fluctuation is
suppressed by the functions $\Delta_{f}(r,k^2,k^{+\,2})$,
which becomes stronger with increasing gap between
$\mu_{gq}^2$ and $k^2$.
The second term on the right hand side corresponds then to the adjoint
probability, that the parton is actually a dressed parton with a substructure,
described by the balance between real and virtual emission and absorption
processes,
while localized within the cell around $r$.
It is obvious, that the spectral densities  of dressed partons, introduced
Sec. 3.3, are identical to the usual {\it parton structure functions} ${\cal
P}_f$,
i.e. the probability densities for finding a dressed parton $f$ in an
intermediate
state containing number of bare partons as virtual and real fluctuations.

{}From (\ref{PP}) with (\ref{dq2}), and using the representation (\ref{Fgq})
of the parton densities $F_{f}$ in terms of the parton structure functions
${\cal P}_{f}^{f'}$,
follows then the final form of the {\it renormalization equations},
\begin{eqnarray}
k^2 \frac{\partial}{\partial k^2}
F_g\left(r; x,k^2\right)
&=&
\;\;\int_0^1 dz\,
\,A(r; k^2,z) \,
\left\{
\;\left[
\frac{1}{z} \,
F_g\left(r; \frac{x}{z},zk^2\right)
\;-\;
\frac{1}{2} F_g\left(r; x,k^2\right)
\right]
\;\Gamma_g^{gg}(z,\epsilon)
\right.
\nonumber \\
& &
\;\;\;\;\;\;\;+
\left.\frac{}{}
2\, N_f\;F_q\left(r; x,k^2\right)
\Gamma_q^{gq}(z,\epsilon)
\;-\;
N_f\;F_g\left(r; x,k^2\right)
\Gamma_g^{qq}(z,\epsilon)
\;\right\}
\;\;\;
\nonumber
\\
& & \nonumber \\
p^2 \frac{\partial}{\partial p^2}
F_q\left(r; x,p^2\right)
&=&
\;\;\int_0^1 dz\,
\,A(r; p^2,z) \,
\left\{
\;\left[
\frac{1}{z} \,
F_q\left(r; \frac{x}{z},zp^2\right)
\;-\;
F_q\left(r; x,p^2\right)
\right]
\;\Gamma_q^{qg}(z,\epsilon)
\right.
\nonumber \\
& &
\;\;\;\;\;\;\;+
\left.\frac{}{}
F_g\left(r; x,p^2\right)
\Gamma_g^{qq}(z,\epsilon)
\;\right\}
\label{R5}
\;,
\end{eqnarray}
which are the space-time generalization of the
DGLAP evolution equations \cite{dok80} that
govern the momentum dependence of the
parton densities.
The effective branching kernels $\Gamma_f^{f'f''}$ \cite{ms1} are related to
the
$\gamma_f^{f''}$ by,
\begin{equation}
\Gamma_f^{f'f''}\;=\; \gamma_f^{f'f''}\;
\left(1\;-\;\frac{F_{f'}}{F_{f'}\,\pm \,1}\right)
\;,
\end{equation}
where the upper (lower) sign stands in the term in brackets is for  gluons
(quarks).
It yields a suppression, when the phase-space density
$F_g$ or $F_q$ becomes large, so that the emission processes $f\rightarrow
f'f''$
are competed significantly by absorption processes
$f'f''\rightarrow f$.
In the limit $F_{f'}\gg 1$, the {\it detailed balance} is established, in
accord with the Bose-Einstein and Fermi-Dirac statistics of the gluon,
respectively
quark densities.
For instance, in thermal equilibrium, $F^{(eq)}_{g(q)}=(e^{-E/T}\mp 1)^{-1}$,
so that $\Gamma = \gamma (1-e^{E/T})$, which tends to zero as the temperature
$T$
becomes large.
\smallskip

Eqs. (\ref{R5}) are the main result of this Section.
They emerge as a direct consequence of the renormalization
equations (\ref{R}) in the short-distance regime of virtual and dissipative
quantum fluctuations,
and ensures unitarity conservation locally in each space-time cell.
They embody the Heisenberg uncertainty principle,
expressing the fact that it is impossible to
localize soft partons in a given cell if their wavelength
exceeds the cell size, which sets the resolution scale.
Last not least, they account for the
balance between real emission and absorption processes
that, in tends to increase (decrease) the effective real emission
rate of gluons (quarks).
\medskip

\noindent {\bf 3.5 Kinetic dynamics and transport equations}
\bigskip

With the dynamical structure of dressed partons quantitatively controlled
by the above renormalization equations (\ref{R5}),
one is now in the position to address the kinetic
space-time evolution of the multi-particle system in terms
of statistical binary scatterings among these dressed partons.
As explained in Sec. 3.1, in order
to obtain  quasi-classical transport equations for the
phase-space distribution functions $F={\cal N}\otimes {\cal P}$,
two key conditions have to be met: First, as before, the
maximal space-time extension
of relevant quantum fluctuations, $\lambda_c = \mu_{gq}^{-1}$, is supposed
to be smaller than the mean free path $\lambda_{mf}$ between scatterings.
Second,  the typical four-momentum transfer $q_\perp\equiv\sqrt{|q^2|}$ in the
scattering of
any two partons, is required to be larger than
the inverse Compton wavelength $\lambda_c^{-1}=\mu_{gq}$. That is (c.f. eq.
(\ref{cond0})),
\begin{equation}
\lambda_{mf}\;\,>\;\, \mu_{gq}^{-1}
\;\;\;,\;\;\;\;\;\;\;\;\;\;\;\;\;\;
q_\perp^2\;\,>\;\, \mu_{gq}^{2} \;=\; \lambda_c^{-2}
\label{scattcond}
\;.
\end{equation}
The first condition ensures that the quantum evolution, taken care of by the
renormalization equations, can be factorized from the scattering
processes. The second condition guarantees that the
scattering is sufficiently hard, i.e., is of short range compared to the
space-time extent of the scattering partons' intrinsic quantum motion,
so that over the duration of the scattering, the dressed partons can be
treated as `frozen' assemblies of bare particles that represent
their instantanous quantum state (the usual sudden approximation).
These two conditions are equivalent to the factorization assumption of
the well established `QCD hard scattering picture'  \cite{hardscatt} for, e.g.,
high energy hadron-hadron collisions, where the colliding hadrons are described
as
conglomerates
of bare partons in terms of their structure functions.
The relation of this hard scattering picture to the present
approach is its adoption to multiple, internetted scattering processes
in a system of stochastically colliding dressed partons, each of them
represented by their own
structure function, or spectral density.

The two requirements (\ref{scattcond}) are the crucial points, which allow in
the following
to cast the kinetic evolution into simple, probabilistic Boltzmann-type
equations,
which however have to be solved self-consistently in conjunction with the
renormalization equations (\ref{R5}).
Here is the key difference from other formulations \cite{baym,henning,rau}
to include quantum effects in a quasiclassical treatment of transport
phenomena,
in which one has only one type of equation, a generalized Boltzmann equation,
that contains local, classical part and a non-local quantum contribution,
containing the space-time history of memory effects.
In the present approach, this is translated to stochastically occurring
(`local'),
hard parton-parton scatterings, linked with the causal quantum evolution
in between scatterings (`non-local'), accounting for renormalization and
dissipation
due to the previously occured scatterings.
The advantage is here, that while quantum effects are included in the
multi-particle
evolution, still a local (in space and time) picture can be maintained, where
memory effects are embodied effectively in the dressed partons' structure
function evolution.

To proceed, recall from Sec. 3.3, that the correlation functions $D_{\mu\nu}^C,
S^C$
are the  quantities which are determined by the transport equations of the form
(\ref{X2}) (explicitly given in Appendix B, eqs. (\ref{BGC}) and (\ref{BQC}).
On account of the presumed conditions (\ref{scattcond}), over
kinetic space-time scales $\lambda_{mf}>\lambda_c$, the quantum
motion decouples, so that the correlation functions are
determined by the collsional self-energies $\Pi_{\mu\nu}^C, \Sigma^C$,
in conjunction with the real parts of the
retarded and advanced functions (c.f.(\ref{X2})).
This means that the collisional
self-energies are to be evaluated
with the renormalized propagators (\ref{ansatz1})  and vertices
(\ref{ansatz2}),
which were obtained from the retarded and advanced self-energies before.

Noting that from (\ref{rac}), one has
\begin{equation}
D_{\mu\nu}^C \;=\; D_{\mu\nu}^> \;+\; D_{\mu\nu}^<
\;\;\;\;\;\;\;\;\;\;
S^C \;=\; S^> \;+\; S^<
\;,
\end{equation}
the transport equations (\ref{T}) now read
(c.f. Appendix B),
\begin{eqnarray}
k\cdot\partial_r\; D_{ab}^{\mu\nu\;\gl} (r,k)
&=&
-\;\frac{i}{2} \;\left(\frac{}{}\Pi^{\gl}\, D^{A}\;+\; \Pi^R\,D^{\gl}
\;-\; D^\gl\, \Pi^{A}\;-\; D^{R}\,\Pi^\gl\right)^{\mu\nu}_{ab}
\label{X10}
\\
\!\!\!\!\!\!\!
i\;\left\{ \gamma\cdot \partial_r\,,\, S_{ij}^\gl(r,p)\right\}_+
&=&
-\;\left(\frac{}{}\left[\gamma\cdot p\,,\, S^\gl \right]_-\right)_{ij}
\;+\;
\left(\frac{}{}\Sigma^{\gl}\, S^{A}\;+\; \Sigma^R\,S^{\gl}
\;-\; S^\gl\, \Sigma^{A}\;-\; S^{R}\,\Sigma^\gl\right)_{ij}
\;,
\nonumber
\end{eqnarray}
which can be rewritten as
\begin{eqnarray}
k\cdot\partial_r\; D^{\mu\nu\;\gl}_{ab} (r,k)
&=&
- \frac{1}{2}\;
\left(\frac{}{}
\left\{ \frac{}{}
\Pi^{>}(r,k)\, , \, D^{<}(r,k) \right\}_+
\;-\;
\left\{ \frac{}{}
\Pi^{<}(r,k)\, , \, D^{>}(r,k) \right\}_+
\right)_{ab}^{\mu\nu}
\nonumber
\\
p\cdot \partial_r\;  S_{ij}^{\gl} (r,p)
&=&
 \frac{1}{2}\;
\left(\frac{}{}
\left\{ \frac{}{}
\Sigma^{>}(r,p)\, , \, S^{<}(r,p) \right\}_+
\;-\;
\left\{ \frac{}{}
\Sigma^{<}(r,p)\, , \, S^{>}(r,p) \right\}_+
\right)_{ij}
\label{T4}
\;.
\end{eqnarray}
These equations correspond to what is usually termed the {\it quasi-particle
approximation}.
In the cellular space-time picture,
the characteristics of the statistical kinetic evolution of the system
are, per design, insensitive to the localized fluctuations
associated with short-distance quantum dynamics inside a space-time cell.
To stress it more precisely,
the space-time variation can be considered homogenous over the range
of the Compton wavelength $\lambda_c= \mu_{gq}^{-1}\le \mu^{-1}(r)$,
so that $G^C(r,k)| \gg |\lambda_c^{2} \partial_r^2 G^C(r,p)|$ and the
derivatives
with respect to $r$ on the left hand side of the original eqs. (\ref{R}) and
(\ref{T}) may be omitted.
In the present context, it emerges as the logical consequence, that
the partons can be described on kinetic space-time scales
as quasi-particles, with
the underlying quantum motion effectively accounted for in the renormalized
propagators and vertices.

The self energies $\Pi_{\mu\nu}^{\gl}$ and $\Sigma^{\gl}$
are obtained from the general expressions (\ref{Pi}) and (\ref{Sigma}),
respectively. The lowest order  non-vanishing contributions are
the two-loop diagrams shown in Fig. 10, which are proportional to $\hbar$
and $O(g_s^4)$. In terms of the
renormalized correlators $D^{\gl}$ and $S^{\gl}$ one finds:
\begin{eqnarray}
\!\!\!\!\!\!\!\!\!
&&
\Pi_{ab}^{\sigma\tau\,\gl}(r,k)
\;=\;
\nonumber
\\
& &
\;=\;
\;\frac{g_s^4}{2}
\int \frac{d^4 k'}{(2\pi)^4i} \frac{d^4 q}{(2\pi)^4i}
\;
f_{aa'c} \lambda_{\mu\mu'\sigma}(-k,k-k',k')\,
D_{cc'}^{\sigma\sigma'\,\gl}(r,k')
\nonumber \\
& &
\;\;\;\;\;\;\;\;\;\;\;\;\;\;
\times\;
f_{c'fe} \lambda_{\sigma'\rho\lambda}(-k',-q,q+k')\,
D_{ff'}^{\rho\rho'\,\gl}(r,-q)\,
f_{e'f''d'} \lambda_{\lambda'\rho'\tau'}(-q-k',q,k')\,
D_{dd'}^{\tau'\tau\,\gl}(r,q+k')\,
\nonumber \\
& &
\;\;\;\;\;\;\;\;\;\;\;\;\;\;
\times\;
f_{db'b} \lambda_{\tau\nu'\nu}(-k',-k+k',k)\,
D_{a'b'}^{\mu'\nu'\,\gl}(r,k-k')\,
\nonumber \\
& &
\;\;\;\;
+\frac{g_s^4}{6}
\int \frac{d^4 k'}{(2\pi)^4i} \frac{d^4 q}{(2\pi)^4i}
\;
v_{ac'a'c}^{\mu\sigma'\mu'\sigma}(-k,q',k-q-q',q)\,
D_{cd}^{\sigma\tau\,\gl}(r,q)\,
D_{c'd'}^{\sigma'\tau'\,\gl}(r,q')\,
\label{Pi2}
\\
& &
\;\;\;\;\;\;\;\;\;\;\;\;\;\;
\times\;
v_{db'd'b}^{\tau\nu'\tau'\nu}(-q,-k+q+q',-q',k)\,
D_{a'b'}^{\mu'\nu'\,\gl}(r,k-q-q')\,
\nonumber \\
& &
\;\;\;\;
+g_s^4\,N_f
\int \frac{d^4 k'}{(2\pi)^4i} \frac{d^4 q}{(2\pi)^4i}
\;
f_{aa'c} \lambda_{\mu\mu'\sigma}(-k,k-k',k')\,
D_{cc'}^{\sigma\sigma'\,\gl}(r,k')
\,\gamma_{\sigma'}T_{ln}^{c'} \,S_{nn'}^{\gl}(r,-q)
\nonumber \\
& &
\;\;\;\;\;\;\;\;\;\;\;\;\;\;
\times\;
\gamma_{\tau'}T_{n'l'}^{d'} \,S^\gl_{l'l}(r,q+k')\,
D_{d'd}^{\tau'\tau\,\gl}(r,k')\,
f_{db'b} \lambda_{\tau\nu'\nu}(-k',-k+k',k)\,
D_{a'b'}^{\mu'\nu'\,\gl}(r,k-k')\,
\nonumber
\\
& &
\;\;\;\;
-g_s^4\,2 N_f
\int \frac{d^4 k'}{(2\pi)^4i} \frac{d^4 q}{(2\pi)^4i}
\;
\gamma_{\mu}T_{li}^{a} \,S_{ln}^{\gl}(r,k-k')
\,\gamma_{\nu} T_{jn}^{b} \,S_{j'j}^{\gl}(r,k')
\nonumber \\
& &
\;\;\;\;\;\;\;\;\;\;\;\;\;\;
\times\;
\gamma_{\tau} T_{n'j'}^{d} \,S_{l'n'}^{\gl}(r,k'-q)
\,\gamma_{\sigma} T_{i'l'}^{c} \,S_{ii'}^{\gl}(r,k')
\,D_{cd}^{\sigma\tau\,\gl}(r,q)
\nonumber
\\
&&
\nonumber \\
\!\!\!\!\!\!\!\!\!
& &
\Sigma_{ij}^{\gl}(r,p)
\;=\;
\nonumber
\\
& &
\;=\;
- \frac{g_s^4}{2}
\int \frac{d^4 k'}{(2\pi)^4 i} \frac{d^4 q}{(2\pi)^4 i}
\;
\gamma_{\sigma}T_{ii'}^{c} \, D_{cc'}^{\sigma\sigma'\,\gl}(r,k')
\nonumber \\
& &
\;\;\;\;\;\;\;\;\;\;\;\;\;\;\;\;\;\;\;\;\;\;\;\;
\times\;
f_{c'fe} \lambda_{\sigma'\rho\lambda}(-k',-q,q+k')\,
D_{ff'}^{\rho\rho'\,\gl}(r,-q)\,
f_{e'f''d'} \lambda_{\lambda'\rho'\tau'}(-q-k',q,k')\,
\nonumber \\
& &
\;\;\;\;\;\;\;\;\;\;\;\;\;\;
\times\;
D_{dd'}^{\tau'\tau\,\gl}(r,q+k')\,
\gamma_{\tau}T_{j'j}^{d} \, S_{i'j'}^\gl(r,p-k')
\nonumber \\
& &
\;\;\;\;
-g_s^4\,N_f
\int \frac{d^4 k'}{(2\pi)^4i} \frac{d^4 q}{(2\pi)^4i}
\;
\gamma_{\sigma}T_{ii'}^{c} \, D_{cc'}^{\sigma\sigma'\,\gl}(r,k')
\nonumber \\
& &
\;\;\;\;\;\;\;\;\;\;\;\;\;\;
\times\;
S_{nn'}^{\gl}(r,-q) \,\gamma_{\sigma'}T_{ln}^{c'} \,
S^\gl_{l'l}(r,q+k')\,\gamma_{\tau'}T_{n'l'}^{d'}
D_{d'd}^{\tau'\tau\,\gl}(r,k')\,
\nonumber \\
& &
\;\;\;\;\;\;\;\;\;\;\;\;\;\;
\times\;
\,\gamma_{\tau}T_{j'j}^{d}\, S_{i'j'}^{\gl}(r,p-k')
\label{Sigma2} \\
& &
\;\;\;\;
+g_s^4
\int \frac{d^4 k'}{(2\pi)^4i} \frac{d^4 q}{(2\pi)^4i}
\;
\gamma_{\sigma}T_{il}^{c} \,  S_{ll'}^{\gl}(r,p-k')
\, D_{ee'}^{\lambda\lambda'\,\gl}(r,q)
\nonumber \\
& &
\;\;\;\;\;\;\;\;\;\;\;\;\;\;
\times\;
\,\gamma_{\lambda}T_{l'm}^{e} \,  S_{mm'}^{\gl}(r,p-k'-q)
\,\gamma_{\tau}T_{nj}^{d} \, D_{cd}^{\sigma\tau\,\gl}(r,k')
\nonumber
\;,
\end{eqnarray}
where $\lambda^{\mu\rho\nu}(p_1,p_2,p_3)$ and
$v_{abcd}^{\mu\sigma\tau\nu}(p_1,p_2,p_3,p_4)$
are the usual 3-gluon and 4-gluon vertices (c.f. Appendix D, eq. (\ref{D34})).
The correlation functions $D^\gl_{\mu\nu}, S^\gl$ are related to the
phase-space densities $F_g, F_q$ by (\ref{SIII}), i.e.,
\begin{eqnarray}
\!\!\!\!\!\!\!\!\!\!\!
D_{\mu\nu}^{\gl}(r,k) &=&
-2 \pi i \, (-d_{\mu\nu}(k)) \,\left[
\frac{}{}
\theta(\pm k^0) \,+ \,F_g(r,k) \right]
\;\delta\left( k^2 \;-\;{\cal M}_g^2(r,k)\right)
\nonumber
\\
\!\!\!\!\!\!\!\!\!\!\!
S^{\gl}(r,p) &=&
-2 \pi i \, \left(\gamma\cdot p \right) \,\left[
\frac{}{}
\theta(\pm p^0) \,-\, F_q(r,p)
\right]
\;\delta\left( p^2 \;-\;{\cal M}_q^2(r,p)\right)
\label{GC2}
\;,
\end{eqnarray}
where the signs $+(-)$ refer to $>(<)$
\footnote{
It must be mentioned that eq. (\ref{GC2}) assumes a spin-symmetric form
for the quark-antiquark spinor products, which means a
neglect of spin-polarization effects. As shown by Elze {\it et al.}
\cite{elze86},
in general  the quark phase-space distribution does
require at least an $8\times 8$-matrix representation.
}.
As repeatedly stressed, the densities $F$ are the distributions of dressed
partons,
with their substructure represented in terms of corresponding assemblies of
bare partons,
that satisfy the condition (\ref{scale3}). Therefore,
the functions $F$  can also be interpreted to measure the number
of bare partons with dynamical invariant masses
$k^2 \ge \mu_{gq}^2$.
Consequently, a binary collision of two dressed partons can be described
in terms of the above `hard scattering picture' as
a scattering of two bare partons, one out of each assembly, picked
statistically from the instantanous quantum state of the
two dressed partons, as given by their  structure functions, or spectral
densities.
The four-momentum transfer $q_\perp^2$ sets hereby the probing scale,  so that
$k^2\approx q_\perp^2$.
Therefore the energy spectra of partons emerging from these scatterings
are - using eq. (\ref{offshell}) - determined by
\begin{eqnarray}
\omega &\equiv&  \pm k^0(r,k) \;=\;
\omega_{(0)}\,\left( 1\;+\;\frac{{\cal
M}_g^2(r,k^+,k^2)}{2\;\omega_{(0)}}\right)
\nonumber \\
E &\equiv& \pm p^0(r,p) \;=\;
E_{(0)}\,\left( 1\;+\;\frac{{\cal M}_q^2(r,p^+,p^2)}{2\;E_{(0)}}\right)
\label{Espectra0}
\;,
\end{eqnarray}
where $k^0=1/2(k^++k^-)\simeq k^+/2$,
$\;p^0\simeq p^+/2$,
and
$\omega_{(0)} = \pm\sqrt{\vec k^{\;2} + \mu_{gq}^2}$,
$E_{(0)} = \pm\sqrt{\vec p^{\;2} + \mu_{gq}^2}$.
Accordingly, one can now write
\begin{equation}
\omega  \;=\; \pm\sqrt{\vec k^{\;2} + q_\perp^2}\;\theta(q_\perp^2-\mu_{gq}^2)
\;\;\;\;\;\;\;\;\;\;\;\;\;
E  \;=\; \pm\sqrt{\vec p^{\;2} + q_\perp^2}\;\theta(q_\perp^2-\mu_{gq}^2)
\label{Espectra}
\;,
\end{equation}
and
\begin{eqnarray}
F_g(r,k) \,\delta(k^2 - q_\perp^2) &=&
\frac{1}{2\omega}
\left[\frac{}{}
F_g(r,\vec k)\,\delta(k^0 - \omega) \;+\;
F_g(r,-\vec k)\,\delta(k^0 + \omega)
\right]
\nonumber \\
F_q(r,p) \,\delta(p^2 - q_\perp^2) &=&
\frac{1}{2E}
\left[\frac{}{}
F_q(r,\vec p)\,\delta(p^0 - E) \;+\;
\overline{F}_q(r,-\vec p)\,\delta(p^0 + E)
\right]
\;,
\end{eqnarray}
which exhibits explicitly
the particle-antiparticle character of the phase-space densities.
In particular,
$F_q(r,\vec p)$ denotes the quark distribution and $\overline{F}_q(r,-\vec p)$
the antiquark distribution.

Using the representations (\ref{GC2}) for $D_{\mu\nu}^{\gl}$, $S^{\gl}$
in the self-energies (\ref{Sigma2}) and (\ref{Pi2}), and substituting into
the equations (\ref{T4}), gives the final form of the transport
equations in the kinetic regime (c.f. Appendix D),
\begin{equation}
k\cdot \partial_r \,F_g(r,k) \;=\;
{\cal I}_g(r,k)
\;\;\;\;\;\;\;\;\;\;\;\;\;
p\cdot \partial_r \,F_q(r,p) \;=\;
{\cal I}_q(r,p)
\;,
\label{T5}
\end{equation}
where
\begin{eqnarray}
{\cal I}_g(r,k)
&=&
\frac{1}{2}\;
\left\{ \frac{}{}
\widehat{\Pi}^{>}(r,k)\, , \, F_g(r,k) \right\}_+
\;-\;
\frac{1}{2}\;
\left\{ \frac{}{}
\widehat{\Pi}^{<}(r,k)\, , \, F_g(r,k) + 1 \right\}_+
\nonumber
\\
{\cal I}_q(r,p)
&=&
\frac{1}{2}\;
\left\{ \frac{}{}
\widehat{\Sigma}^{>}(r,p)\, , \, F_q(r,p) \right\}_+
\;-\;
\frac{1}{2}\;
\left\{ \frac{}{}
\widehat{\Sigma}^{<}(r,p)\, , \, F_q(r,p) - 1 \right\}_+
\;.
\label{colli1}
\;,
\end{eqnarray}
and the `hatted' self-energy functions $\widehat{\Pi}$, $\widehat{\Sigma}$
stand for
\begin{eqnarray}
\widehat{\Pi}^{\gl}(r,k)
&=&
\frac{1}{2i}
\;
\sum_{s=1,2}
\varepsilon^\mu(k,s)  \varepsilon^{\nu\;\ast}(k,s)
\;
{\Pi}^{\gl}_{\mu\nu}(r,k)
\nonumber \\
\widehat{\Sigma}^{\gl}(r,p)
&=&
\frac{1}{2i} \;\sum_{s=1,2} \left[\bar{u}(p,s) \,\Sigma^\gl(r,p) \,u(p,s)
\;+\;\bar{v}(p,s) \,\Sigma^\gl(r,-p) \,v(p,s)\right]
\label{colli2}
\;.
\end{eqnarray}
The collision terms ${\cal I}_g$ and ${\cal I}_q$
on the right hand side of (\ref{T5}) describe the balance of gain and loss
of partons in a phase-space element $d^3r d^4k$, or $d^3d^4p$, within a time
slice around $r_0$.
Their explicit form is obtained as explained in  Appendix D, and
emerges as the result of
applying the usual cutting rules \cite{cut} to the self-energies
(\ref{Pi2}), (\ref{Sigma2}) and averaging (summing) over initial (final)
spin and color degrees of freedom (see Fig. 11). The resulting
structure of the collision terms is as follows:
\begin{eqnarray}
& &
{\cal I}_a(r,p_1)\;\equiv\; \sum_{bcd}
\left(\frac{}{}
- {\cal I}_{cd \rightarrow ab}^{(loss)} (p_1, r) \,+\,
{\cal I}_{ab \rightarrow cd}^{(gain)} (p_1, r)
\right)
\nonumber
\\
& &
\nonumber
\\
& & =\, - \,
\sum_{bcd}
C_{ab} \,C_{cd}
\,
\int \frac{d^3 p_2}{(2 \pi)^3 2 E_2}
\int \frac{d^3 p_3}{(2 \pi)^3 2 E_3}
\int \frac{d^3 p_4}{(2 \pi)^3 2 E_4}
\;(2 \pi)^4 \,\delta^4( p_1 + p_2 - p_3 - p_4) \,
\nonumber
\\
& &
\nonumber
\\
& &
\;\;\;\;\;\;\;\;
\times
\left\{
\frac{}{}
\,
F_a(1) F_b(2)
\;\;
\vert {\cal M}(ab \rightarrow cd) \vert ^2
\;\theta(q_\perp^2-\mu_{gq}^2)\;\;
\left[ 1 \pm F_c(3) \right]
\left[ 1 \pm F_d(4) \right]
\right.
\nonumber
\\
& &
\left.
\;\;\;\;\;\;\;\;\;\;\;
\frac{}{}
\,-\,
\left[ 1 \pm F_a(1) \right]
\left[ 1 \pm F_b(2) \right]
\;\;
\vert {\cal M}(cd \rightarrow ab) \vert ^2
\;\theta(q_\perp^2-\mu_{gq}^2)\;\;
F_c(3) F_d(4)
\,
\frac{}{}
\right\}
\;,
\label{colli3}
\end{eqnarray}
Here the notation is $F_f(i) \equiv F_f (r, p_i)$ for the distribution
functions of
the parton species $f = g, q, \bar q$
with four-momenta $p_i = p, p_2, p_3, p_4$ at space-time point
$r=(r^0,\vec r)$.
The structure of the collision terms in conjunction with
the equations (\ref{T5}) is such that
the squared matrix elements for the various scattering processes
$12\rightarrow 34$ (depicted in Fig. 11, and explicitly given in Appendix D)
are weighted by a distribution function $F_f(i)$
for each of the partons coming into the vertex and a
factor $(1\pm F_{f'}(j))$ for each of the outgoing ones.
The + sign is for gluons so that
$(1+ F_g)$ results in a Bose enhancement, and
the $-$ sign refers to quarks and antiquarks with $(1-F_q)$
indicating Pauli blocking. This is a direct consequence
of the quantum statistical difference between the
gluon and quark propagators (\ref{S00}) and (\ref{S000}).
The factors $C_{ab}\,(C_{cd})$ in front account for the
identical particle effect, if incoming (outgoing) partons
are indistinguishable.
\smallskip

Eqs. (\ref{T5})  are the essential result of this Section.
These Boltzmann-type equations are the final form of the transport equations
for the dressed partons with phase-space densities $F$.
The equations have a a drift term on the left hand side, and a collision
term on the right hand side, which balances the various
processes by which a dressed parton may be gained or lost
in a phase-space element $d^3rd^4p$ around time $r^0\pm\Delta r/2=r^0\pm
1/(2\mu)$.
They describe the  dynamics of the multi-parton
system on kinetic scales, due to statistical, binary collisions,
in which
dressed partons appear as quasi-particles with a dynamical substructure, which
is described in terms of probabilities to find a parton as a state consisting
of
a number of bare gluons and quarks of virtualities $k^2>\mu_{gq}^2$.
These the underlying quantum fluctuations are embodied in
$F={\cal N}\otimes {\cal P}$ through the spectral density, or parton
structure function, ${\cal P}$ and are
determined by the renormalization equations (\ref{R5}).
A scattering between two dressed partons is therefore
described as a `hard scattering' determined
by the probabilities of finding in each of them a hard fluctuation
with $k^2\simeq q_\perp^2>\mu_{gq}^2$ of the order of the momentum transfer
that sets the probing scale.
This is expressed by the collision term on the right hand side,
in which the products of $F$'s involve the convolution
of spectral densities ${\cal P}$, weighted with the squared
matrix elements.
A graphical illustration of this is shown in Fig. 12.
\bigskip

\noindent {\bf 3.6 A Monte Carlo calculation scheme}
\bigskip

A practical calculation scheme to compute the evolution of a multi-parton
system
as governed by the couled renormalization and transport equations may
be outlined with the following concrete example.
Consider the collision of two large nuclei with mass number $A \gg1$ at
ultra-relativistic
center-of-mass energy. Before their contact upon collision, the approaching
nuclei appear as two highly Lorentz contracted discs of coherently bound
gluons and quarks with a coherence length $L_0=1/\mu_0$, where
$\mu_0\simeq g_s \,\rho_\perp$ with $\rho_\perp = N_{gq}/(\pi R_A^2)$ sets the
scale of the typical space-like parton virtuality, and hence $L_0$ their
characteristic
transverse size. McLerran and Venugopalan
\cite{mclerran94} have shown that if $A$ is sufficiently large,
the associated primeval parton distribution
of the nuclei before, and shortly after the collision can be
calculated non-perturbatively from first principles in terms
of coherent quantum fields.
Their conclusion is that, as long
as the very early generation of this parton matter distribution has
$\Delta p \Delta r \sim 1$, it cannot be described by a kinetic
particle picture, which requires $\Delta p \Delta r \gg 1$.
However, after a  time $t_0\simeq 1/\mu_0$ passed  the nuclear contact,
the parton matter has gone through a decoherence stage, so that the
latter condition is be satisfied, and a kinetic
description can be matched to the complex coherent evolution of
the primeval matter.
In other words, at time $t_0$, one may proceed with a probabilistic
description of the parton dynamics in terms of the
interplay between coherent radiative evolution and incoherent binary
interactions,
as suggested in the present work.

At time $t_0$, one starts from the initial multi-parton state, and
the subsequent time evolution of the partons' phase-space densities
$F_{gq}(t,\vec r, p^2,\vec p)$
may be calculated by a Monte Carlo procedure, using the advocated
discretization of space-time with
four-dimensional cells of size $\Delta r = \Delta t \Delta^3 r \simeq
\mu^{-4}$.
\begin{description}
\item[(i)]
The first step consists in evaluating, from the collision kernel of the
transport equations (\ref{T5}),
the probabilities of
scatterings among the initial partons within the time slice
$\Delta t_0=\mu^{-1}(t_0)\equiv \mu_0^{-1}$ between
$t_0$ and $t_0+\Delta t_0$  for each cell centered around $\vec r$.
The essential condition (\ref{scattcond})
provides the possibility of treating the scattering among partons incoherently,
and requires that the impact parameter $b_{ab}$ of any two scattering
candidates
$a$ and $b$ must satisfy $b_{ab}< \mu_0^{-1}$, implying for the momentum
transfer
of the scattering $q_{ab\perp}^2 > \mu_0^2$,
where $\mu_0^2$ is the initial virtuality of the partons at $t_0$, set by the
coherence
length of the colliding nuclei.
The primary parton scatterings $a+b$ that occur within $\Delta t_0$ subject to
this condition
change the phase-space occupation of partons at $t_1 = t_0+\Delta t_0$ in two
ways:
on the one hand their spatial profile is altered due to gain and loss of
deflected scatterers $a'$ and $b'$
in the spatial cells, and on the other hand, the virtualities are reset from
$k_a^2 = k_b^2 =\mu_0^2$ to $k_a^{'\,2}=k_b^{'\,2} = q_{ab\perp}^2 > \mu_0^2$.
\item[(ii)]
In the next step one must now calculate the quantum fluctuations,
virtual plus real emission and absorption processes, that are triggered by
the primary scatterings and the change of virtualities from $k^2$ to
$k^{'\,2}$.
That is, the parton structure functions need to be evolved within $\Delta t_0$
according to the renormalization equations (\ref{R5}), which account for the
associated renormalization and dissipation.
One then obtains a spatial profile of new dressed partons, that defines the
initial
condition for the further evolution, starting at $t_1$.
The  procedure repeats, as before at $t_0$, by evaluating the scattering
probabilities
in the next time slice $\Delta t_1=\mu^{-1}(t_1)$ between
$t_1$ and $t_1+\Delta t_1$, now however subject to the modified incoherence
condition,
that for any scattering of partons $a$ and $b$ their impact parameter must
be $b_{ab}< \max(k_a^2,k_b^2)^{-1/2}$, i.e. $q_{ab\perp}^2 >
\max(k_a^2,k_b^2)$.
\end{description}

It is important to realize that
the condition (\ref{scattcond})  of a well defined separation between quantum
and kinetic scales,
imposes the crucial incoherence requirement for binary scatterings, and
allows a `hard scattering' picture, in which the quantum evolution
and hard scattering of evolving quanta factorize.
The condition defines the range of  validity for a  probabilistic description
in terms of incoherently scattering
particles, and is essentially the uncertainty principle:
the quasi-classical picture holds only, if
the scattering partons may be treated as well distinguishable, incoherent
quasi-particles of size $1/\sqrt{k^2}$, meaning that
at least a `formation time' of $t_k\simeq 1/\sqrt{k^2}$ must have been passed
since their previous
scattering, during which their quantum structure evolves with virtuality $k^2$.
A concrete example of this scheme will be presented elsewhere.
\bigskip
\bigskip

\noindent {\bf 4. CONCLUDING REMARKS}
\bigskip

In this paper I have attempted to lay out a foundation to
obtain from the fundamental quantum field theoretical principles of QCD
a self-consistent kinetic description for the evolution
of a high-energy system of self- and mutual interacting
gluons and quarks.
The main result is a set of two distinct, but coupled equations that govern
the time evolution of the gluon and quark Wigner functions, the quantum
analogues of the
classical phase-space densities:
\begin{description}
\item[(i)]
A renormalization equation, which describes the the momentum dependence of
short-distance
quantum fluctuations due to the partons' self-interactions. It defines
the state of a dressed parton as a quasi-particle with a renormalized mass
and a decay width, corresponding to  virtual and real emission and absorption
processes. The solution of this equation describes, locally in space-time, the
spectral density
in terms of bare partons that are associated with the quantum substructure of
the state of a dressed parton,
and determines the partons' structure functions as well as their dynamical mass
spectrum.
\item[(ii)]
A transport equation, which describes the
space-time evolution of the dressed partons in the kinetic
quasi-particle regime by means of mutual binary collisions. It
determines the time dependence of both the change of the spatial density and
the energy-momentum
distribution of dressed partons due to elastic and inelastic collisions.
Accordingly, it redistributes not only the partons in space, but also
modifies their momentum spectrum and virtualities, which feeds
back into the renormalization equations.
\end{description}
Loosely speaking, the renormalization equation defines the state of dessed
partons, whereas
the transport equation governs the occupation of these states.
The self-consistent solution of the equations provides the means to trace
the dynamical development of the multi-parton system in real-time and full
7-dimensional phase-space $d^3r d^3p dp^2$, spanned by position, momentum and
invariant virtuality. It suggests a probabilistic, causal description, which is
predestined for numerical evaluation by using Monte Carlo simulation
techniques.
\smallskip

The essential steps that lead to this kinetic framework may be summarized as
follows:
\begin{description}
\item[a)]
the path integral quantization of the classical action, using the CTP
formalism with {\it in-in} boundary conditions, including initial state
correlations at time $t_0$ described by the density matrix $\hat \rho(t_0)$;
\item[b)]
the 2-point source approximation, that allows to rewrite
the CTP path integral  as the generating functional
for a possible color background field, and for the 2-point gluon
and quark Green functions defined on the closed-time-path between $t_0$ and
$t_\infty$;
\item[c)]
the derivation of the self-consistent equations of motion
for mean field (Ginzburg-Landau equation) and Green functions
(Dyson-Schwinger equations);
\item[d)]
the transition to kinetic theory  by imposing
the physics-motivated well defined separation between
the {\it quantum scale} that specifies the range of
short-distance quantum fluctuations and the {\it kinetic scale} that
characterizes the range of inter-particle correlations and
stochastic binary interactions;
\item[e)]
the conversion of the Dyson-Schwinger equations for the Green-functions to
the set of renormalization and transport equations for the
corresponding Wigner functions, on the basis of
the separation of quantum and kinetic scales and a cellular space-time
picture.
\item[f)]
the calculation of the spectral density of bare partons, locally within the
cells, from
the renormalization equations, which defines the state of dressed partons in
terms of
their substructure, and the evaluation of the collision kernel of
the transport equations, which determine the statistical occurrence of
scattering processes
among these dressed partons.
\end{description}
\smallskip

This quantum kinetic framework may be extended in straightforward manner to
include effects of a color background field, or gluonic mean field, that acts
as a classical background medium in which the partonic quanta evolve
(e.g. in a QCD plasma, where it
may be generated due to the bulk dynamics of soft gluon modes).
This option has not been considered in the present paper, however
the framework incorporates this possibility by considering a
non-vanishing $\tilde A^\mu$ instead of setting it to zero as in Sec. 3.
The inclusion of such a mean field would extend the set of renormalization
and transport equations for the partons' Wigner functions,
by coupling to the Ginzburg-Landau equation for
the mean field.
Qualitatively, the effect would be twofold: first, the poles
of the Wigner functions would be shifted by a mean field generated mass
$\mu(\tilde A^\mu)$, and second, the transport equations
would acquire an additional force term of Vlasov form.
\smallskip

The future extensions and applications are manifold.
Most important at first,
I believe, is a detailed calculation for a specific situation
where the concepts and formalism presented here may be illustrated
and checked for consistency. For instance, it would be desirable
to study a thermal (or close to thermal) parton system in this
real-time description, and compare it with the well known results
in the imaginary-time formalism of finite-tempaerature QCD.
Such a project is planned to be carried out in the near future.
On the other hand, the probabilistic interpretation of the real-time
evolution of rather general multi-parton system offers the
opportunity to simulate the dynamical development on the
basis of the master equations with Monte Carlo techniques on
a computer \cite{msrep}.
\bigskip
\bigskip

\begin{center}
{\bf ACKNOWLEDGEMENTS}
\end{center}

\bigskip
I would like to thank the following colleagues:
John Ellis for his continuing support and energy;
Thomas Elze for  initiating this work; Berndt M\"uller
for carefully reading a preliminary version of the manuscript and
helping with numerous advices; Bo Andersson, Yuri Dokshitzer, Ulrich Heinz,
Sasha Makhlin, Pino Marchesini, for inspiring discussions and
useful suggestions.
Last not least, I wish to thank the `European Centre For Theoretical Studies in
Nuclear
Physics and Related Areas' (ECT*)
in Trento (Italy)
for the hospitality during the International Workshop on
`QCD and Ultra-relativistic Heavy Ion Collisions', June 1995.

\newpage

\begin{center}
{\bf APPENDIX A:\\ The CTP formalism and the 2-point source generating
functional}
\end{center}
\bigskip

In this Appendix,  I review the CTP functional formalism applied to
the case of QCD. For additional reading concerning the general techniques,
I refer to the extensive literature
\cite{schwinger,keldysh,baym,chou,calzetta,rammer}.
In the  {\it in-in} formulation of quantum field theory,
mentioned in Sec. 2,
the generating functional
is defined as the $in$-vacuum to $in$-vacuum amplitude
$Z[J,\hat{\rho}] = Tr \sum_\varphi\langle 0^{in}|\varphi\rangle_J
\langle \varphi|\hat\rho|0^{in}\rangle_J$,
including
possible initial state correlations represented by the density matrix
$\hat{\rho}$ at $t_0$, and
a sum over a complete set of states $\varphi$ at $t_\infty$.
In the Heisenberg picture it is represented by
\begin{equation}
Z_P[J,\hat{\rho}]\;=\;
\mbox{Tr}\left\{ \frac{}{} T_P\,
\exp\left(i \sum_f \int_P d^4x J^f(x)\phi_f(x)\right)\,\hat{\rho}\right\}
\label{Z0}
\;,
\end{equation}
where
$f=g,u,\bar{u}, d, \bar{d},\ldots$ specifies the gluon and quark field degrees
of freedom,
and $\phi_f=(A^\mu,\psi,\overline{\psi})$.
The symbol $P$ refers to the time integration along
a {\it closed-time path} in the
complex $t$-plane as illustrated in Fig. 2:
the path goes forward from $t_0$ to $t_\infty$ on the positive branch,
and then back from $t_\infty$ to $t_0$ on the negative branch.
The generalized time-ordering $T_P$ is therefore defined such that
any point on the second branch is understood at a later instant
than any point on the first branch.
Utilizing (\ref{U2}), eq. (\ref{Z0}) can be rewritten as
\begin{equation}
Z_P[J_+,J_-,\hat{\rho}]\;=\;
\mbox{Tr}\left\{
U^\dagger_{J_-}(t_0 ,t)\, U_{J_+}(t,t_0)\, \hat{\rho}(t_0)\right\}
\label{Z1}
\;,
\end{equation}
where
$J_+$ ($J_-$) is the source along the positive (negative) branch of Fig. 2a.
In general $J_+\ne J_-$, so that $Z_P$ depends on two different sources. If
these
are set equal, one has
$Z_P(J,J,\rho)=\mbox{Tr} \hat{\rho}$, which is equal to unity in the
absence of initial correlations, being a statement of unitarity.
The derivatives of $Z_P$ with respect to the sources generate the
$n$-point CTP Green functions
\begin{equation}
G_{\alpha_1,\ldots, \alpha_n}^{f_1,\ldots, f_n}(x_1,\ldots, x_n)\;=\;
\frac{\delta^n Z_P[J_+,J_-,\hat{\rho}]}{\delta J_{\alpha_1}^{f_1}(x_1)\ldots
\delta J_{\alpha_n}^{f_n}(x_n)}
\;=\;
(-i)^{n-1}\;
\mbox{Tr}\left\{  T_P\,
\phi_{\alpha_1}^{f_1}(x_1) \ldots \phi_{\alpha_n}^{f_n}(x_n) \,\hat\rho\right\}
\label{G0}
\;,
\end{equation}
where $\alpha_i = \pm$, and the indices $f_i$ label the type of
the $i$-th field.
The functional $Z_P$ can be represented as a path integral by employing the
relation
between the Heisenberg and interaction pictures (\ref{S2}).
One imposes boundary conditions
in terms of complete sets of eigenstates of
the Heisenberg fields $\Phi_H$ at initial time $t=t_0$,
\begin{eqnarray}
\Phi_H(t_0,\vec x) \,|\,\phi^+ (t_0)\,\rangle &=&
\Phi_I(t_0,\vec x) \,|\,\phi^+ (t_0)\,\rangle \;=\;
\phi^+(\vec x) \,|\,\phi^+ (t_0)\,\rangle
\nonumber \\
\Phi_H(t_0,\vec x) \,|\,\phi^-(t_0)\,\rangle &=&
\phi^-(\vec x) \,|\,\phi^- (t_0)\,\rangle
\label{bc1}
\;,
\end{eqnarray}
and in the remote future at $t=t_{\infty}$,
\begin{equation}
\Phi_H(t_\infty,\vec x) \,|\,\varphi(t_{\infty})\,\rangle
\;=\;
\varphi(\vec x) \,|\,\varphi (t_\infty)\,\rangle
\;.
\label{bc2}
\end{equation}
Then, making use of the  completeness of the eigenstates,
one obtains from (\ref{Z1}) the following functional integral representation
for $Z_P$,
\begin{eqnarray}
Z_P[J_+,J_-,\hat{\rho}] &=&
\int \,{\cal D} \phi^+ {\cal D} \phi^-
{\cal D} \varphi
\;
\langle \,\phi^-(t_0)\, | \,U^\dagger_{J_-}(t_0,t_\infty)\,|\,\varphi
(t_\infty)\,\rangle
\nonumber \\
& & \;\;\;\;\;\;\;\;\;
\times \;
\langle \,\varphi(t_\infty)\, | \,U_{J_+}(t_\infty,t_0)\,|\,\phi^+
(t_0)\,\rangle
\langle\,\phi^+(t_0)\, | \,\hat{\rho} \,|\,\phi^+(t_0)\,\rangle
\label{Z3}
\;.
\end{eqnarray}
At this point it is  convenient to
represent $+$ and $-$ by greek indices $\alpha, \beta,\gamma,\ldots$ and
to introduce
a 2$\times$2 matrix $\sigma$ as a `metric tensor',
\begin{equation}
\tau_{\alpha\beta}\;=\; \tau^{\alpha\beta} \;=\; \mbox{diag} (1, -1)
\;,\;\;\;\;\;\;\;\;\;
\alpha,\beta\;=\;\pm
\label{sigma}
\end{equation}
and similarly higher rank tensors
\begin{equation}
u_{\alpha\beta\gamma}\;=\; \delta_{\alpha\beta} \tau_{\beta\gamma}
\;\;,\;\;\;\;\;\;
v_{\alpha\beta\gamma\delta}\;=\; \mbox{sign}(\alpha)\, \delta_{\alpha\beta}
\delta_{\beta\gamma}
\delta_{\gamma\delta}
\;,
\label{sigma2}
\end{equation}
with the usual summation convention over repeated greek indices $\alpha,
\beta,..$.
With this convention one can generalize the classical action (\ref{I1})
to account for all four field orderings on the closed-time path $P$:
\begin{eqnarray}
I[\phi_f^\alpha]&\equiv & I[\phi_f^+]\;-\;I^\ast[\phi_f^-]
\;=\;
{\cal I}^{(0)}[\tau_{\alpha\beta} A^\alpha_\mu A^\beta_\nu]
\;+\;
{\cal I}^{(0)}[\tau_{\alpha\beta} \overline{\psi}^\alpha \psi^\beta]
\nonumber
\\
& &
\;\;\;\;\;\;\;\;\;\;\;\;\;\;
\;+\;
{\cal I}^{(int)}[u_{\alpha\beta\gamma}\overline{\psi}^\alpha
A_\mu^\beta\psi^\gamma ,
\,u_{\alpha\beta\gamma}(\partial_\mu A_{\nu}^\alpha) A_\mu^\beta A_\nu^\gamma ,
\,v_{\alpha\beta\gamma\delta} A_{\mu}^\alpha A_\nu^\beta A_\mu^\gamma
A_\nu^\delta ]
\;,
\label{I2}
\end{eqnarray}
where the correspondance with the terms ${\cal I}$ with the ones of (\ref{I1})
is obvious
(the color indices are suppressed here).
Also, the following shorthand notation for the integration over the space-time
variables
will be used in the functional sense,
\begin{equation}
J\,\phi \;\equiv \; \int_P d^4x \,J(x)\;\phi(x)
\;\;,\;\;\;\;\;
\phi\,K\,\phi \;\equiv \; \int_P d^4x \,\phi(x)\;K(x,y)\;\phi(y)
\;.
\label{conv2}
\end{equation}
Returning to the  functional integral (\ref{Z3}),
I now utilize the above conventions and exploit the fact that
the first two amplitudes are just the ordinary transition matrix elements
in the presence of $J_+$ and $J_-$, whereas the density matrix element
incorporates
the initial state correlations at $t_0$ at the endpoints of the closed-time
path $P$,
one obtains the path integral representation for $Z_P$ in analogy
to usual field theory \cite{cornwall,jordan}
\begin{equation}
Z_P[J_+,J_-,\hat{\rho}]
\;=\;
\int \,\prod_{f}{\cal D} \phi_f^\alpha
\;\, \exp \left[i \left( \frac{}{} I[\phi_f^\alpha] \;+\;
\sum_f J_{f,\,\alpha}\phi_f^\alpha
\right)
\right]
\;{\cal N}[\hat{\rho}]
\label{Z4}
\;.
\end{equation}
Here
${\cal D} \phi_f^\alpha\equiv
{\cal D} \phi_f^+ {\cal D} \phi_f^-$, and
I suppressed the formal presence of the Fadeev-Popov determinant
associated with the gauge freedom, because
for the class of ghost-free gauges (\ref{gauge}) it is equal to unity.
The functional
${\cal N}[\hat{\rho}]$ is the density matrix element containing the
initial state correlations
that may be represented by a non-local source functional $K$ as follows:
\begin{equation}
{\cal N}[\hat{\rho}]\;=\;
\langle\,\phi_+(t_0)\,|\,\hat{\rho} \,|\, \phi_-(t_0)\,\rangle
\;\equiv\; \exp\left(i \sum_f K_f[\phi_+,\phi_-]\right)
\end{equation}
When expanded functionally as
\begin{equation}
K_f[\phi_+,\phi_-]\;=\;
K^f\;+\; K^{f}_\alpha \phi^\alpha
\;+\;\frac{1}{2}\,K^{ff'}_{\alpha\beta} \phi_f^\alpha \phi_{f'}^\beta
\;+\;\frac{1}{6}\,K^{ff'f''}_{\alpha\beta\gamma} \phi_f^\alpha \phi_{f'}^\beta
\phi_{f''}^\gamma
\;+\;\ldots
\;,
\end{equation}
eq. (\ref{Z4}) becomes a functional of an infinite number of non-local sources
\cite{calzetta},
which however contribute {\it only} at $t=t_0$, corresponding to the initial
state
correlations,
\begin{eqnarray}
Z_P[J_+,J_-,\hat{\rho}]
&\equiv&
Z_P[J_\alpha,K_{\alpha\beta},\ldots]
\label{Z5}
\\
&=&
\int \,\prod_{f,\,\alpha}{\cal D} \phi_f^\alpha
\;\, \exp \left[i \left( \frac{}{} I[\phi_f] \;+\;
\sum_f \left( J^f_\alpha\phi_f^\alpha
\;+\; \frac{1}{2} \phi_f^\alpha K^{ff'}_{\alpha\beta} \phi_f^\beta
\;+\; \ldots\right)
\right)
\right]
\;,
\nonumber
\end{eqnarray}
where the constant term
$K^f$ has been absorbed into the normalization
and the local initial state kernel $K^f_\alpha(x)$ has been
combined with the external source term $J^f_\alpha(x)$, i.e.
$J^f_\alpha\equiv J^f_\alpha+K^f_\alpha$.

The corresponding
generating functional for the {\it connected} Green functions $W_P$
as given as usual by the logarithm of $Z_P$,
\begin{equation}
W_P[J^\alpha, K^{\alpha\beta},K^{\alpha \beta \gamma},\ldots]
\;=\; -i\,\ln Z_P[J^\alpha, K^{\alpha\beta}, K^{\alpha\beta\gamma} ,\ldots]
\label{W5}
\;.
\end{equation}
The functional derivatives of $W_P$ with respect to the local
sources $J^\alpha_f(x)$
define gluon and quark mean fields $\tilde{\phi}_f^\alpha(x)$ as
the expectation values of the single field operators, which can arise
either through non-vanishing external sources, or, in the case of
gluons, may be  generated
dynamically by the system itself depending on the initial conditions.
Similarly,
the variation of $W_P$ with respect to the non-local kernels
$K_{ff'}^{\alpha\beta}(x,x'), K_{ff'f''}^{\alpha\beta\gamma}(x,x',x''),\ldots$,
give the the $n$-point Green functions
$G_{ff'}^{\alpha\beta}(x,x')$, $G_{ff'f''}^{\alpha\beta\gamma}(x,x',x'')$,
etc.,
for the different particle species,
which are the expectation values of products of $n$ field operators.
{}From (\ref{Z5}) and (\ref{W5}), one finds
\begin{eqnarray}
\frac{\delta W_P}{\delta J^f_\alpha}&=& \tilde{\phi}_f^\alpha
\nonumber \\
\frac{\delta W_P}{\delta K_{\alpha\beta}^{ff'}}&=&
\frac{1}{2}\,\left( i\,G_{ff'}^{\alpha\beta} \;+\; \tilde{\phi}_f^\alpha
\tilde{\phi}_{f'}^\beta
\right)
\nonumber \\
\frac{\delta W_P}{\delta K_{\alpha\beta\gamma}^{ff'f''}}&=&
\frac{1}{6}\,\left( \,G_{ff'f''}^{\alpha\beta} \;+\;
3i\,G_{ff'}^{\alpha\beta}\tilde{\phi}_{f''}^\gamma
\;+\; \tilde{\phi}_f^\alpha \tilde{\phi}_{f'}^\beta \tilde{\phi}_{f''}^\gamma
\right)
\;\;\;\;\;\;\;
\mbox{etc.}
\label{E1}
\end{eqnarray}
One immediately recognizes the inifinite hierarchy of the
Green functions, the complete knowledge of which would correspond to
the full solution of the theory. Clearly, in practice one must
truncate this infinite series.
In what follows, I will  assume that all
$n$-point sources of order $n\ge 3$ (i.e. the kernels $K^{\alpha\beta\gamma}$
etc.,
are neglegible and thus can be omitted.
Such an approximation is justified if the higher order correlations generated
by the $n\ge3$ source terms are comparably small and the quantum
dynamics of the system can be sufficiently well described by single-particle
distributions corresponding to the 2-point functions. In fact, this is the
very hypothesis of the parton description of QCD at large energies, where
higher order
correlations (`higher twist' effects) are kinematically suppressed
by powers of a large momentum scale $Q^{-2}$ corresponding to an approximate
factorization of dominant short-distance 2-point correlations and
larger distance 3-,4-,..-point correlations associated with multi-particle
effects,
an approximation which becomes exact in the asymptotic limit
\cite{dok80}.
In this approximation the generating functional (\ref{Z5}) then explicitly
reads,
\begin{eqnarray}
Z_P[J^\mu,j,\overline{j}, K^{\mu\nu}, k] &=&
e^{i \,W_P[J^\mu,j,\overline{j}, K^{\mu\nu}, k]}
\nonumber \\
&=&
\int \,{\cal D} A_\alpha^\mu{\cal D} \psi_\alpha{\cal D}
\overline{\psi}_\alpha\,
\;\exp \left[i \left( \frac{}{}
I[A^\mu_\alpha,\psi_\alpha,\overline{\psi}_\alpha]
\right.
\right.
\label{Z6}
\\
& &
\left.
\left.
\frac{}{}
\;+\;\;
J_\mu^\alpha A_\alpha^\mu \;+\; j^\alpha
\overline{\psi}_\alpha\;+\;\overline{j}^\alpha \psi_\alpha
\;+\;
\frac{1}{2} \,A_\alpha^\mu K_{\mu\nu}^{\alpha\beta} A_\beta^\nu
\;+\;
\overline{\psi}_\alpha k^{\alpha\beta} \psi_\beta
\right)
\right]
\nonumber
\;,
\end{eqnarray}
which is the expression I stated in eq. (\ref{Z000}) of Sec. 2.2.
Since the present interest concerns  only cases where no external sources are
present,
one obtains from (\ref{Z6}) the mean fields and the 2-point functions
for gluons and quarks
by  taking  into account the fact in the that absence of external sources the
establishment of a local
colored mean field can only occur for the gluons, but not for
quarks or antiquarks.
Because of their bosonic character gluons the production of gluons
can lead to a dynamically generated coherent field acting as a background
medium, whereas the
production of quarks and antiquarks occurs only in pairs and cannot
yield a coherent mean field.
Furthermore,  a physical gluon mean field is determined the equality
$\tilde{A}^\mu_+= \tilde{A}^\mu_-\equiv \tilde{A}^\mu$.
Hence, one gets from (\ref{Z6}), using (\ref{E1})
\begin{eqnarray}
\frac{\delta W_P}{\delta J^\mu_\alpha(x)}&=& \tilde{A}_\mu(x)
\;,\;\;\;\;\;\;\;\;\;\;\;\;
\frac{\delta W_P}{\delta \overline{j}_\alpha(x)}\;=\;
\frac{\delta W_P}{\delta j_\alpha(x)}\;=\;0
\nonumber
\\
\frac{\delta W_P}{\delta K^{\mu\nu}_{\alpha\beta}(x,y)}&=&
\frac{1}{2}\,\left( i\,D_{\mu\nu}^{\alpha\beta}(x,y) \;+\;
\tilde{A}_\mu^\alpha(x) \tilde{A}_{\nu}^\beta(y)
\right)
\nonumber \\
\frac{\delta W_P}{\delta k_{\alpha\beta}(x,y)}&=&
-\, i\,S^{\alpha\beta}(x,y)
\label{E2b}
\;,
\end{eqnarray}
where, as a reminder $\alpha, \beta =\pm$, not to be confused
with color indices which I suppressed in this Appendix.
\bigskip
\bigskip

\newpage

\begin{center}
{\bf APPENDIX B:\\ From quantum field description to kinetic theory}
\end{center}
\bigskip

In this Appendix I explain steps that lead from the Dyson-Schwinger
equations (\ref{eog1}), (\ref{eog2}) to the kinetic counterparts,
the renormalization equations (\ref{R}) and transport equations (\ref{T}).
A well separation between quantum scale of short-distance fluctuations
and the statistical-kinetic scale, is the essential
requirement for recasting the quantum theoretical problem,
formulated in terms of the 2-point Green functions $G(x,y)$,
into the much simpler form of kinetic teory, employing
Wigner transforms $G(r,p)$.
In the picture of cellular space-time, constructed in Sec. 3.1,
a clearly defined separation between the two scales is controlled by the
characteristic size of the space-time cells with volume $\Omega\simeq
\mu^{-4}$,
by choosing $\mu$ such that
$\Delta r_{qua}\le \mu^{-1} < \Delta r_{kin}$,
i.e.  the cell size is larger than the
range of short-distance quantum fluctuations, $\Delta r_{qua}\le \mu^{-1}$,
but small compared to the mean-free path of the quanta between their
kinetic, statistical interactions,  $\Delta r_{kin}$. This is illustrated in
Figs. 6 and 7.
Therefore the correlation between different cells is negligible
by design, and the only relevant case is when the points $x$ and $y$ in the
argument of $G(x,y)$ lie in the same cell.
In the interior of each cell, one can then assume approximate
translation invariance, because large-distance inhomogenities of space-time
are, by construction,  not resolvable within the small cell volume.
Thus, for each individual cell,
one can Fourier-transform the Green functions over the  cell
volume, and use the common machinery of propagator theory as for
homogenous systems, or the vacuum.
Specifically, one transforms the Green functions
$G\equiv D_{\mu\nu},S$
with respect to the relative coordinate $s=x-y$, whereas
the absolute  coordinate $r=\frac{1}{2}(x+y)$ serves as
a cell label:
\begin{equation}
G(x,y) \;=\;
\int \frac{d^4 p}{(2\pi)^4} \, e^{-i\,p\,\cdot\, (x-y)}\;
G\left( \frac{x+y}{2}, x-y\right)
\;\,=\,\;
\int \frac{d^4 p}{(2\pi)^4} \, e^{-i\,p\,\cdot\,s}\;
G\left( r, s\right)
\;,
\label{B1}
\end{equation}
where $G(r,p)$ is called the Wigner transform of $G(x,y)$,
and similarly for the self-energies ${\cal E}(x,y)\equiv\Pi_{\mu\nu}, \Sigma$,
\begin{equation}
{\cal E}(x,y) \;=\;
\int \frac{d^4 q}{(2\pi)^4} \, e^{-i\,q\,\cdot\, (x-y)}\;
{\cal E}\left( \frac{x+y}{2}, x-y\right)
\;\,=\,\;
\int \frac{d^4 q}{(2\pi)^4} \, e^{-i\,q\,\cdot\,s}\;
{\cal E}\left( r, s\right)
\;.
\label{B2}
\end{equation}
If the separation between quantum and kinetic scales
would be perfect (as in vacuum, where $\Delta r_{kin} = \infty$),
then the $r$-dependence would drop out and the Wigner transforms
reduce to the ordinary Fourier transforms $G(p)$, ${\cal E}(q)$.
In case of moderately inhomogenous media, meaning that the
Green functions and self-energies vary only slowly with $r$
and are strongly peaked around $s=x-y$, as I consider here,
then one can expand the Green functions
$G(x,y)=W\left(r+\frac{1}{2}s, r-\frac{1}{2}s\right)= W(r,s)$
in a series of gradients,
\begin{equation}
W(r+s, s) \;\simeq \; W(r, s) \;+ \;s \,\cdot\, \partial_r\,W(r,s)
\;\,+\;\, \ldots
\label{B3}
\;,
\end{equation}
and analogous for the self-energies ${\cal E}$.
The basic Dyson-Schwinger equations (\ref{eog1}), (\ref{eog2}) remain
the same in terms of the Wigner transforms, for example
for the terms ${\cal E} G\equiv \Pi_{\mu\sigma} D^{\sigma}_\nu$
or ${\cal E}G\equiv \Sigma S$, one has
\begin{eqnarray}
\int d^4x' {\cal E}(x,x') \,G(x',y)
&=&
\int \frac{d^4q}{(2\pi)^4}\frac{d^4p}{(2\pi)^4} \int d^4x'
\label{B4}
 \\
& &
\times
\, e^{-i\,q\,\cdot\, (x-x')}\;
{\cal E}\left(\frac{x+x'}{2}, x-x'\right)
\; e^{-i\,p\,\cdot\, (x'-y)}\;
G\left(\frac{x'+y}{2}, x'-y \right)
\nonumber
\;.
\end{eqnarray}
The integrand will be significantly different from zero only if
$x'$ lies within the same cell as $x$ and $y$, in which  case
$\frac{1}{2}(x+x')\simeq \frac{1}{2}(x'+y) \simeq r$. Therefore
(\ref{B4}) reduces to
\begin{equation}
\int d^4x' {\cal E}(x,x') \,G(x',y)
\;=\;
\int \frac{d^4p}{(2\pi)^4}
\,\left\{
\frac{}{}
\, e^{-i\,p\,\cdot\, (x-y)}\;
{\cal E}\left(r, p\right)
\,
G\left(r, p\right)
\;\,+\;\, \Delta(r,p)
\right\}
\label{B5}
\;,
\end{equation}
where $\Delta(r,p)$ embodies the corrections to the ideal separation
of cells. In terms of the gradient expansion (\ref{B3}), the
first order correction which is $O(\hbar)$, is given by
\begin{eqnarray}
\Delta(r,p)
&=&
\frac{i}{2} \frac{\partial {\cal E}(r,p)}{\partial p_\mu} \,\frac{\partial
G(r,p)}{\partial r^\mu}
\;-\;
\frac{i}{2} \frac{\partial {\cal E}(r,p)}{\partial r^\mu} \,\frac{\partial
G(r,p)}{\partial p^\mu}
\nonumber \\
&\equiv &
\frac{i}{2}\;\left[ (\partial_p {\cal E})\cdot(\partial_r G)
\;-\; (\partial_r {\cal E})\cdot(\partial_p G) \right]
\label{B6}
\;.
\end{eqnarray}
In general, the convolution between two functions $f$ and $g$ is given by
\begin{equation}
\int d^4x' f(x,x') \,g(x',y)
\;\longrightarrow\;
\exp\left[\frac{i}{2}\,\left(
\partial_p^{(f)}\cdot\partial_r^{(g)}\;-\;
\partial_r^{(f)}\cdot\partial_p^{(g)}\right)\right]
\;f(r,p) \,g(r,p)
\;.
\end{equation}
The gradient expansion (\ref{B3}) corresponds to keeping
only the first two terms in the Taylor series of the exponential function,
which gives the following set of  conversion rules:
\begin{eqnarray}
\int d^4x' f(x,x') \,g(x',y)
&\longrightarrow &
f(r,p) \,g(r,p)
\;+\;
\frac{i}{2}\;\left[ (\partial_p f)\cdot(\partial_r g)
\;-\; (\partial_r f)\cdot(\partial_p g) \right]
\nonumber \\
h(x) \,g(x,y)
&\longrightarrow &
h(r) \,g(r,p)
\;-\;
\frac{i}{2}\, (\partial_r h)\cdot(\partial_p g)
\nonumber \\
h(y) \,g(x,y)
&\longrightarrow &
h(r) \,g(r,p)
\;+\;
\frac{i}{2}\, (\partial_r h)\cdot(\partial_p g)
\nonumber \\
\partial_x^\mu f(x,y)
&\longrightarrow &
(-i p^\mu \,+\,\frac{1}{2} \partial_r^\mu)  \,f(r,p)
\nonumber \\
\partial_y^\mu f(x,y)
&\longrightarrow &
(+i p^\mu \,+\,\frac{1}{2} \partial_r^\mu)  \,f(r,p)
\label{B7}
\;.
\end{eqnarray}
If one applies these rules now to the Dyson-Schwinger equations
(\ref{eog1}) and (\ref{eog2}), which upon
setting for simplicity
the mean field contributions $\tilde{\mu}_g=\tilde{\mu}_q = 0$,
read,
\begin{eqnarray}
\stackrel{\rightarrow}{\Box}_{x, \,\mu\rho}
\;  D^{\rho\nu}_{ab}(x,y)
&=&
\delta_{ab} \,g^{\mu\nu} \,\delta^4_P(x,y)
\;-\;\int_P d^4x' \, \Pi^\mu_{\sigma ,\;a,b'}(x,x')\, D^{\sigma
\nu}_{b'b}(x',y)
\nonumber
\\
D^{\rho\nu}_{ab}(x,y)\;
\stackrel{\leftarrow}{\Box}_{y, \,\mu\rho}
&=&
\delta_{ab} \,g_{\mu\nu} \,\delta^4_P(x,y)
\;-\;\int_P d^4x' \, D^\mu_{\sigma ,\;a,b'}(x,x')\, \Pi^{\sigma
\nu}_{b'b}(x',y)
\label{B00a}
\end{eqnarray}
and
\begin{eqnarray}
i \gamma\cdot \stackrel{\rightarrow}{\partial}_x
\; S_{ij}(x,y)
&=&
\delta_{ij} \delta^4_P(x,y)\;+\;
\int_P d^4x' \,\Sigma_{ik}(x,x')\,S_{kj}(x',y)
\nonumber
\\
-S_{ij}(x,y)
\;
i \gamma\cdot \stackrel{\leftarrow}{\partial}_y
&=&
\delta_{ij} \delta^4_P(x,y)\;+\;
\int_P d^4x' \,S_{ik}(x,x')\,\Sigma_{kj}(x',y)
\;,
\label{B00b}
\end{eqnarray}
one finds
on the basis of the gradient expansion (\ref{B3})
a set of corresponding matrix equations
for the Wigner transforms of the gluon and quark Green functions,
\begin{eqnarray}
& &
\left(
\frac{}{}
-k^2 \,+\,\frac{1}{4} \stackrel{\rightarrow}{\partial}_r^2  \,-\,i
k\cdot\stackrel{\rightarrow}{\partial}_r
\right)
\;  D^{\mu\nu}_{ab}(r,k)
\;=\;
\nonumber \\
& &
\;\;\;\;\;\;\;\;\;\;\;
=\;
d^{\mu\nu}(k) \,\delta_{ab}\;\hat 1_P
\;-\;
\left(\frac{}{}
\Pi\; D
\right)_{ab}^{\mu\nu}
\;-\;
\frac{i}{2}\;\left(\frac{}{} (\partial_k \Pi)\cdot(\partial_r D)
\;-\; (\partial_r \Pi^{\mu\sigma})\cdot(\partial_k D_\sigma^\nu)
\right)_{ab}^{\mu\nu}
\nonumber \\
& &
D^{\mu\nu}_{ab}(r,k)
\;
\left(
\frac{}{}
-k^2 \,+\,\frac{1}{4} \stackrel{\leftarrow}{\partial}_r^2  \,+\,i
k\cdot\stackrel{\leftarrow}{\partial}_r
\right)
\;=\;
\label{B8a}
\\
& &
\;\;\;\;\;\;\;\;\;\;\;
=\;
d^{\mu\nu}(k)\,\delta_{ab}\;\hat 1_P
\;-\;
\left(
\frac{}{}
D\;\Pi
\right)_{ab}
\;-\;
\frac{i}{2}\;\left(\frac{}{} (\partial_k D)\cdot(\partial_r \Pi)
\;-\; (\partial_r D^)\cdot(\partial_k \Pi) \right)_{ab}^{\mu\nu}
\nonumber
\end{eqnarray}
and
\begin{eqnarray}
\!\!\!\!\!\!\!\!
\left(\frac{}{}
 \gamma\cdot
( p\,+\,\frac{i}{2} \stackrel{\rightarrow}{\partial_r})
\right)
\; S_{ij}(r,p)
&=&
\delta_{ij}\;
\hat 1_P
\;+\;
\left(\frac{}{}
\Sigma\;S
\right)_{ij}\;+\;
\frac{i}{2}\;\left(\frac{}{} (\partial_p \Sigma)\cdot(\partial_r S)
\;-\; (\partial_r \Sigma)\cdot(\partial_p S) \right)_{ij}
\nonumber
\\
\!\!\!\!\!\!\!\!
S_{ij}(r,p)
\;
\left(\frac{}{}
 \gamma\cdot
( p\,-\,\frac{i}{2} \stackrel{\leftarrow}{\partial_r})
\right)
&=&
\delta_{ij}
\;\hat 1_P
\;+\;
\left(\frac{}{}
S\;\Sigma
\right)_{ij}
\;+\;
\frac{i}{2}\;\left(\frac{}{} (\partial_p S)\cdot(\partial_r \Sigma)
\;-\; (\partial_r S)\cdot(\partial_p \Sigma) \right)_{ij}
\;,\;\;\;\;
\label{B8b}
\end{eqnarray}
where
\begin{equation}
\hat 1_P
\;=\;
\left\{
\begin{array}{cc}
\hat 1 & \;\;\mbox{for} \;F, \;\overline{F} \\
0 & \;\;\mbox{for} \;>, \;<
\end{array}
\right.
\;,
\end{equation}
recalling that
$G\equiv D^{\mu\nu},\,S$ and the self-energies
${\cal E}\equiv \Pi^{\mu\nu},\,\Sigma $  each represent a 2$\times$2 matrix
as defined by (\ref{G22}),
\begin{equation}
G \;=\;
\left(
\begin{array}{cc}
G^F\;  & \; G^> \\
G^<\;  & \; G^{\overline{F}}
\end{array}
\right)
\;\;, \;\;\;\;\;\;\;\;\;\;
{\cal E}\;=\;
\left(
\begin{array}{cc}
{\cal E}^F\;  & \; {\cal E}^> \\
{\cal E}^<\;  & \; {\cal E}^{\overline{F}}
\end{array}
\right)
\;.
\label{B101}
\end{equation}
Adding  the two equations of (\ref{B8a}), respectively of (\ref{B8b}),
yield the imaginary parts as the
the passage to the renormalization equations stated in (\ref{R}),
whereas  subtracting
the two equations of (\ref{B8a}), respectively of (\ref{B8b}),
gives the real parts as the
transport equations (\ref{T}).
For the gluon Wigner functions one obtains
\begin{eqnarray}
\left(k^2 \,-\,\frac{1}{4} \partial_r^2 \right)
\; D^{\mu\nu}_{ab} (r,k)
&=&
-\,d^{\mu\nu}(k)\,\delta_{ab}\,\hat 1_P
\;+\;
\frac{1}{2} \;\left(\frac{}{}\left\{\Pi\,,\, D \right\}_+\right)^{\mu\nu}_{ab}
\;+\; \frac{i}{4}\;{\cal G}^{\mu\nu\, (-)}_{ab}
\nonumber
\\
k\cdot\partial_r\; D_{ab}^{\mu\nu} (r,k)
&=&
-\;\frac{i}{2} \;\left(\frac{}{}\left[\Pi\,,\, D \right]_-\right)^{\mu\nu}_{ab}
\;+\;\frac{1}{4}\;{\cal G}^{\mu\nu\,(+)}_{ab}
\label{B9}
\; ,
\end{eqnarray}
where $[A,B]_- \equiv AB-BA$, $\{A,B\}_+\equiv AB+BA$, and
\begin{equation}
\!\!\!\!\!\!
{\cal G}^{\mu\nu\,(-)}\,=\,
\left[ \partial^\lambda_k \Pi^\mu_{\sigma} \, , \partial_\lambda^r
D^{\sigma\nu}\right]_-
-
\left[ \partial^\lambda_r \Pi^\mu_{\sigma} \, , \partial_\lambda^k
D^{\sigma\nu}\right]_-
\;,\;\;\;
{\cal G}^{\mu\nu\,(+)}\,=\,
\left\{ \partial^\lambda_k \Pi^\mu_{\sigma} \, , \partial_\lambda^r
D^{\sigma\nu}\right\}_+
-
\left\{ \partial^\lambda_r \Pi^\mu_{\sigma} \, , \partial_\lambda^k
D^{\sigma\nu}\right\}_+
\;.
\end{equation}
For the quark Wigner functions the corresponding equations read
\begin{eqnarray}
\!\!\!\!\!\!\!
\!\!\!\!\!\!\!
& &
\frac{1}{2} \;\left\{ \gamma\cdot p\,,\, S_{ij}(r,p)\right\}_+
\;=\;
\delta_{ij}\,\hat 1_P
\;-\;
\frac{i}{2} \;\left(\frac{}{}\left[\gamma\cdot \partial_r\,,\, S
\right]_-\right)_{ij}
\;+\;
\frac{1}{2} \;\left(\frac{}{}\left\{\Sigma\,,\, S \right\}_+\right)_{ij}
\;+\;
\frac{i}{4} \,{\cal F}^{(-)}_{ij}
\nonumber
\\
\!\!\!\!\!\!\!
\!\!\!\!\!\!\!
& &
\frac{1}{2} \;\left\{ \gamma\cdot \partial_r\,,\, S_{ij}(r,p)\right\}_+
\;=\;
\frac{i}{2} \;\left(\frac{}{}\left[\gamma\cdot p\,,\, S \right]_-\right)_{ij}
\;-\;
\frac{i}{2} \;\left(\frac{}{}\left[\Sigma\,,\, S \right]_-\right)_{ij}
\;+\;
\frac{1}{4} \,{\cal F}^{(+)}_{ij}
\;,
\label{B10}
\end{eqnarray}
where
\begin{equation}
{\cal F}^{(-)}\;=\;
\left[ \partial^\lambda_p \Sigma \, , \partial_\lambda^r S\right]_-
-
\left[ \partial^\lambda_r \Sigma \, , \partial_\lambda^p S\right]_-
\;,\;\;\;\;\;\;\;
{\cal F}^{(+)}=
\left\{ \partial^\lambda_p \Sigma \, , \partial_\lambda^r  S\right\}_+
-
\left\{ \partial^\lambda_r \Sigma \, , \partial_\lambda^p  S\right\}_+
\;.
\end{equation}
The equations (\ref{B10}) for quark propagators
can be formally brought in the same form as (\ref{B9}) for the
gluon propagators by multiplying the first equation of (\ref{B8b}) with
$\gamma\cdot(p+\frac{i}{2}\stackrel{\leftarrow}{\partial_r})\delta_{li} +
\Sigma_{li}$
from the left, and the second equation of (\ref{B8b}) by
$\gamma\cdot(p-\frac{i}{2}\stackrel{\leftarrow}{\partial_r})\delta_{jl}+
\Sigma_{jl}$
from the right, and then adding and subtracting the resulting equations:
\begin{eqnarray}
\!\!\!\!\!\!\!\!\!
\left(p^2 \,-\,\frac{1}{4} \partial_r^2 \right)
\; S_{ij} (r,p)
&=&
\left(\gamma\cdot p + \Sigma\right)
\,\delta_{ij}\,\hat 1_P
\;+\;
\frac{1}{2} \;\left(\frac{}{}\left\{\Sigma^2\,,\, S \right\}_+\right)_{ij}
\;+\; \frac{i}{4}\;{\cal A}^{(+)}_{ij}
\;-\; \frac{1}{8}\;{\cal B}^{(-)}_{ij}
\nonumber
\\
\!\!\!\!\!\!\!\!\!
p\cdot\partial_r\; S_{ij} (r,p)
&=&
\frac{1}{2}\,
\,\left(\gamma\cdot \partial _r\right)
\,\delta_{ij}\,\hat 1_P
\;-\;
\frac{i}{2} \;\left(\frac{}{}\left[\Sigma^2\,,\, S \right]_-\right)_{ij}
\;+\;\frac{1}{4} \;{\cal A}^{(-)}_{ij}
\;+\;\frac{i}{8}\;{\cal B}^{(+)}_{ij}
\; ,
\;\;\;\;
\label{B10a}
\end{eqnarray}
where the notation $\Sigma_{ij} =\delta_{ij} \Sigma$ is employed, and
\begin{eqnarray}
& &
{\cal A}^{(\pm)}\;=\;
\frac{1}{2}\left(
\frac{}{}(\gamma\cdot p + \tilde{\Sigma})\;({\cal F}^{(-)} + {\cal F}^{(+)})
\;\pm\;
({\cal F}^{(-)} - {\cal F}^{(+)}) \;(\gamma\cdot p + \tilde{\Sigma})\right)
\nonumber
\\
& &
{\cal B}^{(\pm)}\;=\;
\frac{1}{2}\left(
\frac{}{}
(\gamma\cdot \stackrel{\rightarrow}{\partial_r})\;({\cal F}^{(-)} + {\cal
F}^{(+)}) \;\pm\;
({\cal F}^{(-)} - {\cal F}^{(+)}) \; (\gamma\cdot
\stackrel{\leftarrow}{\partial_r}) \right)
\;.
\end{eqnarray}
Due to the $2\times 2$ matrix character of the equations (\ref{B9}) and
(\ref{B10}), (\ref{B10a}),
the four components $F,\overline{F}, >, <$ of the Green functions $D^{\mu\nu},
S$ and
self-energies $\Pi^{\mu\nu}, \Sigma$ mix, so that each of these equations
actually represent
a non-trivial coupled set of four equations. However, as advertised in Sec.
3.3,
in the {\it physical representation} (\ref{rac}),
\begin{equation}
\breve{G}
\;=\;
\left(
\begin{array}{cc}
0\;  & \; G^A \\
G^R\;  & \; G^C
\end{array}
\right)
\;\;, \;\;\;\;\;\;\;\;\;\;
\breve{{\cal E}}\;=\;
\left(
\begin{array}{cc}
{\cal E}^C\;  & \; {\cal E}^R \\
{\cal E}^A\;  & \; 0
\end{array}
\right)
\;,
\end{equation}
one has the great advantage that the retarded and advanced
functions $G^{R(A)}$ are determined exclusively by
the $R$ and $A$ components, and components, and only
the equation for $G^C$ involves a mixing with these.
Omitting for lucidity the gradient terms ${\cal G}, {\cal F}$,
the equations (\ref{B9}) become in the physical representation
a self-contained set for the retarded and advanced functions,
\begin{eqnarray}
\left(k^2 \,-\,\frac{1}{4} \partial_r^2 \right)
\; D_{ab}^{\mu\nu\;R(A)} (r,k)
&=&
-\,d^{\mu\nu}(k)\,\delta_{ab}
\;+\;
\frac{1}{2} \;\left(\frac{}{}\Pi^{R(A)}\, D^{R(A)}\;+\; D^{R(A)}\,
\Pi^{R(A)}\right)^{\mu\nu}_{ab}
\nonumber
\\
k\cdot\partial_r\; D_{ab}^{\mu\nu\;R(A)} (r,k)
&=&
-\;\frac{i}{2} \;\left(\frac{}{}\Pi^{R(A)}\, D^{R(A)}\;-\; D^{R(A)}\,
\Pi^{R(A)}\right)^{\mu\nu}_{ab}
\label{BGRA}
\end{eqnarray}
and
\begin{eqnarray}
\frac{1}{2} \;\left\{ \gamma\cdot p\,,\, S_{ij}^{R(A)}(r,p)\right\}_+
&=&
\delta_{ij}\;
-\;
\frac{i}{4} \;\left(\frac{}{}\left[\gamma\cdot \partial_r\,,\, S^{R(A)}
\right]_-\right)_{ij}
\;+\;
\frac{1}{2} \;\left(\frac{}{}\left\{\Sigma^{R(A)}\,,\, S^{R(A)}
\right\}_+\right)_{ij}
\nonumber
\\
\frac{1}{2} \;\left\{ \gamma\cdot \partial_r\,,\, S_{ij}^{R(A)}(r,p)\right\}_+
&=&
\frac{i}{2} \;\left(\frac{}{}\left[\gamma\cdot p\,,\, S^{R(A)}
\right]_-\right)_{ij}
\;-\;
\frac{i}{2} \;\left(\frac{}{}\left[\Sigma^{R(A)}\,,\, S^{R(A)}
\right]_-\right)_{ij}
\label{BQRA}
\;
\;,
\end{eqnarray}
plus a set of mixed equations for the correlation functions,
\begin{eqnarray}
\left(k^2 \,-\,\frac{1}{4} \partial_r^2 \right)
\; D_{ab}^{\mu\nu\;\gl)} (r,k)
&=&
-\frac{1}{2} \;\left(\frac{}{}\Pi^{\gl}\, D^{A}\;+\; \Pi^R\,D^{\gl}
\;+\; D^\gl\, \Pi^{A}\;+\; D^{R}\,\Pi^\gl\right)^{\mu\nu}_{ab}
\nonumber
\\
k\cdot\partial_r\; D_{ab}^{\mu\nu\;\gl} (r,k)
&=&
-\;\frac{i}{2} \;\left(\frac{}{}\Pi^{\gl}\, D^{A}\;+\; \Pi^R\,D^{\gl}
\;-\; D^\gl\, \Pi^{A}\;-\; D^{R}\,\Pi^\gl\right)^{\mu\nu}_{ab}
\;\;\;\;\;
\label{BGC}
\end{eqnarray}
with $D_{\mu\nu}^C = D_{\mu\nu}^>+ D_{\mu\nu}^<$, and,
\begin{eqnarray}
\!\!\!\!\!\!\!
\left\{ \gamma\cdot p\,,\, S_{ij}^{\gl}(r,p)\right\}_+
&=&
-\;
\frac{i}{2} \;\left(\frac{}{}\left[\gamma\cdot \partial_r\,,\, S^{\gl}
\right]_-\right)_{ij}
\;+\;
\left(\frac{}{}\Sigma^{\gl}\, S^{A}\;+\; \Sigma^R\,S^{\gl}
\;+\; S^\gl\, \Sigma^{A}\;+\; S^{R}\,\Sigma^\gl\right)_{ij}
\nonumber
\\
\!\!\!\!\!\!\!
i\;\left\{ \gamma\cdot \partial_r\,,\, S_{ij}^\gl(r,p)\right\}_+
&=&
-\;\left(\frac{}{}\left[\gamma\cdot p\,,\, S^\gl \right]_-\right)_{ij}
\;+\;
\left(\frac{}{}\Sigma^{\gl}\, S^{A}\;+\; \Sigma^R\,S^{\gl}
\;-\; S^\gl\, \Sigma^{A}\;-\; S^{R}\,\Sigma^\gl\right)_{ij}
\;,
\label{BQC}
\end{eqnarray}
with $S^C = S^>+ S^<$.
\bigskip
\bigskip

\begin{center}
{\bf APPENDIX C:\\ The renormalization functions and
the spectral densities of partons}
\end{center}
\bigskip

In the following I explain in more detail the steps that lead from the
determining equation for the retarded self-energies $\hat \Pi$ and
$\hat\Sigma$,
eqs. (\ref{dg1}), (\ref{dq1}), to the solution for the renormalization
functions
$\Delta_g$, $\Delta_q$, eqs. (\ref{dq2}), and finally to the
evolution equations for the phase-space densities $F_g$, $F_q$, (\ref{R5}).
I exemplify the procedure for the simpler case of the quark self-energy. The
case of the gluon self-energy is more eleborate, but conceptually it is
completly
analogous.
I confine myself here to the leading log approximation (LLA), referring to
Refs.
\cite{dok80,amati80} for additional reading.

The quantity of interest is hence the quark self-energy
$\Sigma_{ij}= \delta_{ij}p^2\hat{\Sigma}$, eq. (\ref{hatPiS}),
given by (\ref{Pi1l}).
As explained in Sec. 2.1, when studying short-distance dynamics around the
lightcone, it is appropriate and most convenient to work in the
planar axial gauge $n\cdot A = 0$, eqs. (\ref{Enn}), (\ref{gauge}) with the
constant vector $n^\mu$ satisfying $n^2 \ll 1$.
Parametrizing it as $n^\mu = (a+b, 0,0,a-b)$ then requires $n^2 = 4ab \ll 1$.
Without loss of generality, one may set $b=1$ and $a \ll 1/4$, so that
the scalar product of $n$ with some four-vector $q$ is
$n\cdot q= q^+ + a q^- \simeq  q^+$
with
$q^\pm= q_0\pm q_3$,
$q^+ q^-= q^2 - q_\perp^2 \simeq q^2 \ll q^{+\;2}$.
Let me then proceed with eq. (\ref{dq1}) for the variation of scalar
quark self-energy function $\hat \Sigma(r,p)$ of a quark with momentum $p$
and virtuality $p^2 \ll p^{+\;2}$, within a given space-time
cell of volume $\Omega(r)=\Delta r^0 \Delta^3 r = \mu^{-4}(r)$ around $r$ (c.f.
Sec. 3.1):
\begin{eqnarray}
& &
\!\!\!\!\!\!\!\!\!
p^2 \frac{\partial}{\partial p^2} \hat{\Sigma}^{R(A)}(r,p)
\;=\;
-\;g_s^2\,C_F \,(2\pi i) \;\int \frac{d^4 p'}{(2\pi)^4 i}
\int_{r_0-1/(2\mu)}^{r_0+1/(2\mu)} d\tau\,{\cal T}(p'\tau)
\label{C1}
\\
& &
\;\times\;
\;
\frac{\partial \Delta_q(r,p')}{\partial p'^{\,2}}
\;
\frac{\partial}{\partial k''^{\,2}}
\left(\frac{}{}
V^2_{qqg}(r;\, p^2,p'^2,k''^2, \kappa_p,\kappa_{p'}, \kappa_{k''})
\,
\Delta_g(r,k'')
\right)
\times
{\cal U}_q^{qg}(p',k'',n)
\nonumber
\end{eqnarray}
with
\begin{equation}
\kappa_p \;=\; \frac{n\cdot p}{n^2} = p^{+\;2} \,
+\,\frac{1}{2} \,(p^2 - p_\perp^2) \;\simeq\;\,p^{+\;2}
\,
\end{equation}
and
\begin{equation}
{\cal U}_q^{qg}(p',k'',n) \;=\;
-2 \;
\frac{(p'\cdot k'') (\gamma \cdot n) + (p'\cdot n) (\gamma\cdot k'')}{n\cdot
k''}
\;+\; O(n^2)
\label{C2}
\;.
\end{equation}
Here, $p$ is the four-momentum of
the incoming quark which branches into $p'$ and $k''$
of the outgoing quark and gluon, respectively.
Because $n^2 \ll 1$, the  terms $O(n^2)$ are negligible
and will be omitted in the following.

The time integral $\int d\tau$ in (\ref{C1}) over the finite
time slice $\mu^{-1}\equiv\mu^{-1}(r)$ of the space-time cell around $r$,
is weighted by the function ${\cal T}$, satisfying
\begin{equation}
\int_0^\infty d\tau \,{\cal T}(p'\tau)\;=\; 1
\;,
\end{equation}
e.g. ${\cal T} = p'\exp(-p'\tau)$, and the relation between
momentum and space-time is  determined by the uncertainty constraint,
which limits the range of virtualities $p^{'\,2}$ such that
within the finite time slice $\Delta r_0=\mu^{-1}$ only those fluctuations
$k\leftrightarrow k'+k''$ are resolvable that are sufficiently short-living,
with proper life-time $\tau_0\simeq 1/k'$ and $\gamma\tau_0 \simeq k^+/k{'\,2}
< \Delta r_0$.
(for details see Ref. \cite{ms17})
\begin{equation}
\tau(p)\;=\; \gamma \,\tau_0(p) \;=\;\frac{p^+}{p^{'\,2}}
\;< \;\frac{1}{\mu(r)}
\;,
\end{equation}
where $\tau_0\simeq 1/p'$ is the proper life-time of the virtual parton,
to be understood in the averaged sense.

To perform the integral (\ref{C1}), the procedure is as follows.
First, due to the kinematic ordering condition
$p^2 \gg p^{'\,2}, k^{''\,2}$
in the LLA, the
gluon momentum $k''$ can be decomposed as
\begin{eqnarray}
k'' &=&\; (1-z) \,p \;-\;\left(\frac{1}{2}-z\right) \, \frac{n}{n\cdot p}\;p^2
\;+\; k''_\perp
\;+\;O(n^2)
\nonumber
\\
z&=& \frac{p^{'\,+}}{p^+} \;\;,\;\;\;\;\;\;k''_\perp\cdot p\;=\;k''_\perp\cdot
n\;=\;0
\label{C3}
\;.
\end{eqnarray}
Then one can rewrite (\ref{C2}) in the form
\begin{equation}
{\cal U}_q^{qg}(p',k'',n) \;=\;
2 z \,(\gamma\cdot p) \;+\; (1-z + 2 z^2) \;\frac{(\gamma\cdot n)}{(n\cdot
k'')}\;p^2
\label{C4}
\;.
\end{equation}
Next, one rewrites the the integration measure in (\ref{C1}) as
\begin{equation}
d^4 p' \;=\; \frac{\pi}{2} \,dp^{'\,2} dk^{''\,2} \, dz \;
\theta\left( p^2 - \frac{p^{' \,2}}{z} - \frac{k^{''\,2}}{1-z}\right)
\;,
\label{C5}
\end{equation}
where the $\theta$-function accounts for the aforementioned
ordering of virtualities and acts as a kinematic constraint
that limits the integration range, such that
\begin{equation}
k^{''\,2}\;\le\; (1-z)\left(p^2 \;-\; \frac{p^{'\,2}}{z}\right)
\label{C6}
\;.
\end{equation}
Finally, to simplify the analysis, I approximate the $\tau$ integral
by \cite{ms17}
\begin{equation}
\int_{r_0-1/(2\mu)}^{r_0+1/(2\mu)} d\tau \,{\cal T}(p'\tau)
\;\approx\; \theta\left( \frac{1}{\mu(r)} - \frac{p^+}{p^{'\,2}}\right) \;=\;
\theta\left( \frac{z p^2}{p^+}\,r_0 \,-\,\mu(r) \right)
\;,
\end{equation}
where $\mu(r)$ is characterizes the size of the space-time intervall
of the localized quantum fluctuations (c.f. Sec. 3.1, eq. (\ref{scale1}).
Using the above formulae, the integration over $k^{''\,2}$
now gives
\begin{eqnarray}
p^2 \frac{\partial}{\partial p^2} \hat{\Sigma}^{R(A)}(r,p)
&=&
\frac{g_s^2\,C_f}{16 \pi^2} \;\int dz \int^{zp^2}
\int d^4 p^{'\,2}
\;\left[
2 z \,(\gamma\cdot p) \;+\; (1-z + 2 z^2) \left(\frac{p^2 \,(\gamma\cdot
n)}{(n\cdot k'')}\right)
\right]
\nonumber \\
& &
\times \;
V^2_{qqg}\left(r;\, p^2,p'^2,(1-z)\left(p^2 - \frac{p^{'\,2}}{z}\right);
p^{+\,2},zp^{+\,2}, (1-z)p^{+\,2}
\right)
\nonumber \\
& &
\times\;
\Delta_g\left(r;(1-z)\left(p^2 - \frac{p^{'\,2}}{z}\right)\right)
\;
\frac{\partial \Delta_q(r,p')}{\partial p^{'\,2}}
\;\,
\theta\left( \frac{z p^2}{p^+}\,r_0 \,-\,1 \right)
\nonumber
\;.
\label{C8}
\end{eqnarray}
Next one integrates over  $p^{'\,2}$, which yields for the last two
factors in (\ref{C6}) an
`effective vertex function' under the remaining $z$-integral
\begin{equation}
V^2_{eff}\;\equiv\;
\frac{g_s^2}{4\pi}
V^2_{qqg}\left(r;\, p^2,zp^2,(1-z)p^2;
p^{+\,2},zp^{+\,2}, (1-z)p^{+\,2}
\right)
\;
\Delta_q(r; z p^2)
\;
\Delta_g(r; (1-z) p^2)\;
\label{C9}
\;,
\end{equation}
which in the LLA has been shown \cite{dok80,amati80} to generate the running of
the
coupling $\alpha_s=g_s^2/(4\pi)$, in the present case however modified by
the finite time slize effect,
\begin{equation}
V^2_{eff}\;=\;
\Delta^{-1}_q(r;p^2) \;\alpha_s\left( (1-z) p^2\right)
\;\theta\left( \frac{z p^2}{p^+}\,r_0 \,-\,1 \right)
\;.
\label{C10}
\end{equation}
Employing this identification,  inserting the decomposition
\begin{equation}
\hat{\Sigma}^{R(A)}(r,p)\;=\;
\hat{\Sigma}_1 \,(\gamma\cdot p)
\;+\;
\hat{\Sigma}_2 \,\frac{p^2 (\gamma \cdot n)}{n\cdot p}
\label{C11}
\;,
\end{equation}
into (\ref{C8}),
and solving for $\Delta_q$ and $\tilde{\Delta}$ in the
parametrization of the self-energy (\ref{ansatz1})
\begin{equation}
\Delta_q(r,p) \;=\;
\frac{1}{1 + \hat{\Sigma}_1 + \hat{\Sigma}_2}
\;\;\;\;\;\;\;\;\;\;\;\;
\tilde{\Delta}_q(r,p) \;=\;
\Delta_q (r;p^2) \;\frac{\hat{\Sigma}_2}{1+\hat{\Sigma}_1}
\label{C12}
\end{equation}
one finds the form of the quark renormalization function as stated in Sec. 3.5:
\begin{equation}
\Delta_q(r,p^2,p^{+\,2})
\;=\;
\exp\left\{
-\,\int_{p^2}^{p^{+\,2}}
\frac{d p^{'\,2}}{p^{'\,2}}
\int_0^1 dz \;
A(r,z,p^{'\,2})
\,\gamma_q^{qg}(z,\epsilon)
\right\}
\;,
\label{C13}
\end{equation}
where
\footnote{A detailed derivation of the effective coupling function $A(r,p^2,z)$
can be found in Ref. \cite{ms17}.
}
\begin{equation}
A(r,p^2,z)\;\simeq\; \frac{\alpha_s\left((1-z)p^2\right)}{2\pi}\;
\theta\left( \frac{z p^2}{p^+}\,-\,\mu(r) \right)
\label{C14}
\end{equation}
and
\begin{equation}
\gamma_{q}^{q g} (z,\epsilon) \;=\;
C_F\; \left( \frac{1 + z^2}{1 - z +\epsilon(p^{+\,2})} \right)
\;,
\label{C15}
\end{equation}
with
\begin{equation}
\epsilon(p^{+\,2})\;=\; \frac{p^2 n^2}{4 (p\cdot n)^2}\;=\;
\frac{p^2}{p^{+\,2}}\;\ll\; 1
\;.
\end{equation}
By repeating the same analysis for the gluon case one obtains
\begin{equation}
\Delta_g(r,k^2,k^{+\,2})
\;=\;
\exp\left\{
-\,\int_{k^2}^{k^{+\,2}}
\frac{d k^{'\,2}}{k^{'\,2}}
\int_0^1 dz \;
A(r,z,k^{'\,2})
\;\left(
\,\frac{1}{2} \gamma_g^{gg}(z,\epsilon)
\,+\,
\,\gamma_g^{qq}(z,\epsilon)
\right)
\right\}
\end{equation}
where
\begin{equation}
\gamma_{g}^{gg} (z,\epsilon) \;=\;
2\,C_A\;\left( \frac{z}{1-z+\epsilon(k^{+\,2})} + \frac{1-z}{z} + z ( 1 - z )
\right)
\;,\;\;\;\;\;
\\
\gamma_{g}^{qq} (z,\epsilon) \;=\;
\frac{1}{2} \, \left( z^2 + (1 - z)^2 \right)
\;.
\end{equation}

The renormalization functions $\Delta_f$ ($f=g,q$) determine the form
of the gluon and quark structure functions ${\cal P}_{f}^{f'}(r,p)$ defined
in Sec. 3.1, eq. (\ref{Pgq}), as
the spectral densities of the parton phase-space-distributions $F_{f}(r,p)$:
\begin{eqnarray}
& &
{\cal P}_{f}^{f'}(r; x,p^2) \;=\;
\delta_f^{f'}\;\delta(1-x)\,\delta\left(p^2 - \mu_{gq}^2\right)
\;\Delta_{f}(r;\mu_{gq}^2,p^{+\,2})
\nonumber
\\
& &
\;\;\;+
\;\;
\Delta_{f}(r,\mu_{gq}^2,p^{+\,2})
\;\,\sum_{f''}
\int_{\mu_{gq}^2}^{p^2}\frac{d p^{'\,2}}{p^{'\,2}}
\int_0^1 dz \,
\left\{
A(r,p^2,z) \,\gamma_f^{f''f'}(z,\epsilon)
\;{\cal P}_{f''}^{f'}\left(\frac{x}{z},p^{'\,2};p^{'\,+\,2}\right)
\right.
\nonumber
\\
& &
\left.
\;\;\;\;\;\;\;\;\;\;\;\;\;\;
\;\;\;\;\;\;\;\;\;\;\;\;\;\;
\;\;\;\;\;\;\;\;\;\;
\times\;
\Delta_f^{-1} \left(r,\frac{p^{'\,2}}{z},p^{'\,+\,2}\right)
\right\}
\;.
\end{eqnarray}
Because  $F_f = {\cal N}_f\otimes {\cal P}_f$,
eq. (\ref{Fgq}),
the variation of $F_{f}(r,p)$ with the parton momentum p
(more precisely, with virtuality $p^2$)
therefore reflects the parton's changing  gluon-quark substructure
as dictated by the renormalization functions $\Delta_{g(q)}$, also called
the Sudakov formfactor of a gluon (quark).
This connection between $\Delta_{f}$ and $F_{f}$ emerges as follows.
Treating gluons and quarks on the same footing,
the differentiation of $\Delta_f^{-1}$ with respect to $p^2$,
the incoming partons's virtuality, yields
\begin{equation}
p^2 \frac{\partial}{\partial p^2} \Delta_f^{-1}(r;p^2,p^{+\,2})
\;=\;
- \Delta_f^{-1}(r;p^2,p^{+\,2})
\;\sum_{f''}\int dz \,A(r; p^2,z) \,\gamma_f^{f'f''}(z,\epsilon)
\label{C15a}
\;,
\end{equation}
where the sum over  $f''=g,q$ automatically fixes $f'$ due to the symmetry
properties of the kernels $\gamma_f^{f'f''}$ under interchange of $f'$ and
$f''$ \cite{dok80}.
On account of the momentum sum rule (\ref{X6}) for the parton structure
functions
${\cal P}(r,p) = {\cal P}(r; x,p^2)$, i.e.
the momenta $p^+=xP^+(r)$ of all partons in a given
cell around $r$ add up to the total cell momentum $P^+(r)$,
one has
\begin{equation}
1\;=\;\sum_{ff'} \int dx\; x\; {\cal P}_f^{f'}(r; x,p^2)
\;,
\label{C16a}
\end{equation}
and hence
\begin{equation}
\sum_{ff'}
\int dx\; x\;
p^2 \frac{\partial}{\partial p^2} {\cal P}_f^{f'}(r; x,p^2)
\;=\;0
\;,
\label{C17}
\end{equation}
and therefore can rewrite (\ref{C15}) in the following form:
\begin{eqnarray}
& &
\left[
p^2 \frac{\partial}{\partial p^2} \Delta_f^{-1}(r;p^2,p^{+\,2})
\right]
\;{\cal P}_f^{f'}(r; x,p^2)
\;+\;
\Delta_f^{-1}(r;p^2,p^{+\,2})
\left[
p^2 \frac{\partial}{\partial p^2} {\cal P}_f^{f'}(r; x,p^2)
\right]
\nonumber \\
& &
\;\;\;\;\;\;\;\;\;\;\;\;\;\;\;
\;=\;
- \Delta_f^{-1}(r;p^2,p^{+\,2})
\sum_{f''} \int dz\,
 \,A(r; p^2,z) \,\gamma_f^{f'f''}(z,\epsilon)
\,\frac{1}{z} \,
{\cal P}_{f''}^{f'}\left(r; \frac{x}{z},zp^2\right)
\label{C18}
\;.
\end{eqnarray}
Employing eq. (\ref{C15a}), then yields
\begin{eqnarray}
& &
\Delta_f^{-1}(r;p^2,p^{+\,2})
\;p^2 \frac{\partial}{\partial p^2}
{\cal P}_f^{f'}\left(r; x,p^2\right)
\;=\;
\\
& &
\;\;\;\;\;\;\;\;\Delta_q^{-1}(r;p^2,p^{+\,2})
\;
\sum_{f''} \int dz\,
 \,A(r; p^2,z) \,\gamma_f^{f'f''}(z,\epsilon)
\;
\left\{
{\cal P}_{f}^{f'}\left(r; x,p^2\right)
\;-\;
\,\frac{1}{z} \,
{\cal P}_{f''}^{f'}\left(r; \frac{x}{z},zp^2\right)
\right\}
\nonumber
\;.
\end{eqnarray}
The final evolution equation for the parton phase-space densities
$F_f(r,p)$ is obtained by
(i) multiplying with $\Delta_f$, (ii) convoluting the resulting
equation according to eq. (\ref{Fgq})
with the local parton density ${\cal N}_f(r,p)$, i.e. the number
of dressed partons in a given cell around $r$,
and, (iii) accounting for the competition between real emission
and reverse absorption processes \cite{ms1}, using the fact that the squared
matrix-elements
$\propto \gamma_a^{bc}$ are invariant under reversal
$a\rightarrow bc$ and $bc \rightarrow a$.
The extended result is:
\begin{eqnarray}
k^2 \frac{\partial}{\partial k^2}
F_g\left(r; x,k^2\right)
&=&
\;\;\int_0^1 dz\,
\,A(r; k^2,z) \,
\left\{
\;\left[
\frac{1}{z} \,
F_g\left(r; \frac{x}{z},zk^2\right)
\;-\;
\frac{1}{2} F_g\left(r; x,k^2\right)
\right]
\;\Gamma_g^{gg}(z,\epsilon)
\right.
\nonumber \\
& &
\;\;\;\;\;\;\;+
\left.\frac{}{}
2\, N_f\;F_q\left(r; x,k^2\right)
\Gamma_q^{gq}(z,\epsilon)
\;-\;
N_f\;F_g\left(r; x,k^2\right)
\Gamma_g^{qq}(z,\epsilon)
\;\right\}
\;\;\;
\nonumber \\
& & \nonumber \\
p^2 \frac{\partial}{\partial p^2}
F_q\left(r; x,p^2\right)
&=&
\;\;\int_0^1 dz\,
\,A(r; p^2,z) \,
\left\{
\;\left[
\frac{1}{z} \,
F_q\left(r; \frac{x}{z},zp^2\right)
\;-\;
F_q\left(r; x,p^2\right)
\right]
\;\Gamma_q^{qg}(z,\epsilon)
\right.
\nonumber \\
& &
\;\;\;\;\;\;\;+
\left.\frac{}{}
F_g\left(r; x,p^2\right)
\Gamma_g^{qq}(z,\epsilon)
\;\right\}
\label{C20}
\;,
\end{eqnarray}
where
\begin{equation}
\Gamma_f^{f'f''}\;=\; \gamma_f^{f'f''}\;
\left(1\;-\;\frac{F_{f'}}{F_{f'}\,\pm\,1}\right)
\label{C21}
\end{equation}
represents the net emission probability,
being a manifestation of the principle of detailed balance
between the rate of emission of parton $f'$ from a parton $f$, and the
rate of absorption of a quantum $f'$ in the phase-space proximity of parton
$f$.
The net rate $\Gamma_f^{f'f''}$ results in a suppression
when $F_g$ or $F_q$ becomes large, and thus reflect correctly
the Bose-Einstein and Fermi-Dirac statistics (+ for gluons, $-$ for quarks).
On the other hand, when $F_g$ and $F_q$ are small compared to 1, the usual
branching kernels $\gamma_f^{f'f''}$ are recovered.
\bigskip
\bigskip

\begin{center}
{\bf APPENDIX D:\\ Derivation of drift term  and collison kernel of the
transport equations}
\end{center}
\bigskip

This Appendix explains the derivation of the  transport
equations (\ref{R5}), (\ref{T5}) that govern the kinetic,
dispersive dynamics of dressed gluons and quarks.
First of all, one observes that from the matrix representation
of the CTP Dyson-Schwinger equations
(\ref{eom2a}), (\ref{eom2b}) in terms of the Green functions
$G^F,G^{\overline{F}},G^>, G^<$ and self-energies ${\cal E}^F,{\cal
E}^{\overline{F}},
{\cal E}^>,{\cal E}^<$, where $G\equiv D_{\mu\nu},S$ and ${\cal
E}=\Pi_{\mu\nu}, \Sigma$,
 follow immediately the corresponding equations for the retarded,
advanced, and correlation functions
$G^R,G^A,G^C$ and ${\cal E}^R,{\cal E}^A, {\cal E}^C$, as given by
(\ref{DSE5}):
\begin{eqnarray}
\left( D_{R(A)}^{-1}\right)^{\mu\nu}(r,k)
&=&
\left( D_{(0)\,R(A)}^{-1}\right)^{\mu\nu}
\;-\;
\left(\Pi_{R(A)}\right)^{\mu\nu}
\nonumber \\
 S_{R(A)}^{-1}(r,p)
&=&
 S_{(0)\,R(A)}^{-1}
\;-\;
\Sigma_{R(A)}
\label{D1}
\end{eqnarray}
and
\begin{eqnarray}
D_{C}^{\mu\nu} (r,k)
&=&
-\,D_{R}^{\mu\mu'}
\left[ \frac{}{}
\left( D_{(0)\,C}^{-1}\right)^{\mu'\nu'}
-
\left(\Pi_{C}\right)^{\mu'\nu'}
\right]
\,D_{A}^{\nu'\nu}
\nonumber \\
S_{C}(r,p)
&=&
-\,S_{R}
\;\left[ \frac{}{}
 S_{(0)\,C}^{-1}
\;-\;
\Sigma_{C}
\right]
\;S_{A}
\label{D2}
\end{eqnarray}
for the quarks.
It is convenient to introduce scalar and dimensionless self-energy functions
$\hat{\Pi}$ and $\hat{\Sigma}$ through
\begin{equation}
\Pi^{\mu\nu}_{ab}(r,k)\;=\;
\delta_{ab}\,
\left(k^\mu k^\nu \,- \,g^{\mu\nu}\,k^2\right) \;\hat{\Pi}(r,k)
\;\;\;\;\;\;\;\;\;\;\;\;\;
\Sigma_{ij}(r,p)\;=\;
\delta_{ij}\;p^2\;\hat{\Sigma}(r,p)
\label{Y1}
\;,
\end{equation}
so that the propagators can be written as
\begin{equation}
\!\!\!\!\!\!\!\!\!
D_{ab}^{\mu\nu\;R\,(A)}(r,k)\;=\;
\delta_{ab}\;(-d_{\mu\nu}(k))\;\frac{1}{\pi_0\,\pm\,i\,\pi_1}
\;\;\;\;\;\;\;
S_{ij}^{R\,(A)}(r,p)\;=\;
\delta_{ij}\;(\gamma\cdot p)\;\frac{1}{\sigma_0\,\pm\,\sigma_1}
\;,
\label{Y2}
\end{equation}
and the correlation functions as
\begin{equation}
\!\!\!\!\!\!\!\!\!
D_{ab}^{\mu\nu\;C}(r,k)
\;=\;
\delta_{ab}\;(-d_{\mu\nu}(k))\;\frac{-2 i \,\pi_2}{\pi_0^2\,+\,\pi_1^2}
\;[1\,+\,F_g]
\;\;\;\;\;\;
S_{ij}^{C}(r,p)\;=\;
\delta_{ij}\;(\gamma\cdot p)\;\frac{-2 i \sigma_2}{\sigma_0\,\pm\,\sigma_1}
\;[1\,-\,F_q]
\;,
\label{Y3}
\end{equation}
where $d_{\mu\nu}(k)$ is defined in (\ref{prop2}), and
$F_g$ and $F_q$ are the phase-space densities of dressed gluons and quarks as
defined by (\ref{Fgq}).
This separation of real and imaginary
contributions uniquely determines the Green functions in terms
of the three functions $\pi_i$, respectively $\sigma_i$, as can be
shown rigorously \cite{chou}.
The real parts correspond to the dispersive and wavefunction renormalization
piece, whereas the imaginary parts give rise to dissipation and decay.
Formally,
\begin{eqnarray}
\pi_0 &=& k^2\;(1\,-\,\mbox{Re}\hat{\Pi})
\;\;\;\;\;\;\;\;\;\;
\mbox{Re}\hat{\Pi}\;=\;
\frac{1}{2} \left( \hat{\Pi}^R\,+\,\hat{\Pi}^A\right)
\nonumber \\
\pi_1 &=& - k^2\;\mbox{Im}\hat{\Pi}
\;\;\;\;\;\;\;\;\;\;
\mbox{Im}\hat{\Pi}\;=\;
\frac{i}{2} \left( \hat{\Pi}^R\,-\,\hat{\Pi}^A\right)
\label{Y5} \\
\sigma_0 &=& p^2\;(1\,-\,\mbox{Re}\hat{\Sigma})
\;\;\;\;\;\;\;\;\;\;
\mbox{Re}\hat{\Sigma}\;=\;
\frac{1}{2} \left( \hat{\Sigma}^R\,+\,\hat{\Sigma}^A\right)
\nonumber \\
\sigma_1 &=& - p^2\;\mbox{Im}\hat{\Sigma}
\;\;\;\;\;\;\;\;\;\;
\mbox{Im}\hat{\Sigma}\;=\;
\frac{i}{2} \left( \hat{\Sigma}^R\,-\,\hat{\Sigma}^A\right)
\label{Y6}
\;,
\end{eqnarray}
and
\begin{equation}
\pi_2 \;=\; \frac{i}{2} \left( \hat{\Pi}^<\,+\,\hat{\Pi}^>\right)
\;\;\;\;\;\;\;\;\;\;\;
\sigma_2 \;=\; \frac{i}{2} \left( \hat{\Sigma}^<\,+\,\hat{\Sigma}^>\right)
\label{Y7}
\;.
\end{equation}
Next, recalling that the correlations among different dressed partons
determine their mutual interactions at kinetic space-time scales,
one focuses on the correlation functions $D^C_{\mu\nu}$ and $S^C$.
Noting that $D_{\mu\nu}^A=D_{\mu\nu}^{R\,\dagger}$, $S^A=S^{R\,\dagger}$,
and employing the representations (\ref{SIII}),
\begin{eqnarray}
D_{\mu\nu}^{C}(r,k)&=&
- 2\pi i \; \left(-d_{\mu\nu}(k)\right)\; \left[ 1\;+\; 2\,F_g(r,k)\right]
\;\delta\left( k^2 \;-\;{\cal M}_g^2(r,k)\right)
\nonumber \\
S^{C}(r,p)&=&
- 2\pi i \;  \left(\gamma\cdot p \right)\;\left[ 1\;-\; 2\,F_q(r,p)\right]
\;\delta\left( p^2 \;-\;{\cal M}_q^2(r,p)\right)
\label{D6}
\;,
\end{eqnarray}
where color indices are suppressed,
eqs. (\ref{Y2})-(\ref{Y7}) may be combined to write
\begin{eqnarray}
-d^{\mu\nu}(k)\;D_{\mu\nu}^{C}(r,k)&=&
\left(\pi_0\,+\,i\,\pi_1\right)^{-1}
\;H_g \;-\; H_g
\left(\pi_0\,-\,i\,\pi_1\right)^{-1}
\;=\;
-\, 2 i  \;
\frac{\pi_{2}}{\pi_{0}^2 \,+\,\pi_{1}^2}
\nonumber \\
\gamma\cdot p \; S^{C}(r,p)&=&
\left(\sigma_0\,+\,i\,\sigma_1\right)^{-1}
\;H_q \;-\; H_q
\left(\sigma_0\,-\,i\,\sigma_1\right)^{-1}
\;=\;
-\, 2 i  \;
\frac{\sigma_{2}}{\sigma_{0}^2 \,+\,\sigma_{1}^2}
\;,
\label{D7}
\end{eqnarray}
where (c.f eq. (\ref{GCF}))
\begin{equation}
H_g(r,k) \;=\; 1\;+\;2\,F_g (r,k)
\;\;\;\;\;\;\;\;\;\;\;
H_q(r,p) \;=\; 1\;-\;2\,F_q(r,p)
\label{D8}
\;.
\end{equation}
Hence, on account of (\ref{D1}) and (\ref{D2}) one finds
\begin{eqnarray}
-d^{\mu\nu}(k)\;
\left(
D_{(0)}^{C\;-1}\,-\,\Pi^C\right)_{\mu\nu}
&=&
- \left(
H_g\,\pi_{0} \;-\; \pi_{0}\;H_g
\right)
\;+\;
i\,\left(
H_g\,\pi_{1} \;+\; \pi_{1}\;H_g
\right)
\nonumber \\
\gamma\cdot p \;
\left(S_{(0)}^{C\;-1}\,-\,\Sigma^C\right)
&=&
-\left(
H_q\,\sigma_{0} \;-\; \sigma_{0}\;H_q
\right)
\:+\;
i\,\left(
H_q\,\sigma_{1} \;+\; \sigma_{1}\;H_q
\right)
\label{D9}
\;.
\end{eqnarray}
Then, by inserting the expressions (\ref{Y6}), (\ref{Y7}) for
$\pi_i$ and $\sigma_i$, one obtains
\begin{eqnarray}
\!\!\!\!\!\!\!\!\!
F_g\,\pi_{0}-\pi_{0} \,F_g
&=&
\frac{1}{2}
\left(\frac{}{}
(1+F_g)\;\hat{\Pi}^<\;+\; \hat{\Pi}^< \;(1+F_g) \;-\;
F_g\;\hat{\Pi}^>\;-\; \hat{\Pi}^> \;F_g \right)
\;\;\equiv\;\;{\cal C}_g
\nonumber
\\
\!\!\!\!\!\!\!\!\!
F_q\,\sigma_{0}-\sigma_{0} \,F_q
&=&
-\frac{1}{2}
\left(\frac{}{}
(1-F_q)\;\hat{\Sigma}^<\;+\; \hat{\Sigma}^< \;(1-F_q) \;+\;
F_q\;\hat{\Sigma}^>\;+\; \hat{\Sigma}^> \;F_q \right)
\;\;\equiv\;\;{\cal C}_q
\;,\;\;
\label{D10}
\end{eqnarray}
with the self-energies $\hat{\Pi}^{\gl}$ and $\hat{\Sigma}^{\gl}$ given by
(\ref{Pi2}) and (\ref{Sigma2}) together with (\ref{Y1}).
Finally, as argued in Sec. 3.5, the presumed clear separation
of quantum and kinetic space-time scales allows one
to treat the kinetic dynamics quasi-classically, by expanding
both sides of (\ref{D10}) in terms of $\hbar$ and keeping only the
lowest order contributions. The lowest order non-zero terms on the
right hand sides of (\ref{D10}) correspond then to the
Born collision terms which are of order $\hbar$,
\begin{eqnarray}
{\cal C}_{g}(r,k) &=&
\;
\left(1\,+\,F_g\right)\;\hat{\Pi}^< \;-\; F_g\;\hat{\Pi}^>
\;\,+\;\,O(\hbar^3)
\nonumber \\
{\cal C}_{q} (r,p)&=&
-\left(1\,-\,F_q\right)\;\hat{\Sigma}^< \;+\; F_q\;\hat{\Sigma}^>
\;\,+\;\,O(\hbar^3)
\label{D10a}
\;,
\end{eqnarray}
whereas the lowest
order non-zero terms on the left hand sides of (\ref{D10})
result in the classical Poisson brackets which are also of order $\hbar$,
\begin{eqnarray}
& &
\left(\frac{}{}
F_g\,\pi_{0}-\pi_{0} \,F_g
\right)(r,k)
\;=\;
-i\;\left(\frac{}{}
(\partial_r F_g)\cdot(\partial_k \pi_0) \;-\;
(\partial_k F_g)\cdot(\partial_r \pi_0)
\right)
\nonumber \\
&&
\;\;\;\;\;\;\;\;\;\;
=\;
-i \left(\frac{\partial \pi_0(r,k)}{\partial k^2}\right)_{k^2=\mu_{gq}^2}
\;
\left(
k\cdot\partial_r F_g(r,k)
\,-\,\frac{1}{2}
\partial_r k_0^2 \cdot \partial_k F_g(r,k)
\right)
\;\,+\;\, O(\hbar^3)
\nonumber \\
& &
\left(\frac{}{}
F_q\,\sigma_{0}-\sigma_{0} \,F_q
\right)(r,p)
\;=\;
-i\;\left(\frac{}{}
(\partial_r F_q)\cdot(\partial_p \sigma_0) \;-\;
(\partial_p F_q)\cdot(\partial_r \sigma_0)
\right)
\label{D11}
\\
&&
\;\;\;\;\;\;\;\;\;\;
=\;
-i\left(\frac{\partial \sigma_0(r,p)}{\partial p^2}\right)_{p^2=\mu_{gq}^2}
\;
\left(
p \cdot\partial_r F_q(r,p)
\,-\,\frac{1}{2}
\partial_r p_0^2 \cdot \partial_p F_q(r,p)
\right)
\;\,+\;\, O(\hbar^3)
\nonumber
\;,
\end{eqnarray}
where as before $\partial_r = \partial/\partial r^\mu$,
$\partial_k = \partial/\partial k^\mu$, etc., and the dot
denotes a scalar product of four-vectors.
The latter equalities in these two equations are obtained by
using the fact that the solutions of the dressed partons'
energy spectra (\ref{Espectra0}) are strongly peaked around
momentum transfers $q_\perp^2 \simeq \mu_{gq}^2$, i.e.
$k_0(r,\vec k)\simeq \sqrt{\vec k^{\,2} +\mu_{gq}^2}$ and
$p_0(r,\vec p)\simeq \sqrt{\vec p^{\,2} +\mu_{gq}^2}$, because
of the well known QCD specific logarithmic behaviour of
the spectral densities $\propto \alpha_s \ln (q_\perp^2/\mu_{gq}^2)$,
and the power law form of the scattering cross-sections
$\propto \alpha_s^2 q_\perp^{-n}$ ($n\simeq 4$).

Finally, using
$\partial_r k_0 =  \partial_r p_0 \approx 0$,
and equating (\ref{D10}) and (\ref{D11}),
one obtains the transport equations of Boltzmann type, stated in Sec. 3.5, eqs.
(\ref{T4}),
\begin{equation}
k\cdot \partial_r \,F_g(r,k) \;=\;
{\cal I}_g(r,k)
\;\;\;\;\;\;\;\;\;\;\;\;\;
p\cdot \partial_r \,F_q(r,p) \;=\;
{\cal I}_q(r,p)
\;,
\label{D12}
\end{equation}
where the Lorentz invariant collision terms ${\cal I}$ on the right hand side
are defined by
\begin{eqnarray}
{\cal C}_g(r,k)
&=&
-i \left(\frac{\partial \pi_0(r,k)}{\partial k^2}\right)_{k^2=\mu_{gq}^2}
\;{\cal I}_g(r,k)
\nonumber \\
{\cal C}_q(r,p)
&=&
-i\left(\frac{\partial \sigma_0(r,p)}{\partial p^2}\right)_{p^2=\mu_{gq}^2}
{\cal I}_q(r,p)
\;.
\label{D12a}
\end{eqnarray}
In (\ref{D11}) and (\ref{D12a}), the derivatives with respect to the virtuality
$k^2$
are to be taken
at $\mu_{gq}^2$, which, according to (\ref{scale3}) and (\ref{norm1}),
defines the scale at which a dressed parton appears as a quasi-particle,
with the renormalization and dissipation effects taken into account in the
spectral
densities ${\cal P}_g$ and ${\cal P}_q$, eq. (\ref{PP}).
Their form is determined by the renormalization equations (\ref{R5}), and
therefore
the parton distributions $F={\cal N}\otimes {\cal P}$ contain implicitely
the short-distance quantum effects.
The derivatives of $\pi_0$ and $\sigma_0$ are related to the renormalization
functions
$\Delta_g$ and $\Delta_q$, respectively, via the correspondence of the
representations (\ref{Y2}) and (\ref{ansatz1}), and one finds
\begin{eqnarray}
-i \left(\frac{\partial \pi_0(r,k)}{\partial k^2}\right)_{k^2=\mu_{gq}^2}
&=&
\Delta_g^{-1}(r,k^2,k^{+\,2}) \left.\right|_{k^2=\mu_{gq}^2} \;\;=\;\;1
\nonumber \\
-i \left(\frac{\partial \sigma_0(r,p)}{\partial p^2}\right)_{p^2=\mu_{gq}^2}
&=&
\Delta_q^{-1}(r,p^2,p^{+\,2}) \left.\right|_{p^2=\mu_{gq}^2} \;\;=\;\;1
\;,
\end{eqnarray}
where the latter equality results from the normalization condition
(\ref{norm1}).

The explicit forms of the collision integrals
${\cal I}$ is obtained by substituting the
correlation functions (\ref{D6}) into the two-loop expressions
(\ref{Pi2}), (\ref{Sigma2}) for the self-energies $\hat \Pi$,
$\hat\Sigma$, and then inserting those into eqs. (\ref{D10}).
Applying the standard cutting rules \cite{cut}
to the resulting self-energies,
as symbolically represented in Fig. 11,
yields the  different
binary collision processes
$ab\leftrightarrow cd$ by which a parton of type $a$ may be gained or lost
in a phase-space element,
namely $gg\leftrightarrow gg$,
$gg\leftrightarrow q\bar q$,
$gq\leftrightarrow g q$,
$qq\leftrightarrow q q$,
$q\bar q\leftrightarrow q \bar q$.
The corresponding collision integrals ${\cal I}_a$ may be compactly represented
in the generic form of (\ref{colli3}):
\begin{eqnarray}
& &
{\cal I}_a(r,p_1)\;\equiv\; \sum_{bcd}
\left(\frac{}{}
- {\cal I}_{cd \rightarrow ab}^{(loss)} (p_1, r) \,+\,
{\cal I}_{ab \rightarrow cd}^{(gain)} (p_1, r)
\right)
\nonumber
\\
& &
\nonumber
\\
& & =\, - \,
\sum_{bcd}
C_{ab} \,C_{cd}
\,
\int \frac{d^3 p_2}{(2 \pi)^3 2 E_2}
\int \frac{d^3 p_3}{(2 \pi)^3 2 E_3}
\int \frac{d^3 p_4}{(2 \pi)^3 2 E_4}
\;(2 \pi)^4 \,\delta^4( p_1 + p_2 - p_3 - p_4) \,
\nonumber
\\
& &
\nonumber
\\
& &
\;\;\;\;\;\;\;\;
\times
\left\{
\frac{}{}
\,
F_a(1) F_b(2)
\;\;
\vert {\cal M}(ab \rightarrow cd) \vert ^2
\;\theta(q_\perp^2-\mu_{gq}^2)\;\;
\left[ 1 \pm F_c(3) \right]
\left[ 1 \pm F_d(4) \right]
\right.
\nonumber
\\
& &
\left.
\;\;\;\;\;\;\;\;\;\;\;
\frac{}{}
\,-\,
\left[ 1 \pm F_a(1) \right]
\left[ 1 \pm F_b(2) \right]
\;\;
\vert {\cal M}(cd \rightarrow ab) \vert ^2
\;\theta(q_\perp^2-\mu_{gq}^2)\;\;
F_c(3) F_d(4)
\,
\frac{}{}
\right\}
\;,
\label{D14}
\end{eqnarray}
where the $F_\alpha (i)
\equiv F_\alpha (p_i, r)$ denote the distribution functions of
parton species $\alpha = a, b, c, d$ and corresponding
four-momenta $p_i = p_1, p_2, p_3, p_4$ at space-time point
$r=(r^0,\vec r)$.
As a consequence of the representations (\ref{D6}), the
squared matrix elements $|{\cal M}|^2$ for the processes $ab\leftrightarrow cd$
(which contain the $2\rightarrow 2$ kinematics, color and spin structure, as
given below)
are weighted by a distribution function $F_\alpha$
for each of the particles coming into the interaction vertex and a
factor $\left[ 1 \pm F_\alpha \right]$ for each of the outgoing ones,
with the $+$ sign refering gluons
and the $-$ sign to quarks and antiquarks.
The factors $S_{ab}=(1 + \delta_{ab})^{-1}$ and $S_{cd}\equiv (1 +
\delta_{cd})^{-1}$
account for the cases where the two incoming and/or outgoing
partons are identical.
Using the identities
\footnote{
Note that in contrast to the standard normalization for
fermions $\propto \sqrt{m/p^0}$, here the normalization is chosen
commonly for both gluons and quarks $\propto 1/(2p^0)$.
},
\begin{eqnarray}
d_{\mu\nu}(k) &=&
\sum_{s=1,2}\varepsilon_\mu(k,s) \cdot \varepsilon_{\nu}^\ast(k,s)
\nonumber \\
\gamma \cdot p  &=&
2 p_0\, \sum _{s=1,2}\overline{u}(p,s)\, u(p,s)
\;=\;
2 p_0 \,\sum _{s=1,2}\overline{v}(p,s) \,v(p,s)
\label{D15}
\;,
\end{eqnarray}
finally, the  squared matrix elements
are obtained  by evaluating the amplitudes illustrated in Fig. 11,
squaring those, averaging over initial
colors and spins, summing over final colors and spins, and summing over
quark flavors.  The resulting expressions are standard and given by:
\begin{eqnarray}
& &
|{\cal M}(g_ag_b\rightarrow g_cg_d)|^2
\;=\;
\nonumber \\
& &
\;\;\;\;\;\;\;\;\;
\;=\;
\frac{g_s^4}{(8\cdot 2)^2}
\sum_{color, \,spin}
\;
\left|
\frac{g_{\tau\tau'}f_{aed}f_{ebc}}{(p_1-p_4)^2}
\;\lambda^{\rho \tau\sigma}(-p_1, p_1-p_4,p_4)
\;\lambda^{\tau' \mu\nu}(p_2-p_3, -p_2,p_3)
\right.
\nonumber \\
& &
\;\;\;\;\;\;\;\;\;
\;\;\;\;\;\;\;\;\;
\;\;\;\;\;\;\;\;\;
+\;
\frac{g_{\tau\tau'} f_{aec}f_{ebd}}{(p_1-p_3)^2}
\;\lambda^{\rho \tau\nu}(-p_1, p_1-p_3,p_3)
\;\lambda^{\tau' \mu\sigma}(p_2-p_4, -p_2,p_4)
\label{M1} \\
& &
\;\;\;\;\;\;\;\;\;
\;\;\;\;\;\;\;\;\;
\;\;\;\;\;\;\;\;\;
+\;
\frac{g_{\tau\tau'}f_{dbe}f_{ecd}}{(p_1+p_2)^2}
\;\lambda^{\rho \mu \tau}(-p_1, -p_2,p_1+p_2)
\;\lambda^{\tau'\nu\sigma}(-p_3-p_4, p_3,p_4)
\nonumber \\
& &
\;\;\;\;\;\;\;\;\;
\;\;\;\;\;\;\;\;\;
\;\;\;\;\;\;\;\;\;
\left.
\frac{}{}
+\;
v_{abcd}^{\rho\mu\nu\sigma}(p_1,p_2,-p_3,-p_4)
\;\right|^2
\nonumber
\\
& &
|{\cal M}(g_ag_b\rightarrow \bar q_i q_j)|^2
\;=\;
\nonumber \\
& &
\;\;\;\;\;\;\;\;\;
\;=\;
\frac{g_s^4}{(8\cdot 2)}
\sum_{f}^{N_f}
\sum_{color, \,spin}
\;
\left|
\overline{u}_j(p_4)\,\left(
T_{ik}^a T_{kj}^b \gamma\cdot\varepsilon(p_2)
\frac{\gamma\cdot(p_1-p_3)}{(p_1-p_3)^2} \gamma\cdot\varepsilon(p_1)
\right)
\,v_i(p_3)
\right.
\nonumber \\
& &
\;\;\;\;\;\;\;\;\;
\;\;\;\;\;\;\;\;\;
\;\;\;\;\;\;\;\;\;
+\;
\overline{u}_j(p_4)\,\left(
T_{ik}^b T_{kj}^a \gamma\cdot\varepsilon(p_1)
\frac{\gamma\cdot(p_2-p_3)}{(p_2-p_3)^2} \gamma\cdot\varepsilon(p_2)
\right)
\,v_i(p_3)
\label{M2}
\\
& &
\left.
\;\;\;\;\;\;\;\;\;
\;\;\;\;\;\;\;\;\;
\;\;\;\;\;\;\;\;\;
+\;
\overline{u}_j(p_4)\,\left(
i f^{abc} T^c_{ij}
\frac{\varepsilon^\mu(p_1)\varepsilon^\nu(p_2)\gamma^\rho}{(p_1+p_2)^2}
\;\lambda_{\mu\nu\rho}(-p_1,-p_2, p_1+p_2)
\right)
\,v_i(p_3)
\right|^2
\nonumber\\
& &
|{\cal M}(g_aq_i\rightarrow g_bq_j)|^2
\;=\;
\nonumber \\
& &
\;\;\;\;\;\;\;\;\;
\;=\;
\frac{g_s^4}{(8\cdot 2)(3\cdot 2)}
\sum_{f}^{N_f}
\sum_{color, \,spin}
\;
\left|
f^{cab} T^c_{ij} \frac{\varepsilon^\mu(p_1)\varepsilon^\nu(p_3)}{(p_1-p_3)^2}
\;\lambda_{\rho\mu\nu}(p_1-p_3, -p_1,p_3)
\;\overline{u}_j(p_4)\gamma^\rho u_i(p_2)
\right.
\nonumber \\
& &
\;\;\;\;\;\;\;\;\;
\;\;\;\;\;\;\;\;\;
\;\;\;\;\;\;\;\;\;
-\;
i T^b_{ik} T^a_{kj}\,\overline{u}_j(p_4)
\gamma\cdot\varepsilon(p_1)\frac{\gamma\cdot(p_2-p_3)}{(p_2-p_3)^2}
\gamma\cdot\varepsilon(p_3)\,u_i(p_2)
\label{M3} \\
& &
\left.
\;\;\;\;\;\;\;\;\;
\;\;\;\;\;\;\;\;\;
\;\;\;\;\;\;\;\;\;
-\;
i T^a_{il} T^b_{lj}\,\overline{u}_j(p_4)
\gamma\cdot\varepsilon^\mu(p_3)\frac{\gamma\cdot(p_2+p_1)}{(p_2+p_1)^2}
\gamma\cdot\varepsilon(p_1)\,u_i(p_2)
\right|^2
\nonumber
\\
& &
|{\cal M}(q_iq_k\rightarrow q_jq_l)|^2
\;=\;
\nonumber \\
& &
\;\;\;\;\;\;\;\;\;
\;=\;
\frac{g_s^4}{(3\cdot 2)^2}
\sum_{f_1,f_2}^{N_f}
\sum_{color, \,spin}
\;
\left|
T_{ij}^a T_{kl}^a \,\overline{u}_j(p_4) \gamma_\mu u_i(p_1)
\frac{1}{(p_1-p_4)^2}
\overline{u}_l(p_3) \gamma^\mu u_k(p_2)
\right.
\nonumber \\
& &
\left.
\;\;\;\;\;\;\;\;\;
\;\;\;\;\;\;\;\;\;
\;\;\;\;\;\;\;\;\;
-\;
\delta_{f_1f_2}
\;
T_{il}^b T_{kj}^b \,\overline{u}_j(p_4) \gamma_\nu u_k(p_2)
\frac{1}{(p_1-p_3)^2}
\overline{u}_l(p_3) \gamma_\nu u_i(p_1)
\right|^2
\label{M4}
\\
& &
|{\cal M}(q_i\bar q_k\rightarrow q_j \bar q_l)|^2
\;=\;
\nonumber \\
& &
\;\;\;\;\;\;\;\;\;
\;=\;
\frac{g_s^4}{3\cdot 2\cdot 2}
\sum_{f_1,f_2}^{N_f}
\sum_{color, \,spin}
\;
\left|
\delta_{f_1f_4}\delta_{f_2f_3}\;
T_{ij}^a T_{lk}^a \,\overline{u}_j(p_4) \gamma_\mu u_i(p_1)
\frac{1}{(p_1-p_4)^2}
\overline{v}_k(p_2) \gamma^\mu v_l(p_3)
\right.
\nonumber \\
& &
\left.
\;\;\;\;\;\;\;\;\;
\;\;\;\;\;\;\;\;\;
\;\;\;\;\;\;\;\;\;
-\;
\delta_{f_1f_2}
\delta_{f_4f_3}
\;
T_{ik}^b T_{li}^b \,\overline{u}_j(p_4) \gamma_\nu v_l(p_3)
\frac{1}{(p_1-p_3)^2}
\overline{v}_k(p_2) \gamma_\nu u_i(p_1)
\right|^2
\label{M5}
\\
& &
|{\cal M}(\bar q_i q_j\rightarrow g_ag_b)|^2
\;=\;
\frac{64}{9}
\;|{\cal M}(g_ag_b\rightarrow \bar q_i q_j|^2
\label{M6}
\;.
\end{eqnarray}
Here
\begin{eqnarray}
\lambda^{\mu\rho\nu}(p_1,p_2,p_3) &\equiv&
(p_1-p_2)^\nu g^{\mu\rho}\;+\;
(p_2-p_3)^\mu g^{\rho\nu}\;+\;
(p_3-p_1)^\rho g^{\mu\nu}
\nonumber
\\
v_{abcd}^{\mu\sigma\tau\nu}(p_1,p_2,p_3,p_4) &\equiv&
\;f_{abe}f_{cde}\,( g^{\rho\nu} g^{\mu\sigma}\,-\,g^{\rho\sigma} g^{\mu\nu} )
\;+\;
f_{ace}f_{bde}\,( g^{\rho\mu} g^{\nu\sigma}\,-\,g^{\rho\sigma} g^{\mu\nu} )
\nonumber \\
& &+\;
f_{ade}f_{cde}\,( g^{\rho\nu} g^{\mu\sigma}\,-\,g^{\rho\mu} g^{\sigma\nu} )
\; \;,
\label{D34}
\end{eqnarray}
are the usual 3-gluon vertex function, and
the 4-gluon vertex, respectively.
The shorthand notation
suppressing spinor and polarizarion indices,
$u(p_1)\equiv u(p_1,s_1)_\alpha$, $\varepsilon(p_2)\equiv
\varepsilon(p_2,s_2)$, etc.,
is employed,
and in (\ref{M4}) and (\ref{M5}), $\delta_{ff'}$ is equal to 1, if the
flavor of the two quarks are of the same flavor, and is zero otherwise.

\newpage

{\bf FIGURE CAPTIONS}
\bigskip

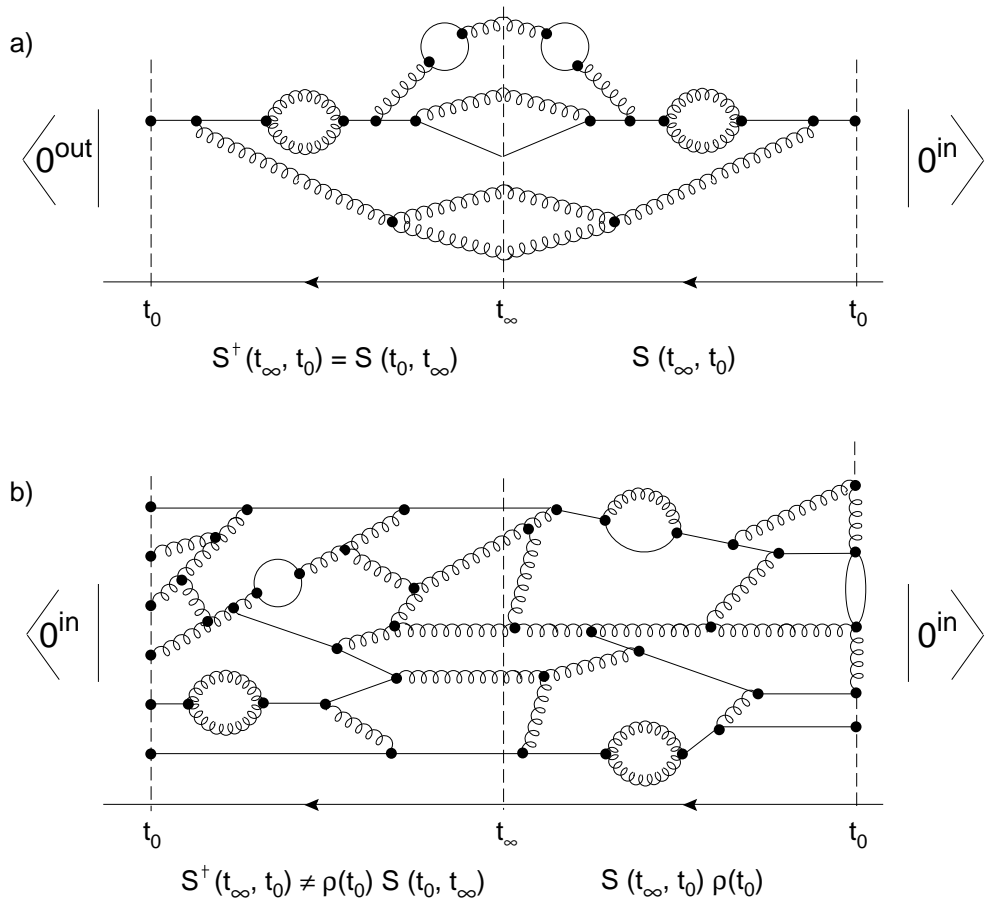
\noindent {\bf Figure 1:}

Illustration of the difference of expectation values
in the {\it in-out} and the {\it in-in} formalism,
corresponding to the time-ordered product of field operators.

{\bf a)}
In the usual $S$-matrix formalism
with a trivial (diagonal) density matrix $\hat\rho(t_0)=\hat 1$
and $|0^{in}\rangle = |0^{out}\rangle$, it suffices to
calculate $\langle 0^{out}|\ldots|0^{in}\rangle$, because of the
symmetry of the time paths $(t_0,t_\infty)$ and $(t_\infty,t_0)$.

{\bf b)}
In the general case of a non-trivial initial state with
multi-particle correlations described by $\hat \rho(t_0) \ne \hat 1$,
one must account for the complete time evolution
on a closed-time-path from $t_0$ to
$t_\infty$ and back to $t_0$ by calculating
$\langle 0^{in}|\ldots \hat\rho |0^{in}\rangle$.
\bigskip

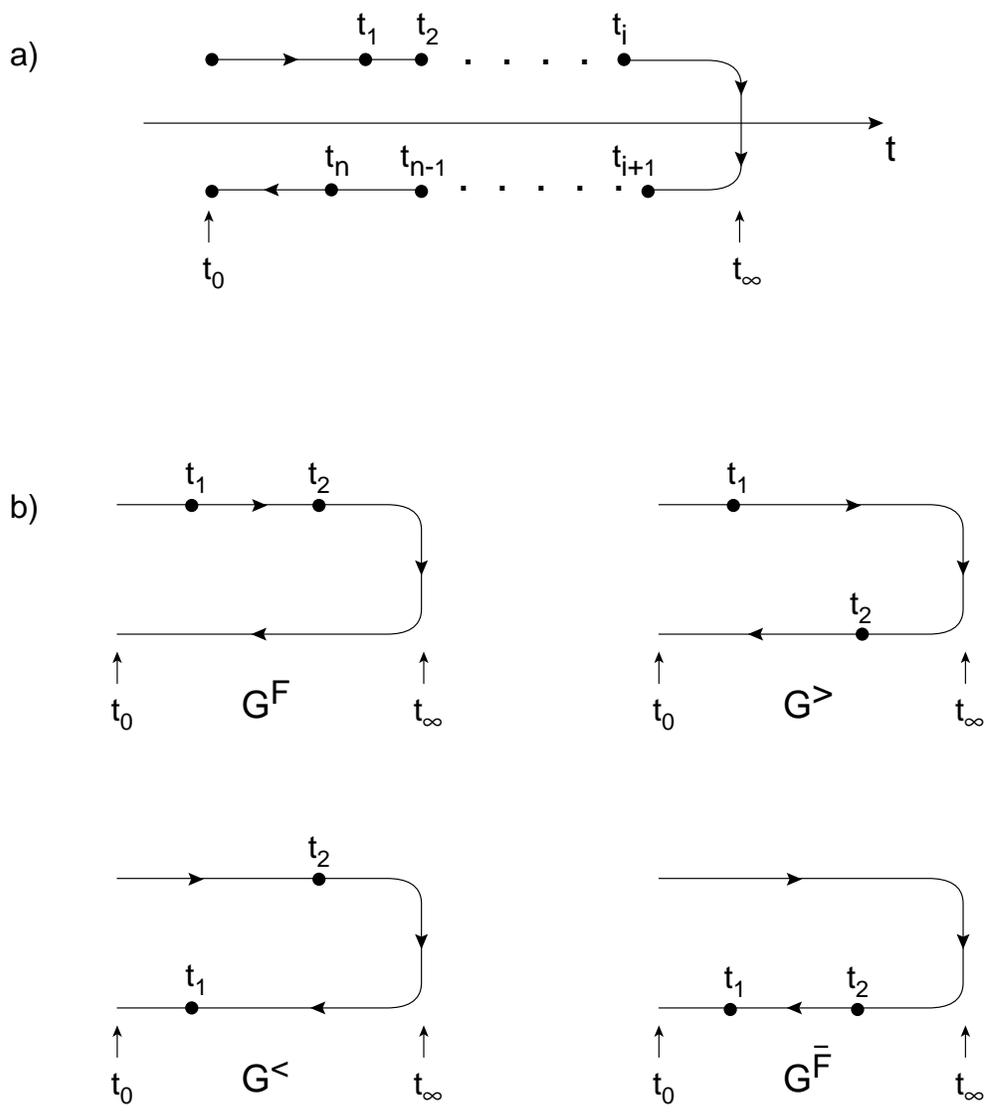
\noindent {\bf Figure 2:}

{\bf a)}
The close-time-path in the complex $t$-plane
for the evolution of operator expectation
values in an arbitrary initial state.
Any point on the  forward, positive branch $t_0\rightarrow t_\infty$
is understood at an earlier instant than any point on the
backward, negative branch $t_\infty\rightarrow t_0$.

{\bf b)}
The four different possible time orderings $(t_1,t_2)$
in the arguments of the 2-point Green functions
$G(x,y)= G(t_1,\vec x;t_2,\vec y)$, corresponding to
$G^F, G^>, G^<,G^{\overline{F}}$.
\bigskip

\noindent {\bf Figure 3:}

Matrix representation of the CTP 2-point functions:
{\bf a)} The Green function $G(x,y)$, and {\bf b)} the
self-energy function $\hat \Sigma(x,y)\propto [G(x,y)]^2$.
\bigskip

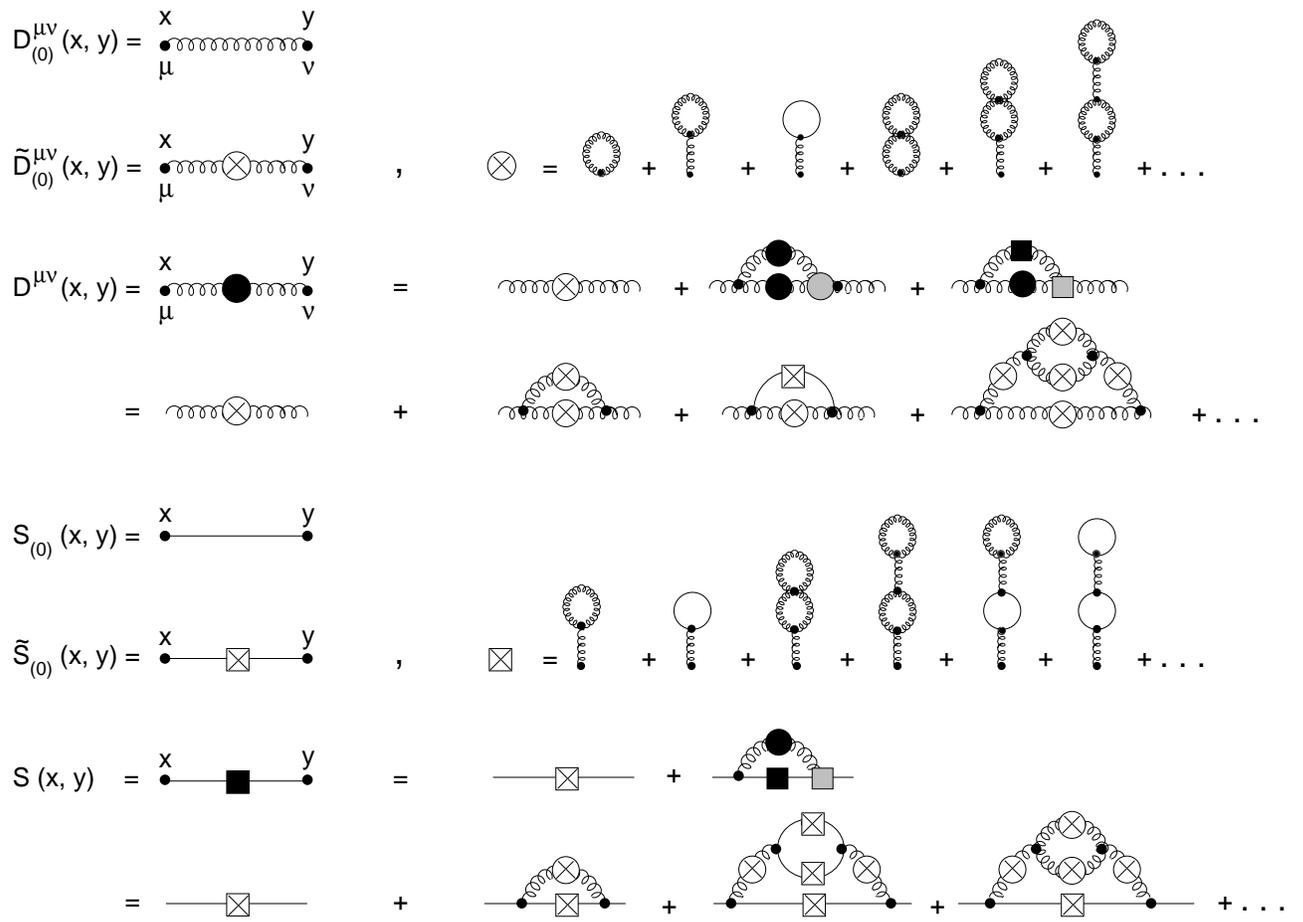
\noindent {\bf Figure 4:}

Diagrammatic representation of the Green functions
(i) $G_{(0)}(x,y)$, the  of the bare propagators,
(ii) $\tilde{G}_{(0)}(x,y)$, including the effect of a mean field
by dressing the bare propagators with a dynamical mass,
and (iii) $G(x,y)$,
the full propagators, dressed by both local mean field and
non-local quantum self-interactions (Dyson-Schwinger equations).
\bigskip

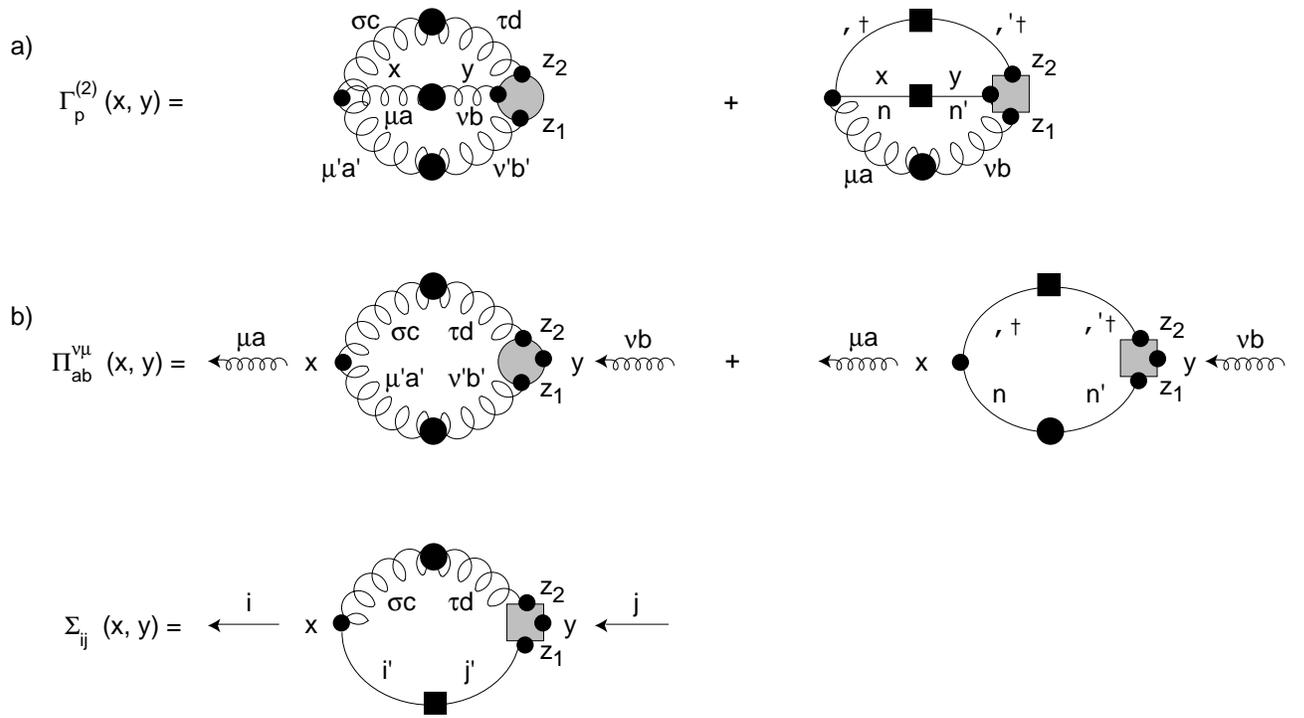
\noindent {\bf Figure 5:}

{\bf a)}
Diagram of the  function $\Gamma_P^{(2)}$, eq. (\ref{Gamma2}),
representing the sum of all two-particle irreducible graphs of order
$\hbar^2,\hbar^3,\ldots$, with fully dressed propagators $D^{\mu\nu}$ and $S$.

{\bf b)}
Illustration of the  self-energies
$\Pi^{\mu\nu}$ and $\Sigma$, eqs. (\ref{Pi}) and (\ref{Sigma}),
which derive from $\Gamma_P^{(2)}$ by functional differentiation with respect
to
$D^{\mu\nu}$ and $S$.
\bigskip

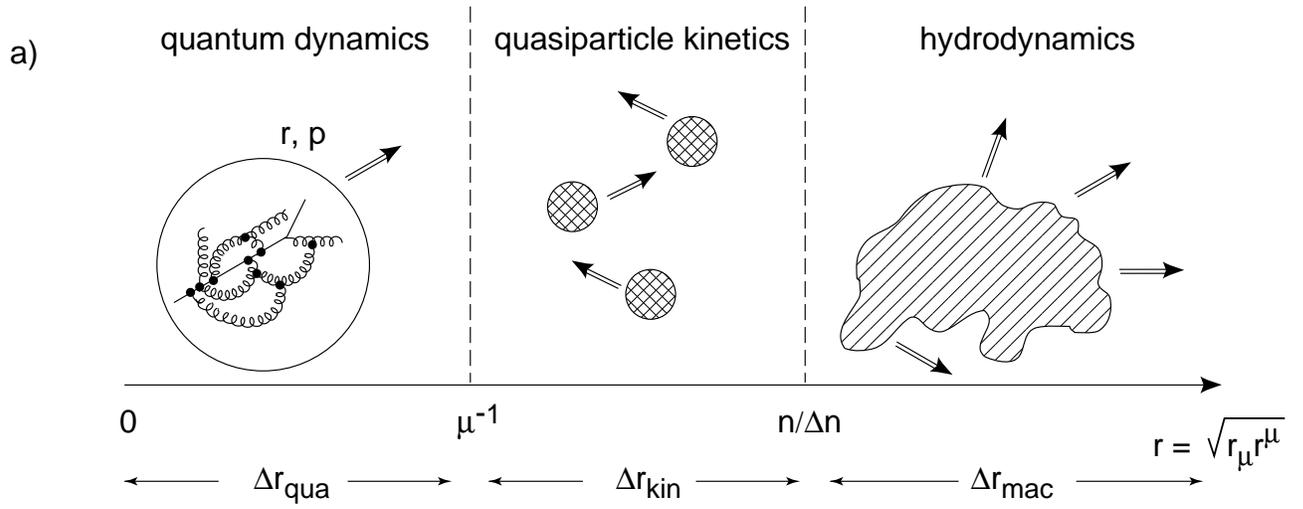
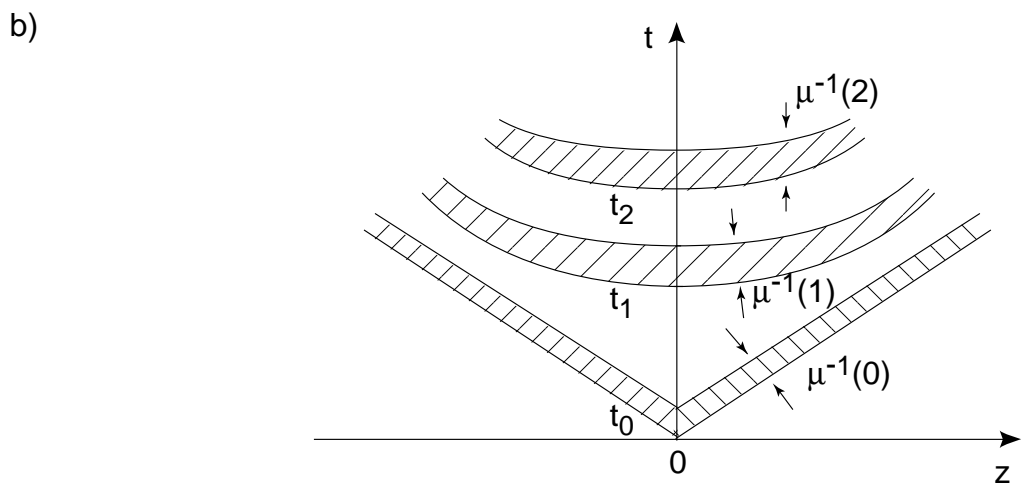
\noindent {\bf Figure 6:}

{\bf a)}
Classification of the different scales of relevance:
(i) the quantum scale $\Delta r_{qua}$, of the order of the spatial extent of
quantum fluctuations associated with the `radiative' self-energies, and
defining
a dressed parton state as a quasi-particle;
(ii) the kinetic scale $\Delta r_{kin}$, measuring the range of
correlations and binary interactions between these quasi-particles,
with the `collisional' self-energies;
and (iii) the `macroscopic' scale $\Delta r_{mac}$,
where the dynamics can be described in terms of bulk thermodynamic variables
or hydrodynamics.

{\bf b)}
The quality of separation of quantum and kinetic scales is controlled by the
choice of
$\mu(r)$. Because the the multi-particle dynamics of the system in general may
change
the scale of separation in space-time, one may choose $\mu(r)$
variable to optimize the kinetic description.
\bigskip

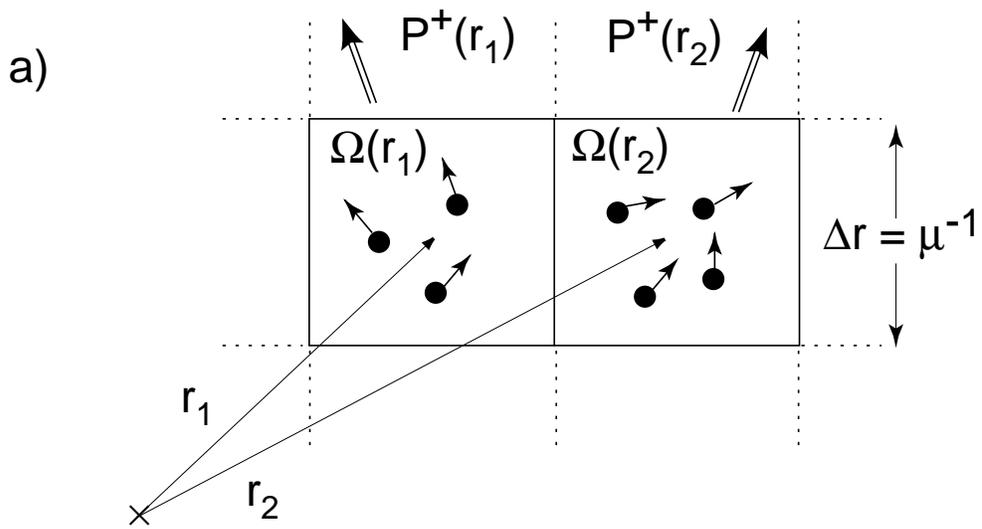
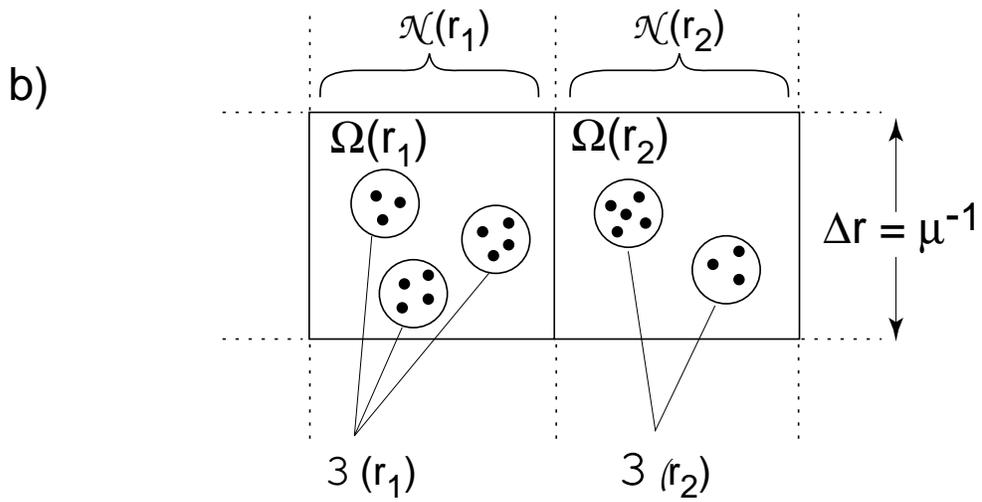
\noindent {\bf Figure 7:}

{\bf a)}
Illustration of the cellular space-time picture, with cell size
chosen intermediate between quantum and kinetic scales such that
the separation between the two scales is optimal, so that
short-distance quantum correlation between different cells are negligible.

{\bf b)}
Representation of the partons' phase-space densities
$F= {\cal N}\otimes {\cal P}$ as a convolution of the statistical density of
dressed partons ${\cal N}$ with the spectral density ${\cal P}$
of each dressed parton, describing its intrinsic density of bare parton states
as its quantum substructure.
\bigskip

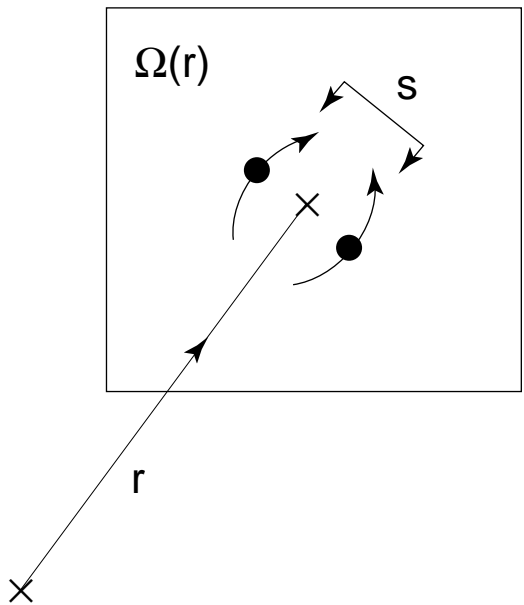
\noindent {\bf Figure 8:}

In the cellular space-time picture,
the `absolute' coordinate $r$ labels the kinetic space-time dependence
$O(\Delta r_{kin}$),
whereas the `relative' coordinate $s$ measures the quantum space-time distance
$O(\Delta r_{qua}$).
\bigskip

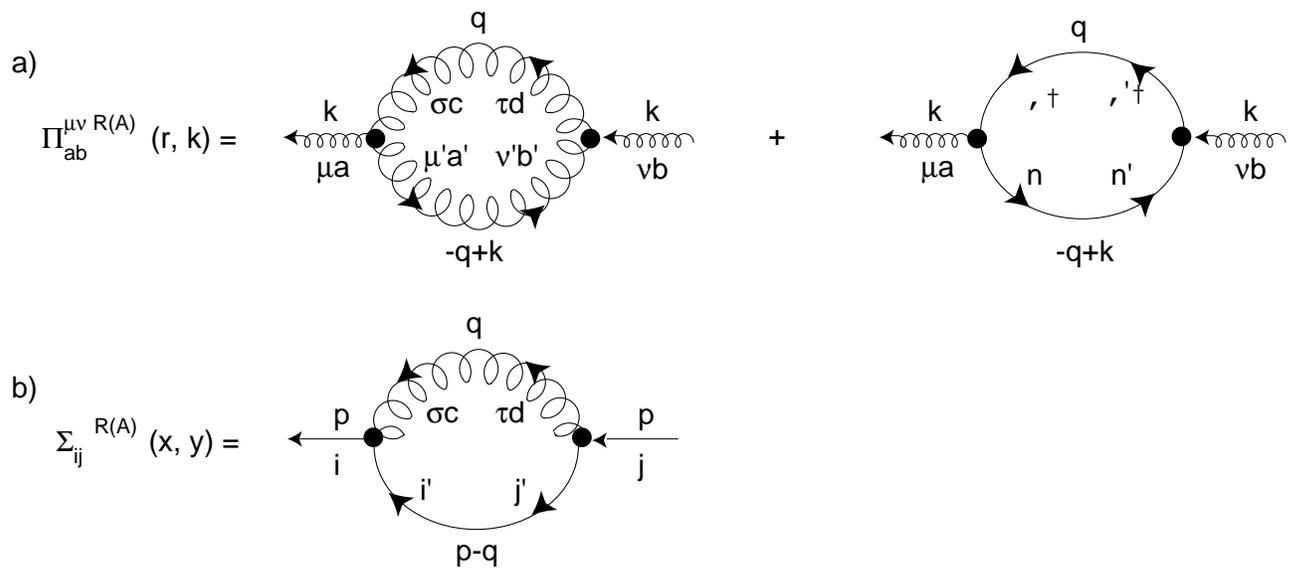
\noindent {\bf Figure 9:}

The `radiative' self-energies in one-loop approximation, eqs. (\ref{Pi1l}):
{\bf a)} the retarded (advanced) gluon self-energies $\Pi_{\mu\nu}^{R(A)}$,
{\bf b)} the retarded (advanced) quark self-energies $\Sigma^{R(A)}$.
\bigskip

\noindent {\bf Figure 10:}

The `collisional' self-energies in two-loop approximation, eqs. (\ref{Pi2}) and
(\ref{Sigma2}):
{\bf a)} the contributions to the gluon correlation functions
$\Pi_{\mu\nu}^{\gl}$,
{\bf b)} the contributions to the quark correlation functions $\Sigma^{\gl}$.
\bigskip

\noindent {\bf Figure 11:}

Cutting the `collisional' two-loop self-energies, gives
the  different binary $2\rightarrow 2$ collision processes,
namely {\bf a)} the gluon terms
$gg\leftrightarrow gg$,
$gq\leftrightarrow g q$,
$gg\leftrightarrow q\bar q$,
and {\bf b)} the quark terms,
$qg\leftrightarrow q g$,
$q\bar q\leftrightarrow q \bar q$,
$qq\leftrightarrow q q$.
\bigskip

\newpage

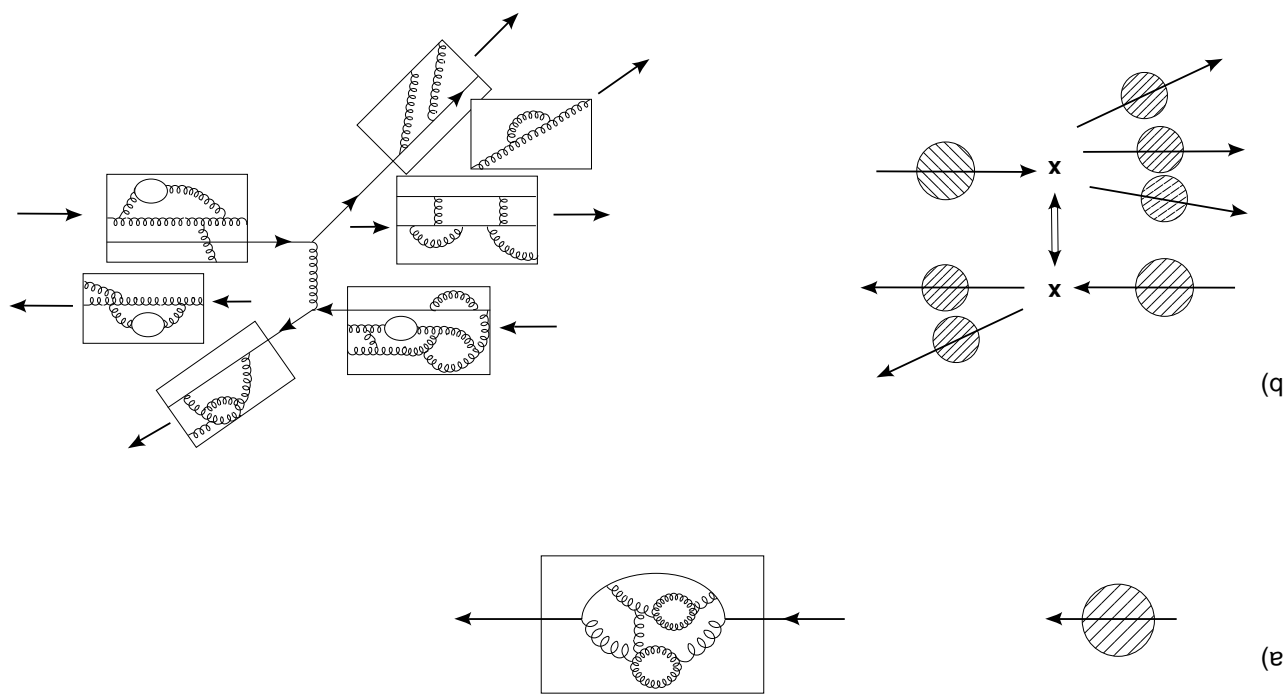
\noindent {\bf Figure 12:}

Illustration of the `hard scattering picture',
for the evolution of a multi-parton
system on the basis of the coupled renormalization and transport equations:

{\bf a)}
A dressed parton is described  as a quasi-particle with a dynamical
substructure,
corresponding to an instantanous state consisting of
a number of bare gluons and quarks (its radiative cloud).
These the underlying quantum fluctuations are embodied in
the spectral densities, or parton structure functions, which
are determined by the renormalization equations.

{\bf b)}
A binary collision between two dressed partons is
described as a statistically occurring `hard scattering', determined
by the local density of dressed partons, and convoluted with their
spectral densities at the `hard scattering scale'
of the order of the momentum transfer.
This is described by the transport equations.

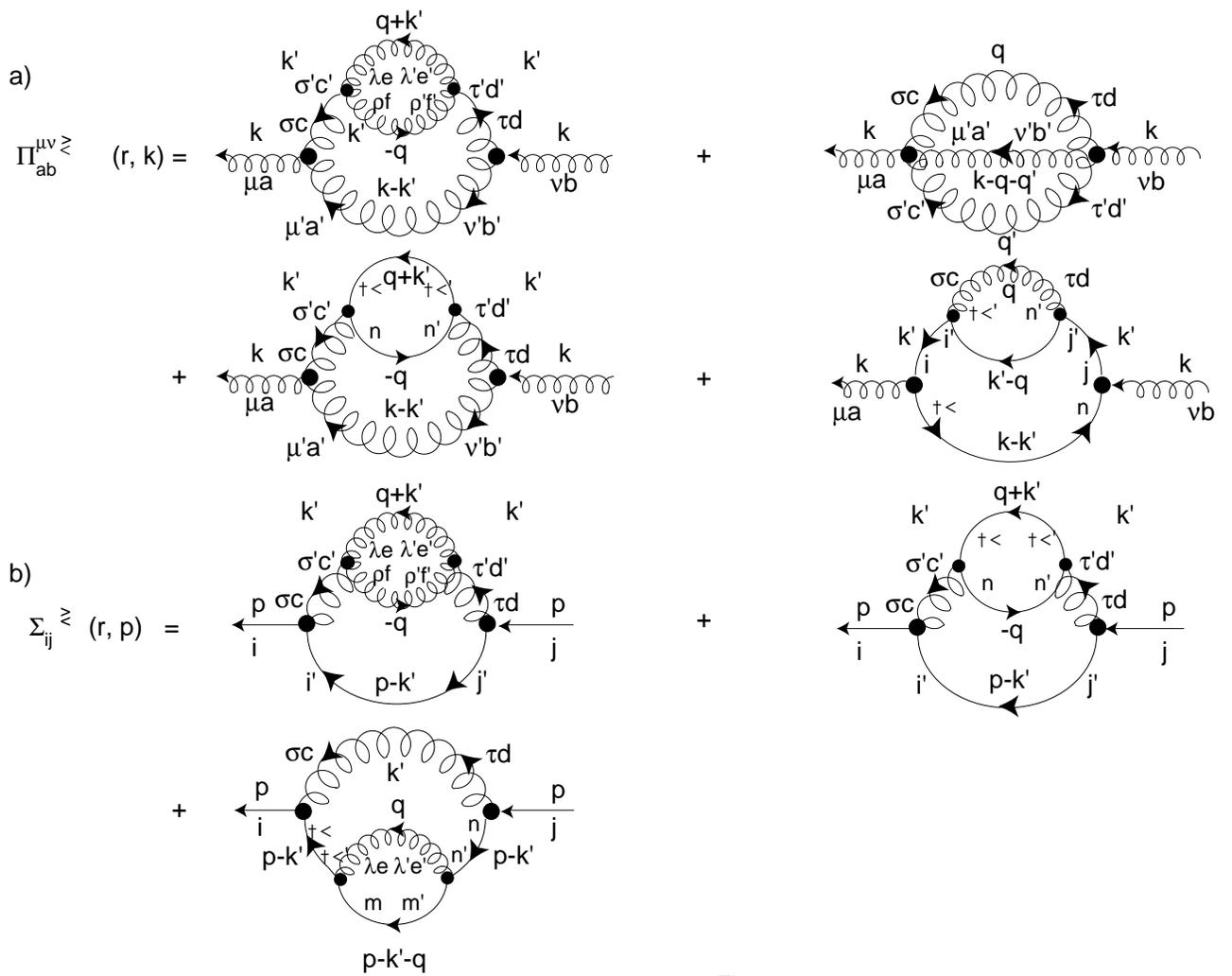
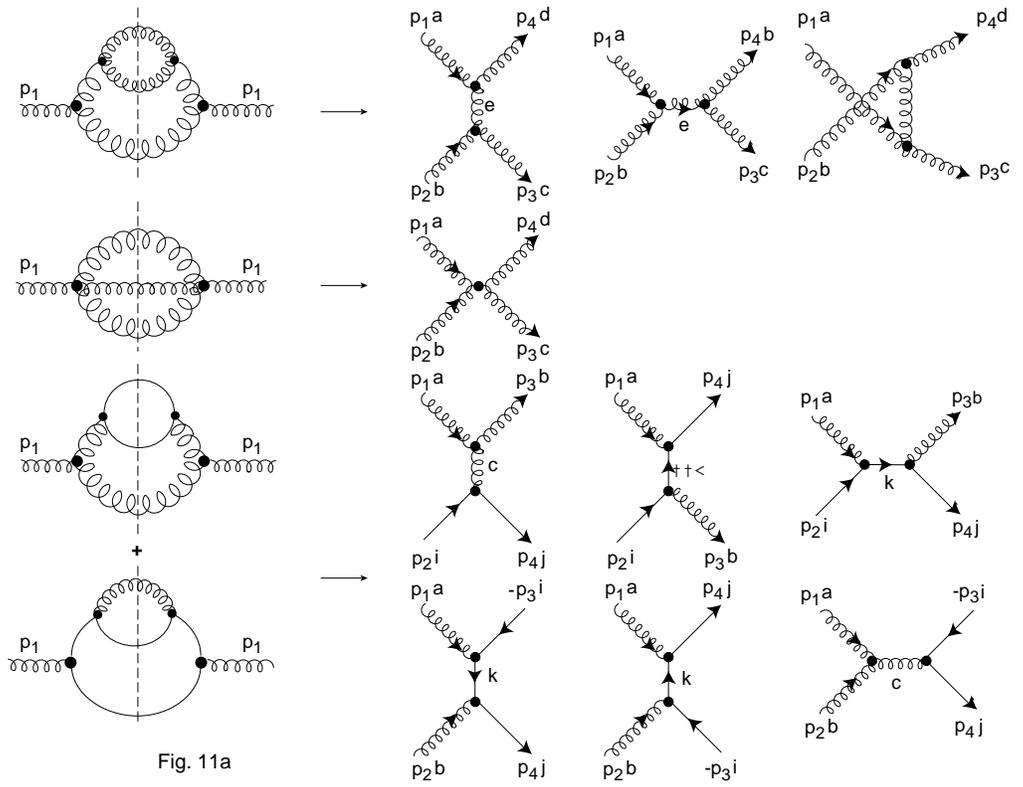
\end{document}